\documentclass[reprint,showpacs,groupedaddress,amsfonts,amsmath,amssymb,aps,prd]{revtex4}
\usepackage{graphicx}
\makeatletter
\newcommand*\bigcdot{\mathpalette\bigcdot@{.5}}
\newcommand*\bigcdot@[2]{\mathbin{\vcenter{\hbox{\scalebox{#2}{$\m@th#1\bullet$}}}}}
\makeatother
\usepackage{latexsym}
\usepackage{amssymb}
\usepackage{mathrsfs}
\usepackage{amsmath}
\usepackage{amsthm}
\usepackage{color}
\usepackage{dcolumn}
\usepackage{bm}

\definecolor{dyellow}{rgb}{1.,0.8,.0}
\definecolor{myblue}{rgb}{.1,.1,.7}
\definecolor{dcyan}{rgb}{.0,.6,.6}
\definecolor{dmagenta}{rgb}{0.6,0.0,0.6}
\definecolor{brown}{rgb}{0.6,0.2,0.}
\definecolor{darkblue}{rgb}{.0,.0,0.5}
\definecolor{darkred}{rgb}{0.75,0.0,0.0}
\definecolor{orange}{rgb}{1.,.6,.0}
\definecolor{dorange}{rgb}{0.8,.4,.0}
\definecolor{darkgreen}{rgb}{0.0,0.6,0.0}
\definecolor{purple}{rgb}{.4,.0,.4}
\definecolor{grey}{rgb}{0.5,0.5,0.5}

\begin{document}
\hyphenpenalty=1000
\preprint{APS/123-QED}
\title{Analytical analysis on the orbits of Taiji spacecrafts to infinite order of the orbital eccentricity}

\newcommand*{\PKU}{School of Physical Sciences, University of Chinese Academy of Sciences, Beijing 100049, China}\affiliation{\PKU}
\newcommand*{\INFN}{Institute of High Energy Physics,
Chinese Academy of Sciences, Beijing, 100049, China}\affiliation{\INFN}
\newcommand*{\CICQM}{CAS Center for Excellence in Particle Physics, Beijing 100049, China}\affiliation{\CICQM}
\newcommand*{\CHEP}{}\affiliation{\CHEP}

\author{Bofeng Wu}\email{bofengw@pku.edu.cn}\affiliation{\PKU}
\author{Chao-Guang Huang}\email{huangcg@ihep.ac.cn}\affiliation{\INFN,\PKU}
\author{Cong-Feng Qiao}\email{ qiaocf@ucas.ac.cn}\affiliation{\PKU,\CICQM}


\begin{abstract}
The dual configuration of the original one is proposed for the orbit design of Taiji spacecrafts. In terms of these two configurations of Taiji, an algorithm is devised to expand the unperturbed Keplerian orbits of spacecrafts to infinite order of $e$, the orbital eccentricity, in the heliocentric coordinate system. Further, based on the algorithm, all the kinematic indicators of Taiji triangles, say three arm-lengths and their corresponding rates of change, and three vertex angles, in both configurations are also be expanded to infinite order of $e$, and
it is proved that both configurations of Taiji possess the same symmetry: At every order, three components of every kinematic indicator of Taiji triangle are identical to each other up to a phase shift of $2\pi/3$, which is independent on the tilt angle of Taiji plane relative to the ecliptic plane. Finally, the above algorithm is  slightly modified, and with it, by adjusting the tilt angle around $\pi/3$ to any order of $e$, the orbits of Taiji spacecrafts in each configuration can be optimized.
\end{abstract}
\maketitle
\section{Introduction}
Because of seismic noise, the ground-based detectors like LIGO and
Virgo~\cite{TheLIGOScientific:2016agk,TheLIGOScientific:2016src,TheLIGOScientific:2016htt,GBM:2017lvd} are powerless to detect the low frequency gravitational waves (GWs) below $0.1$ Hz~\cite{Danzmann:1997hm,Adhikari:2013kya,Harms:2013raa}, which is why the space-based GW detectors like LISA~\cite{Dhurandhar:2004rv,Nayak:2006zm} or later Taiji~\cite{xuefei2011,Gong:2014mca,Hu:2017mde,Wu:2018clg} have been given serious consideration. Taiji takes a similar formation as LISA. Three spacecrafts (SCs) orbit the Sun and form an equilateral triangle with the arm-length about $3\times10^6$ km, and by using coherent laser beams exchanged between SCs, Taiji observes GWs covering the range from $0.1$ mHz to $1.0$ Hz.

The configuration in Refs.~\cite{Dhurandhar:2004rv,Nayak:2006zm,Dhurandhar:2008yu,Pucacco:2010mn} designed for LISA was used as one part of the prestudy of Taiji in our previous paper~\cite{Wu:2019thj}, where the relationship between the inclination $\varepsilon$ of the orbits of SCs with respect to the ecliptic plane and the orbital eccentricity $e$ is the key content. In this paper, we find that there exists the dual relationship between $\varepsilon$ and $e$, and propose a new configuration for the orbits of SCs. In these two configurations of Taiji, the orbits of SC$\kappa\ (\kappa=1,2,3)$ at every order are symmetric about either $z$ axis or $x$-$y$ plane in the heliocentric coordinate system, which embodies the duality between them. Taiji can follow or precede the Earth by $\pi/9$ from the viewpoint of the Sun, and in each case, Taiji has two choices for orbits of SCs, which are symmetric about the ecliptic plane. Therefore, in practice, the new configuration provides four new feasible orbit designs for Taiji.

As shown in Refs.~\cite{Dhurandhar:2004rv,Nayak:2006zm,Wu:2019thj}, the fundamental treatment for analytically analysing the orbits of SCs is to expand them in the small orbital eccentricity $e$. For the original configuration of Taiji, the unperturbed Keplerian orbits of SCs have been expanded to $e^3$ order, which is sufficient for analysing the main perturbative effect of the Earth on SCs~\cite{Wu:2019thj}. However, in the future, the actual operation of Taiji probably requires to further consider the post-Newtonian effects of the Sun's gravitational field and the perturbative effects of the other celestial bodies like Jupiter and the Moon etc., where these effects on SCs are much weaker than that of the Earth. If one wants to discuss these effects, the unperturbed Keplerian orbits of SCs need to be expanded to higher order. For both configurations of Taiji, we devise an algorithm, in the present paper, to expand the unperturbed Keplerian orbits of SCs to infinite order of $e$ in the heliocentric coordinate system. When the above effects on SCs are considered, the unperturbed Keplerian orbits of SCs should be truncated to necessary order and then be viewed as the zeroth-order approximation of the corresponding perturbative solution. For example, as shown in our previous paper~\cite{Wu:2019thj}, because the magnitude of the perturbative parameter of the Earth is about $7.258\times10^{-5}\sim e^2$, the unperturbed Keplerian orbits of Taiji SCs are only truncated to $e^0$ order, and then, by viewing them as the zeroth-order solution, the leading order ($e^2$ order) perturbative solution of the Earth on SCs is obtained. Therefore, the algorithm lays the foundation for discussion of relativistic and perturbative effects on Taiji.

Next, based on above algorithm, all the kinematic indicators of Taiji triangles in both configurations are also expanded to infinite order of $e$, where as a preliminary example, the expressions of arm-lengths and their rates of change to $e^5$ order and the expression of vertex angles to $e^4$ order are presented when $\phi^{\pm}=\pi/3$ with $\phi^{+}$ and $\phi^{-}$, shown in FIG.~\ref{fig1}, as the tilt angles of Taiji planes relative to the ecliptic plane at $t=0$ in both configurations, respectively. These results show that two Taiji triangles have the following feature: \emph{Their shapes depend on $\phi^{\pm}$, and when $\phi^{\pm}=\pi/3$, they are equilateral up to the leading order terms of all the kinematic indicators, where to the higher order terms,
they undergo the inherent variations.} Like LISA, the instability of Taiji formation
may lower its sensitivity~\cite{Dhurandhar:2008yu}, so the inherent variation of Taiji triangle is significant in the data analysis, e.g., the inherent variations of arm-lengths need to be deducted so as to acquire their accurate variations induced by GWs. It is according to the above algorithm that an accurate knowledge of the inherent variations of Taiji triangles in both configurations can be obtained. Note that the word ``inherent'' here denotes the variation of Taiji triangle only induced by the Sun in the Newtonian framework. As mentioned above, the relativistic effect of the Sun's gravitational field and the perturbative effects of some celestial bodies may need to be taken into account in the future, which results in the variation of Taiji triangle as well. In this paper, we only focus our attention on the inherent variation of Taiji triangle, and the other part will be left in the future task. In Ref.~\cite{Rubbo:2003ap},
a special model is studied for the spaced-based GW detector in triangular configuration, where the inclination $\varepsilon$ of the orbits of SCs with respect to the ecliptic plane is assumed to be $\sqrt{3}e$, and three arm-lengths in this model are identical to each other up to a phase shift of $2\pi/3$ up to $e^1$ order. We will generalize this conclusion for both configurations of Taiji in the present paper, and with the help of the above algorithm, it is proved that both configurations possess the same symmetry: \emph{At every order, three components of every kinematic indicator of Taiji triangle are identical to each other up to a phase shift of $2\pi/3$, which is independent on the tilt angle of Taiji plane relative to the ecliptic plane}.

Like LISA~\cite{Nayak:2006zm}, Taiji also needs to suppress the laser frequency noise by time-delay interferometry (TDI). The instability of Taiji triangle may result in that the first generation TDI works unsuccessfully, since it is only applicable for the stationary configuration. One way to deal with this difficulty is to turn to modified first generation TDI or further, the second generation TDI~\cite{Dhurandhar:2008yu,Tinto:2003vj,Vallisneri:2005ji,Tinto:2014lxa}. The application of the second generation TDI involves the complex non-commuting time-delay operators, which could possibly cause difficulty in the data analysis~\cite{Dhurandhar:2008yu}, and therefore, as the case of original LISA
(presented in Refs.~\cite{Dhurandhar:2004rv,Nayak:2006zm,Dhurandhar:2008yu}), the reasonably optimized model of Taiji could contribute to selecting a simpler TDI technique. What needs to be pointed out is that because the orbital eccentricity of Taiji SCs is smaller than that of the original LISA SCs, the more stable formation of Taiji triangle means that Taiji has more chance than original LISA to consider a simpler TDI strategy by the optimization of orbits of SCs. Moreover, optimizing the orbits of SCs also helps to reduce the adverse effect brought about by the Doppler shift of the laser frequency. For the original configuration of Taiji~\cite{Wu:2019thj}, by adjusting the tilt angle $\phi^{+}$ around $\pi/3$ at $e^1$ order,  the orbits of SCs are optimized at the next leading orders of all the kinematic indicators. This result can be generalized, in this paper, by slightly modifying the above algorithm, i.e., by adjusting $\phi^{\pm}$ around $\pi/3$ to any order of $e$,  the orbits of SCs in both configurations of Taiji can be optimized, respectively, and that is to say, Taiji triangles can become as stable as possible with the different specific problem involved. As a preliminary example, the results of optimizing all the kinematic indicators in both configurations by adjusting $\phi^{\pm}$ around $\pi/3$ to $e^4$ order are provided in the present paper. In the future, if the post-Newtonian effects of the Sun's gravitational field and the perturbative effects of some celestial bodies are considered, the above algorithm can be readily generalized so that the more stable formation of Taiji can be obtained.

The paper is arranged as follows. In the next section, the new configuration of Taiji is designed. Both configurations of Taiji to infinite order of $e$ are analysed in Sec.~III.  In Sec.~IV, we shall make some concluding remarks. No summation is taken for repeated indices in the present paper.
\section{New Configuration Of Taiji}
We will discuss the orbit design of SCs in the heliocentric coordinate system $(x,y,z)$, which is defined as
the right-handed Cartesian coordinates with the center of mass of the Sun as the origin and the ecliptic plane as the $x$-$y$ plane. Consider the ellipse in the $x$-$y$ plane,
\begin{figure}
\centering
\includegraphics[scale=0.4]{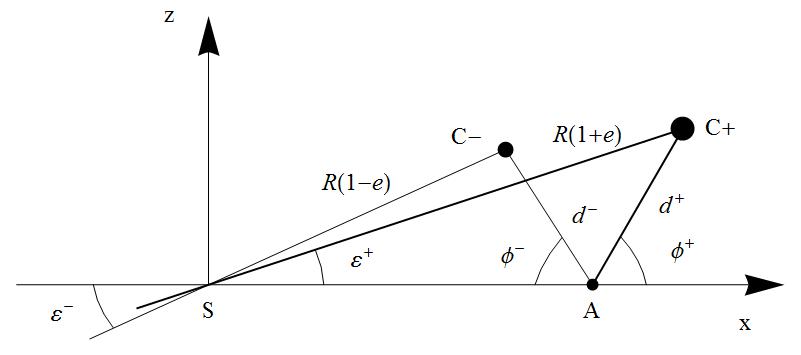}
\caption{\label{fig:1} Plot of the geometry of Taiji in both configurations.
S denotes the Sun, and the coordinates of A are $(R,0,0)$.
C+ and C$-$ denote SC1 at $t=0$ in the original configuration and in the new configuration, respectively.}
\label{fig1}
\end{figure}
\begin{equation}\label{equ1}
\frac{x^2}{R^2}+\frac{y^2}{(R\sqrt{1-e^2}\ )^2}=1,
\end{equation}
where its semimajor axis $R$ is equal to that of the Earth's orbit, and its eccentricity is $e$. Translating this ellipse $eR$ in the positive direction and in the negative direction along the $x$ axis gives two ellipses,
and then, rotating them by $\varepsilon^{\pm}$ about $y$ axis, respectively, provides two choices for the orbit of SC{$1$}. Thus, the obtained two radial vectors of SC{$1$} are $\boldsymbol{r}_{1}^{\pm}=(x_1^{\pm},y_1^{\pm},z_1^{\pm})$ with
\begin{equation}\label{equ2}
\left\{\begin{array}{lll}
\displaystyle x_{1}^{\pm}&=&\displaystyle R\left(\cos\psi_{1}^{\pm}\pm e\right)\cos\varepsilon^{\pm},\smallskip\\
\displaystyle y_{1}^{\pm}&=&\displaystyle R\sqrt{1-e^2}\sin\psi_{1}^{\pm},\smallskip\\
\displaystyle z_{1}^{\pm}&=&\displaystyle R\left(\cos\psi_{1}^{\pm}\pm e\right)\sin\varepsilon^{\pm}.
\end{array}\right.
\end{equation}
If the eccentric anomaly $\psi_{1}^{+}$ satisfies the Kepler's equation
\begin{equation}\label{equ3}
\psi_{1}^{+}+e\sin\psi_{1}^{+}=\Omega t
\end{equation}
with $\Omega$ as the average angular velocity of SC1, $\boldsymbol{r}_{1}^{+}$ manifestly represents the orbit of SC{$1$} in the original configuration~\cite{Wu:2019thj}, and in this case, SC1 is at the aphelion when $t=0$, namely, the point C+ presented in FIG.~\ref{fig1}. From $\triangle$SAC+,
\begin{equation}\label{equ4}
d^{+}=R\left(\sqrt{e^2+2e+\cos^2\phi^{+}}-\cos\phi^{+}\right).
\end{equation}
Defining dimensionless parameter $\alpha^{+}:=\sqrt{3}d^{+}/(2R)$, and then, the relationship between
inclination $\varepsilon^{+}$ of the orbit of SC1 with respect to the ecliptic plane and orbital eccentricity $e$ can be expressed as
\begin{equation}\label{equ5}
\left\{\begin{array}{lll}
\displaystyle \cos\varepsilon^{+}&=&\displaystyle\frac{\sqrt{3}}3 \frac{\sqrt{3}+2\alpha^{+}\cos{\phi^{+}}}{1+e},\smallskip\\
\displaystyle \sin\varepsilon^{+}&=&\displaystyle\frac{\sqrt{3}}3\frac{2\alpha^{+} \sin{\phi^{+}}}{1+e}.
\end{array}\right.
\end{equation}
Eq.~(\ref{equ5}) plays a key role in the expansion of the Keplerian orbits of
SCs, and it is the core content of the original configuration of Taiji.

We propose a new configuration of Taiji, in which, the orbit of SC1 is represented by $\boldsymbol{r}_{1}^{-}$, and the corresponding eccentric anomaly $\psi_{1}^{-}$ satisfies the Kepler's equation
\begin{equation}\label{equ6}
\psi_{1}^{-}-e\sin\psi_{1}^{-}=\Omega t,
\end{equation}
which shows that SC1 is at the perihelion when $t=0$, namely, the point C$-$ presented in FIG.~\ref{fig1}.
Similarly, $\triangle$SAC$-$ gives
\begin{equation}\label{equ7}
d^{-}=R\left(-\sqrt{e^2-2e+\cos^2\phi^{-}}+\cos\phi^{-}\right),
\end{equation}
and then, with the help of the dimensionless parameter $\alpha^{-}:=\sqrt{3}d^{-}/(2R)$, the relationship between
inclination $\varepsilon^{-}$ and orbital eccentricity $e$ can be derived easily,
\begin{equation}\label{equ8}
\left\{\begin{array}{lll}
\displaystyle \cos\varepsilon^{-}&=&\displaystyle\frac{\sqrt{3}}3 \frac{\sqrt{3}-2\alpha^{-}\cos{\phi^{-}}}{1-e},\smallskip\\
\displaystyle \sin\varepsilon^{-}&=&\displaystyle\frac{\sqrt{3}}3\frac{2\alpha^{-} \sin{\phi^{-}}}{1-e},
\end{array}\right.
\end{equation}
which is also the core content of the new configuration of Taiji. According to the above derivation,
we declare that the relationships between inclinations $\varepsilon^{\pm}$ and orbital eccentricity $e$ are dual for both configurations of Taiji.

For the original configuration of Taiji, rotating the orbit of SC1 by $2\pi/3$ and $4\pi/3$ about the $z$ axis, respectively, gives those of SC2 and SC3, where their phases need to be adjusted correspondingly~\cite{Dhurandhar:2004rv,Nayak:2006zm}. The expressions of $\boldsymbol{r}_{2}^{+}$ and $\boldsymbol{r}_{3}^{+}$, representing the orbits of SC2 and SC3, respectively, have been obtained in our previous paper~\cite{Wu:2019thj}. As for the new configuration of Taiji, above method can also be applied to derive the orbits of SC2 and SC3, denoted by $\boldsymbol{r}_{2}^{-}$ and $\boldsymbol{r}_{3}^{-}$, respectively. Here, we directly present the results together for both configurations: $\boldsymbol{r}_{2}^{\pm}=(x_2^{\pm},y_2^{\pm},z_2^{\pm})$ with
\begin{equation}\label{equ9}
\left\{\begin{array}{lll}
\displaystyle x_{2}^{\pm}&=&\displaystyle R\left(\cos\psi_{2}^{\pm}\pm e\right)\cos\varepsilon^{\pm}\cos{\frac{2\pi}{3}}
-R\sqrt{1-e^2}\sin\psi_{2}^{\pm}\sin{\frac{2\pi}{3}},\smallskip\\
\displaystyle y_{2}^{\pm}&=&\displaystyle R\left(\cos\psi_{2}^{\pm}\pm e\right)\cos\varepsilon^{\pm}\sin{\frac{2\pi}{3}}
+R\sqrt{1-e^2}\sin\psi_{2}^{\pm}\cos{\frac{2\pi}{3}},\smallskip\\
\displaystyle z_{2}^{\pm}&=&\displaystyle R\left(\cos\psi_{2}^{\pm}\pm e\right)\sin\varepsilon^{\pm}
\end{array}\right.
\end{equation}
and $\boldsymbol{r}_{3}^{\pm}=(x_3^{\pm},y_3^{\pm},z_3^{\pm})$ with
\begin{equation}\label{equ10}
\left\{\begin{array}{lll}
\displaystyle x_{3}^{\pm}&=&\displaystyle R\left(\cos\psi_{3}^{\pm}\pm e\right)\cos\varepsilon^{\pm}\cos{\frac{4\pi}{3}}
-R\sqrt{1-e^2}\sin\psi_{3}^{\pm}\sin{\frac{4\pi}{3}},\smallskip\\
\displaystyle y_{3}^{\pm}&=&\displaystyle R\left(\cos\psi_{3}^{\pm}\pm e\right)\cos\varepsilon^{\pm}\sin{\frac{4\pi}{3}}
+R\sqrt{1-e^2}\sin\psi_{3}^{\pm}\cos{\frac{4\pi}{3}},\smallskip\\
\displaystyle z_{3}^{\pm}&=&\displaystyle R\left(\cos\psi_{3}^{\pm}\pm e\right)\sin\varepsilon^{\pm},
\end{array}\right.
\end{equation}
where their corresponding eccentric anomalies $\psi_{\kappa}^{\pm}\ (\kappa=2,3)$ satisfy
\begin{eqnarray}
\label{equ11}\psi_{\kappa}^{\pm}\pm e\sin\psi_{\kappa}^{\pm}=\sigma_{\kappa}:=\Omega t-(\kappa-1)\frac{2\pi}{3}.
\end{eqnarray}
Obviously, Eq.~(\ref{equ11}) holds for the case of $\kappa=1$ as well.

As illustrated in FIG.~\ref{fig1}, $\phi^{\pm}$ are the tilt angles of Taiji planes relative to the ecliptic plane at $t=0$ in both configurations, respectively. For the original configuration~\cite{Wu:2019thj}, $\phi^{+}$ can take values of $\pm|\phi^{+}|$, and thus, there are two choices for orbits of SCs, which are symmetry about the ecliptic plane. Further, seeing that Taiji can follow or precede the Earth by $\pi/9$ from the viewpoint of the Sun, the original configuration, in fact, provides four feasible orbit designs for Taiji. Similarly, for the new configuration, two values $\pm|\phi^{-}|$ of $\phi^{-}$ can provide another four feasible orbit designs for Taiji, and consequently, eight kinds of potential orbit schemes are available for Taiji SCs. In the following, in order to highlight the main content about these two configurations of Taiji and simplify the related derivations, we set $\phi^{\pm}>0$.
\section{Orbit Analysis On Both Configurations Of Taiji To Infinite order of $e$}
In this section, for both configurations of Taiji, we devise an algorithm to expand the unperturbed Keplerian orbits of SCs and all the kinematic indicators of Taiji triangles to infinite order of $e$ in the heliocentric coordinate system,
which lays the foundation for further discussing relativistic and perturbative effects on SCs, and provides an accurate knowledge of the inherent variations of Taiji triangles. One of the most significant application of these results is that they contribute to acquiring the accurate variations of arm-lengths of Taiji triangle induced by GWs in the data analysis by deducting the inherent counterparts. Moreover, by the way, based on the algorithm, we prove that both configurations of Taiji possess the same symmetry: \emph{At every order, three components of every kinematic indicator of Taiji triangle are identical to each other up to a phase shift of $2\pi/3$, which is independent on the tilt angle of Taiji plane relative to the ecliptic plane.} Finally, by slightly modifying the above algorithm,
the orbits of SCs in each configuration of Taiji are optimized by adjusting the tilt angle of Taiji plane relative to the ecliptic plane around $\pi/3$ to any order of $e$, which helps to consider a simpler TDI strategy and reduce the adverse effect brought about by the Doppler shift of the laser frequency.

\subsection{Expansions of the orbits of Taiji SCs}

The general idea is originated from the fact that the Kepler's Eq.~(\ref{equ11}) can be expanded to infinite order of $e$ when $e\approx5.789\times10^{-3}\ll1$ for Taiji~\cite{Wu:2019thj} according to the method of Lagrange~\cite{moulton1960}, and then, the combination of Eqs.~(\ref{equ2}), (\ref{equ4}), (\ref{equ5}), and (\ref{equ7})---(\ref{equ10}) can bring about the expansions of the unperturbed Keplerian orbits of SC$\kappa\ (\kappa=1,2,3)$, denoted by $\boldsymbol{r}_{\kappa}^{\pm}=(x_{\kappa}^{\pm},y_{\kappa}^{\pm},z_{\kappa}^{\pm})$, in both configurations of Taiji to infinite order of $e$.  The detailed derivation is put in Appendix A, and here, we only show the expansions of $x_{\kappa}^{\pm},y_{\kappa}^{\pm},$ and $z_{\kappa}^{\pm}$:
\begin{equation}\label{equ12}
\left\{\begin{array}{lll}
\displaystyle x_{\kappa}^{\pm}&=&\displaystyle R\cos(\Omega t)+R\sum_{n=1}^{\infty}(\mp1)^n Q\left(x_{\kappa}^{\pm},n\right) e^n,\smallskip\\
\displaystyle y_{\kappa}^{\pm}&=&\displaystyle R\sin(\Omega t)+R\sum_{n=1}^{\infty}(\mp1)^n Q\left(y_{\kappa}^{\pm},n\right) e^n,\smallskip\\
\displaystyle z_{\kappa}^{\pm}&=&\displaystyle R\sum_{n=1}^{\infty}(\mp1)^{n-1} Q\left(z_{\kappa}^{\pm},n\right)e^n,
\end{array}\right.
\end{equation}
where
\begin{equation}\label{equ13}
\left\{\begin{array}{lll}
\displaystyle Q\left(x_{\kappa}^{\pm},n\right)&:=&\displaystyle C_{h}^{\pm}(n)\cos\rho_{\kappa}+\sum_{\substack{k=0\\ k\neq1}}^{n}\sum_{j=0}^{\left[\frac{n-k}{2}\right]}\Big(f_s^{\pm}(n,k,j)\cos\left((n-k+1-2j)\sigma_{\kappa}-\rho_{\kappa}\right)\smallskip\\
\displaystyle&&\displaystyle\qquad\qquad\qquad\qquad\qquad+g_s^{\pm}(n,k,j)\cos\left((n-k+1-2j)\sigma_{\kappa}+\rho_{\kappa}\right)\Big),\smallskip\\
\displaystyle Q\left(y_{\kappa}^{\pm},n\right)&:=&\displaystyle C_{h}^{\pm}(n)\sin\rho_{\kappa}-\sum_{\substack{k=0\\ k\neq1}}^{n}\sum_{j=0}^{\left[\frac{n-k}{2}\right]}\Big(f_s^{\pm}(n,k,j)\sin\left((n-k+1-2j)\sigma_{\kappa}-\rho_{\kappa}\right)\smallskip\\
\displaystyle&&\displaystyle\qquad\qquad\qquad\qquad\qquad-g_s^{\pm}(n,k,j)\sin\left((n-k+1-2j)\sigma_{\kappa}+\rho_{\kappa}\right)\Big),\smallskip\\
\displaystyle Q\left(z_{\kappa}^{\pm},n\right)&:=&\displaystyle -C_{v}^{\pm}(n)-\sum_{k=1}^{n}\sum_{j=0}^{\left[\frac{n-k}{2}\right]}h_s^{\pm}(n,k,j)\cos\left((n-k+1-2j)\sigma_{\kappa}\right)
\end{array}\right.
\end{equation}
with $\rho_{\kappa}:=(\kappa-1)2\pi/3$, $[(n-k)/2]$ as the integer part of $(n-k)/2$,
\begin{equation}\label{equ14}
\left\{\begin{array}{lll}
\displaystyle C_{h}^{\pm}(n)&:=&\displaystyle \frac{3}{2}(-1)^n Q\left(\cos\varepsilon^{\pm},n-1\right),\smallskip\\
\displaystyle C_{v}^{\pm}(n)&:=&\displaystyle \frac{3}{2}(-1)^n Q\left(\sin\varepsilon^{\pm},n-1\right),
\end{array}\right.
\end{equation}
and
\begin{equation}\label{equ15}
\left\{\begin{array}{lll}
\displaystyle f_s^{\pm}(n,k,j)&:=&\displaystyle \left(C_p^{\pm}(n,k,j)+C_m(n,k,j)\right)C_{n-k+1}^j(n-k+1-2j)^{n-k-1},\smallskip\\
\displaystyle g_s^{\pm}(n,k,j)&:=&\displaystyle \left(C_p^{\pm}(n,k,j)-C_m(n,k,j)\right)C_{n-k+1}^j(n-k+1-2j)^{n-k-1},\smallskip\\
\displaystyle h_s^{\pm}(n,k,j)&:=&\displaystyle C_t^{\pm}(n,k,j)C_{n-k+1}^j(n-k+1-2j)^{n-k-1}.
\end{array}\right.
\end{equation}
Here, $C_{n-k+1}^j$ is the binomial coefficients,
\begin{equation}\label{equ16}
\left\{\begin{array}{lll}
\displaystyle C_p^{\pm}(n,k,j)&:=&\displaystyle \frac{(-1)^{j+k}}{2(2n-2k)!!}Q\left(\cos\varepsilon^{\pm},k\right),\smallskip\\
\displaystyle C_m(n,k,j)&:=&\displaystyle \frac{(-1)^{j}(k-3)!!(n-k+1-2j)}{k!!(2n-2k+2)!!}\cos^2\frac{k\pi}{2},\smallskip\\
\displaystyle C_t^{\pm}(n,k,j)&:=&\displaystyle \frac{(-1)^{j+k}}{(2n-2k)!!}Q\left(\sin\varepsilon^{\pm},k\right),
\end{array}\right.
\end{equation}
and the expressions of $Q(\cos\varepsilon^{\pm},n-1),Q(\sin\varepsilon^{\pm},n-1),Q(\cos\varepsilon^{\pm},k)$, and $Q(\sin\varepsilon^{\pm},k)$ refer to Eq.~(\ref{equA6}). Note that in this paper, $Q(A,\bigcdot)$ represents the mapping related with a quantity $A$, where only $``\bigcdot"$ is the argument, and
$A$ is used to denote the mapping itself. Moreover the following rule is needed to understand Eqs.~(\ref{equ12})---(\ref{equ16}) accurately: The upper (lower) symbol of $``\pm"$ on the left-hand side of one equation corresponds to upper (lower) symbol of $``\pm"$ or $``\mp"$ on its right-hand side, and the rule applies to the full text of this paper. It is easy to check that the expansion of $\boldsymbol{r}_{\kappa}^{+}\ (\kappa=1,2,3)$ to $e^3$ order in Eq~(\ref{equ12}) is the same as that in our previous paper~\cite{Wu:2019thj}.

Obviously, Eq~(\ref{equ12}) shows that at $e^0$ order, the orbits of all SCs in both configurations are the circle in the ecliptic plane with the Sun as center and $R$ as radius, and thus, the trajectories of the barycenters of three SCs at $e^0$ order in these two configurations are also this circle, which is the basis for establishing the Clohessy-Wiltshire system~\cite{Pucacco:2010mn}. The complete expressions for the trajectories of the barycenters of three SCs in both configurations are obtained by Eq.~(\ref{equ12}) in Appendix A, and by using them, one can discuss the actual trailing angle of Taiji constellation following the Earth from the viewpoint of the Sun~\cite{Wu:2019thj}. Moreover, Eq.~(\ref{equ12}) implies that the orbits of SC$\kappa\ (\kappa=1,2,3)$ at every order in both configurations are symmetric about either $z$ axis or $x$-$y$ plane, which embodies the duality between these two configurations. Although Eq.~(\ref{equ12}) is expressed in the form of series, it is
the complete unperturbed Keplerian orbits of SCs. When the post-Newtonian effects of the Sun's gravitational field and the perturbative effects of some celestial bodies are further considered in the future, Eq.~(\ref{equ12}) truncated to the necessary order should be viewed as the zeroth-order approximation of the corresponding perturbative solution. Hence, Eq.~(\ref{equ12}) is the basis for discussion of relativistic and perturbative effects on SCs.

\subsection{Expansions of all the kinematic indicators of Taiji triangles\label{sec3.2}}

All the kinematic indicators of Taiji triangles, say three arm-lengths and their corresponding rates of change, and three vertex angles, depend on the relative radial vectors of SCs, namely, $\boldsymbol{r}_{\mu\nu}^{\pm}:=\boldsymbol{r}_{\mu}^{\pm}-\boldsymbol{r}_{\nu}^{\pm}=(x_{\mu\nu}^{\pm},y_{\mu\nu}^{\pm},z_{\mu\nu}^{\pm})\ (\mu,\nu=1,2,3,\mu\neq \nu)$. By using Eq.~(\ref{equ12}), expanding $\boldsymbol{r}_{\mu\nu}^{\pm}$ is easy, namely,
\begin{equation}\label{equ17}
\left\{\begin{array}{lll}
\displaystyle x_{\mu\nu}^{\pm}&=&\displaystyle R\sum_{n=1}^{\infty}(\mp1)^n Q\left(x_{\mu\nu}^{\pm},n\right)e^n,\smallskip\\
\displaystyle y_{\mu\nu}^{\pm}&=&\displaystyle R\sum_{n=1}^{\infty}(\mp1)^n Q\left(y_{\mu\nu}^{\pm},n\right)e^n,\smallskip\\
\displaystyle z_{\mu\nu}^{\pm}&=&\displaystyle R\sum_{n=1}^{\infty}(\mp1)^{n-1} Q\left(z_{\mu\nu}^{\pm},n\right)e^n
\end{array}\right.
\end{equation}
with
\begin{equation}\label{equ18}
\left\{\begin{array}{lll}
\displaystyle Q\left(x_{\mu\nu}^{\pm},n\right)&:=&\displaystyle Q\left(x_{\mu}^{\pm},n\right)-Q\left(x_{\nu}^{\pm},n\right),\smallskip\\
\displaystyle Q\left(y_{\mu\nu}^{\pm},n\right)&:=&\displaystyle Q\left(y_{\mu}^{\pm},n\right)-Q\left(y_{\nu}^{\pm},n\right),\smallskip\\
\displaystyle Q\left(z_{\mu\nu}^{\pm},n\right)&:=&\displaystyle Q\left(z_{\mu}^{\pm},n\right)-Q\left(z_{\nu}^{\pm},n\right).
\end{array}\right.
\end{equation}
$\boldsymbol{r}_{\mu\nu}^{\pm}$ can be used to define the arm-lengths between SC$\mu$ and SC$\nu$  and their rates of change, respectively,
\begin{eqnarray}
\label{equ19}l_{\mu\nu}^{\pm}:=\sqrt{\left(\boldsymbol{r}_{\mu\nu}^{\pm}\right)^2},\qquad v_{\mu\nu}^{\pm}:=\frac{d l_{\mu\nu}^{\pm}}{dt},
\end{eqnarray}
which shows that it is necessary to first deal with
\begin{eqnarray}
\label{equ20}\left(\boldsymbol{r}_{\mu\nu}^{\pm}\right)^2=\left(x_{\mu\nu}^{\pm}\right)^2+\left(y_{\mu\nu}^{\pm}\right)^2+\left(z_{\mu\nu}^{\pm}\right)^2.
\end{eqnarray}
Expansions of $(\boldsymbol{r}_{\mu\nu}^{\pm})^2$ are readily derived with above Eqs.~(\ref{equ17}) and (\ref{equ18}):
\begin{eqnarray}
\label{equ21}\left(\boldsymbol{r}_{\mu\nu}^{\pm}\right)^2=R^2\sum_{n=2}^{\infty}(\mp1)^n Q\left(\left(\boldsymbol{r}_{\mu\nu}^{\pm}\right)^2,n\right)e^n,
\end{eqnarray}
where
\begin{eqnarray}
\label{equ22}Q\left(\big(\boldsymbol{r}_{\mu\nu}^{\pm}\big)^2,n\right):=\sum_{k=1}^{n-1}\Big(Q\big(x_{\mu\nu}^{\pm},n-k\big)Q\big(x_{\mu\nu}^{\pm},k\big)
+Q\big(y_{\mu\nu}^{\pm},n-k\big)Q\big(y_{\mu\nu}^{\pm},k\big)+Q\big(z_{\mu\nu}^{\pm},n-k\big)Q\big(z_{\mu\nu}^{\pm},k\big)\Big).
\end{eqnarray}
$Q((\boldsymbol{r}_{\mu\nu}^{\pm})^2,n)$ are clearly rewritten as the functions of $t$ in Eq.~(\ref{equB1}), from which,
one can find that they possess the following symmetry:
\begin{equation}\label{equ23}
Q\left(\big(\boldsymbol{r}_{\mu\nu}^{\pm}\big)^2,n\right)=F\left(\theta_{\mu\nu}(\Omega t)\right)\qquad \text{with}\quad \theta_{\mu\nu}(\Omega t):=
\left\{\begin{array}{lll}
\displaystyle \Omega t-\frac{\pi}{3},  &\displaystyle\quad\text{for} &\displaystyle\quad\{\mu,\nu\}=\{1,2\},\smallskip\\
\displaystyle \Omega t-\pi,            &\displaystyle\quad\text{for} &\displaystyle\quad\{\mu,\nu\}=\{2,3\},\smallskip\\
\displaystyle \Omega t-\frac{5\pi}{3}, &\displaystyle\quad\text{for}
&\displaystyle\quad\{\mu,\nu\}=\{3,1\},
\end{array}\right.
\end{equation}
where $F$ is the corresponding function of a single variable. The proof is easy. From the definitions of $\boldsymbol{r}_{\mu\nu}^{\pm}$ and Eq.~(\ref{equ21}),
$Q((\boldsymbol{r}_{\mu\nu}^{\pm})^2,n)=Q((\boldsymbol{r}_{\nu\mu}^{\pm})^2,n)$, which means that one only needs to consider $(\mu,\nu)\in\{(1,2),(2,3),(3,1)\}$ in Eq.~(\ref{equ23}). The first term in Eq.~(\ref{equB1}) keeps the same when $(\mu,\nu)$ takes the above three groups of values, and every remaining term contains $\mathcal{C}^{F}_{\mu\nu}(\Omega t;\epsilon,\eta)$ or $\mathcal{S}^{F}_{\mu\nu}(\Omega t;\epsilon,\eta)$ whose
expressions are shown in Eq.~(\ref{equB2}), where both $\epsilon$ and $\eta$ are integers, and $\epsilon-\eta$ is even. A direct calculation gives
\begin{eqnarray*}
\mathcal{C}^{F}_{\mu\nu}(\Omega t;\epsilon,\eta)&=&
\cos\left(-\epsilon\frac{\pi}{3}\right)\cos\left(\eta\theta_{\mu\nu}(\Omega t)\right),\\
\mathcal{S}^{F}_{\mu\nu}(\Omega t;\epsilon,\eta)&=&
\sin\left(-\epsilon\frac{\pi}{3}\right)\sin\left(\eta\theta_{\mu\nu}(\Omega t)\right),
\end{eqnarray*}
when $(\mu,\nu)\in\{(1,2),(2,3),(3,1)\}$, which implies that Eq.~(\ref{equ23}) holds.

Expansions of arm-lengths $l_{\mu\nu}^{\pm}$ and their rates of change $v_{\mu\nu}^{\pm}$ need to resort to the related results in Appendix C. From Eqs.~(\ref{equ19}) and (\ref{equ21}),
\begin{eqnarray}
\label{equ24}l_{\mu\nu}^{\pm}:=R\sqrt{\sum_{n=2}^{\infty}a_{\mu\nu}^{\pm}(n)e^n}\qquad \text{with}\quad a_{\mu\nu}^{\pm}(n):=(\mp1)^n Q\left(\big(\boldsymbol{r}_{\mu\nu}^{\pm}\big)^2,n\right),
\end{eqnarray}
which can also be rewritten as the following forms:
\begin{eqnarray}
\label{equ25}l_{\mu\nu}^{\pm}=eR\sqrt{a_{\mu\nu}^{\pm}(2)}\left(\sum_{p=0}^{\infty}b_{\mu\nu}^{\pm}(p) e^p\right)^{\frac{1}{2}}\qquad \text{with}\quad b_{\mu\nu}^{\pm}(p):=\frac{a_{\mu\nu}^{\pm}(p+2)}{a_{\mu\nu}^{\pm}(2)},
\end{eqnarray}
by $p:=n-2$, and plugging Eq.~(\ref{equB1}) into the definitions of $a_{\mu\nu}^{\pm}(2)$ in Eq.~(\ref{equ24}) gives
\begin{eqnarray}
\label{equ26}a_{\mu\nu}^{\pm}(2)=\frac{15}{2}+\frac{3}{2}\tan^2\phi^{\pm}+\left(\frac{9}{2}-\frac{3}{2}\tan^2\phi^{\pm}\right)\cos\left(2\theta_{\mu\nu}(\Omega t)\right)>0.
\end{eqnarray}
By use of Eqs.~(\ref{equC11})---(\ref{equC13}) and (\ref{equC15}), $(\sum_{p=0}^{\infty}b_{\mu\nu}^{\pm}(p) e^p)^{1/2}$ in Eq.~(\ref{equ25}) can be expanded, namely,
\begin{eqnarray*}
\left(\sum_{p=0}^{\infty}b_{\mu\nu}^{\pm}(p) e^p\right)^{\frac{1}{2}}=\sum_{p=0}^{\infty}\left(\delta_{0p}+\sum_{k=1}^{p}(-1)^{k-1}\frac{(2k-3)!!}{(2k)!!}b_{\mu\nu}^{(k)\pm}(p)\right)e^p
\end{eqnarray*}
with
\begin{equation}\label{equ27}
b_{\mu\nu}^{(k)\pm}(p)=\left\{\begin{array}{l}
\displaystyle b_{\mu\nu}^{\pm}(p),\quad k=1,\smallskip\\
\displaystyle \sum_{j_{k-1}=k-1}^{p-1}\ \sum_{j_{k-2}=k-2}^{j_{k-1}-1}\cdots\sum_{j_{2}=2}^{j_{3}-1}\sum_{j_{1}=1}^{j_{2}-1}
b_{\mu\nu}^{\pm}(p-j_{k-1})b_{\mu\nu}^{\pm}(j_{k-1}-j_{k-2})\cdots b_{\mu\nu}^{\pm}(j_{2}-j_{1})b_{\mu\nu}^{\pm}(j_{1}),\quad k\geq2,
\end{array}\right.
\end{equation}
where $\delta_{0p}$ is Kronecker symbol, $\sum_{k=1}^{0}(\cdots):=0$, and $(-1)!!:=1$.
Substituting above result to Eq.~(\ref{equ25}) by $p\rightarrow p-1$ gives
\begin{eqnarray}
\label{equ28}l_{\mu\nu}^{\pm}=R\sum_{p=1}^{\infty}Q\left(l_{\mu\nu}^{\pm},p\right)e^p
\end{eqnarray}
with
\begin{eqnarray}
\label{equ29}Q\left(l_{\mu\nu}^{\pm},p\right):=\sqrt{a_{\mu\nu}^{\pm}(2)}\left(\delta_{1p}+\sum_{k=1}^{p-1}(-1)^{k-1}\frac{(2k-3)!!}{(2k)!!}
b_{\mu\nu}^{(k)\pm}(p-1)\right).
\end{eqnarray}
Further, the rates of change of $l_{\mu\nu}^{\pm}$ are trivially obtained,
\begin{eqnarray}
\label{equ30}v_{\mu\nu}^{\pm}=R\Omega\sum_{p=1}^{\infty}Q\left(v_{\mu\nu}^{\pm},p\right)e^p\qquad \text{with}\quad Q\left(v_{\mu\nu}^{\pm},p\right)=\frac{1}{\Omega}\frac{d}{dt}Q\left(l_{\mu\nu}^{\pm},p\right).
\end{eqnarray}

With the above results, the inherent variations of arm-lengths and their rates of change
in both configurations of Taiji can be discussed, and then, an important symmetry of them is directly obtained from Eqs.~(\ref{equ23})---(\ref{equ27}): \emph{For each configuration of Taiji, three components of arm-lengths and their rates of change are identical to each other up to a phase shift of $2\pi/3$ at every order, which is independent on the tilt angle of Taiji plane relative to the ecliptic plane}. From Eqs.~(\ref{equ29}) and (\ref{equ30}),
\begin{eqnarray}
\label{equ31}Q\left(l_{\mu\nu}^{\pm},1\right)=\sqrt{a_{\mu\nu}^{\pm}(2)},\qquad Q\left(v_{\mu\nu}^{\pm},1\right)=\frac{1}{\Omega}\frac{d}{dt}\sqrt{a_{\mu\nu}^{\pm}(2)},
\end{eqnarray}
and then, Eq.~(\ref{equ26}) shows that when $\phi^{\pm}=\pi/3$, $Q(l_{\mu\nu}^{\pm},1)=2\sqrt{3}\Leftrightarrow Q(v_{\mu\nu}^{\pm},1)=0$, which means that in this case, Taiji triangles in both configurations are equilateral at the leading order terms of arm-lengths and their rates of change. However, at their higher order terms, further calculations of $Q(l_{\mu\nu}^{\pm},p)$ and $Q(v_{\mu\nu}^{\pm},p)\ (p\geq2)$ show that even $\phi^{\pm}=\pi/3$, Taiji triangles in both configurations still undergo the inherent variations. Like LISA, this instability of Taiji formation 
may lower its sensitivity~\cite{Dhurandhar:2008yu}, which requires that an accurate analysis on the inherent variation of Taiji triangle should be made in the data analysis. According to the above algorithm, one can acquire an accurate knowledge of the inherent variations of arm-lengths and their rates of change in two configurations of Taiji when $\phi^{\pm}=\pi/3$, and as a preliminary example, here, we present the expansions of $l_{\mu\nu}^{\pm}$ and $v_{\mu\nu}^{\pm}$ to $e^5$ order, namely,
\begin{equation}\label{equ32}
\left\{\begin{array}{lll}
\displaystyle l_{\mu\nu}^{\pm}&=&\displaystyle R\Big(2\sqrt{3}e+Q\left(l_{\mu\nu}^{\pm},2\right)e^2+Q\left(l_{\mu\nu}^{\pm},3\right)e^3+Q\left(l_{\mu\nu}^{\pm},4\right)e^4+Q\left(l_{\mu\nu}^{\pm},5\right)e^5\Big),\smallskip\\
\displaystyle v_{\mu\nu}^{\pm}&=&\displaystyle R\Omega\Big(0e+Q\left(v_{\mu\nu}^{\pm},2\right)e^2+Q\left(v_{\mu\nu}^{\pm},3\right)e^3+Q\left(v_{\mu\nu}^{\pm},4\right)e^4+Q\left(v_{\mu\nu}^{\pm},5\right)e^5\Big),
\end{array}\right.
\end{equation}
where from Eqs.~(\ref{equB1}), (\ref{equ24}), (\ref{equ25}), and (\ref{equ27}), Eqs.~(\ref{equ29}) and (\ref{equ30}) give
\begin{eqnarray*}
Q\left(l_{\mu\nu}^{\pm},2\right)&=&\mp\frac{15\sqrt{3}}{8}\mp\frac{15 \sqrt{3}}{16}\cos\theta_{\mu\nu}(\Omega t)\pm\frac{15\sqrt{3}}{8}\cos(2\theta_{\mu\nu}(\Omega t))\mp\frac{\sqrt{3}}{16}\cos(3\theta_{\mu\nu}(\Omega t)),\\
Q\left(l_{\mu\nu}^{\pm},3\right)&=&\frac{5489\sqrt{3}}{1024}+\frac{1095\sqrt{3}}{256}\cos\theta_{\mu\nu}(\Omega t)-\frac{16239 \sqrt{3}}{2048}\cos(2\theta_{\mu\nu}(\Omega t))-\frac{285\sqrt{3}}{512}\cos(3\theta_{\mu\nu}(\Omega t))\\
&&-\frac{441\sqrt{3}}{1024}\cos(4\theta_{\mu\nu}(\Omega t))+\frac{15\sqrt{3}}{512}\cos(5\theta_{\mu\nu}(\Omega t))-\frac{\sqrt{3}}{2048}\cos(6\theta_{\mu\nu}(\Omega t)),\\
Q\left(l_{\mu\nu}^{\pm},4\right)&=&\mp\frac{1656729\sqrt{3}}{65536}\mp\frac{1165791\sqrt{3}}{65536}\cos\theta_{\mu\nu}(\Omega t)\pm\frac{288213 \sqrt{3}}{8192}\cos(2\theta_{\mu\nu}(\Omega t))\\
&&\pm\frac{40947\sqrt{3}}{32768}\cos(3 \theta_{\mu\nu}(\Omega t))\pm\frac{49515\sqrt{3}}{16384}\cos(4\theta_{\mu\nu}(\Omega t))\pm\frac{3999\sqrt{3}}{32768}\cos(5\theta_{\mu\nu}(\Omega t))\\
&&\pm\frac{1635\sqrt{3}}{8192}\cos(6\theta_{\mu\nu}(\Omega t))
\mp\frac{2697\sqrt{3}}{131072}\cos(7\theta_{\mu\nu}(\Omega t))\pm\frac{45\sqrt{3}}{65536}\cos(8\theta_{\mu\nu}(\Omega t))\\
&&\mp\frac{\sqrt{3}}{131072}\cos(9\theta_{\mu\nu}(\Omega t)),
\end{eqnarray*}
\begin{eqnarray*}
Q\left(l_{\mu\nu}^{\pm},5\right)&=&\frac{2318991805\sqrt{3}}{16777216}+\frac{180178337\sqrt{3}}{2097152}\cos\theta_{\mu\nu}(\Omega t)-\frac{754445953 \sqrt{3}}{4194304}\cos(2\theta_{\mu\nu}(\Omega t))\\
&&+\frac{3801087\sqrt{3}}{2097152}\cos(3\theta_{\mu\nu}(\Omega t))-\frac{3237903959\sqrt{3}}{167772160}\cos(4\theta_{\mu\nu}(\Omega t))+\frac{4280999 \sqrt{3}}{4194304}\cos(5\theta_{\mu\nu}(\Omega t))\\
&&-\frac{92332561\sqrt{3}}{41943040}\cos(6\theta_{\mu\nu}(\Omega t))+\frac{68505\sqrt{3}}{4194304}\cos(7\theta_{\mu\nu}(\Omega t))-\frac{1944549\sqrt{3}}{16777216}\cos(8\theta_{\mu\nu}(\Omega t))\\
&&+\frac{67125\sqrt{3} }{4194304}\cos(9\theta_{\mu\nu}(\Omega t))
-\frac{6753\sqrt{3}}{8388608}\cos(10\theta_{\mu\nu}(\Omega t))+\frac{75\sqrt{3}}{4194304}\cos(11\theta_{\mu\nu}(\Omega t))\\
&&-\frac{5\sqrt{3}}{33554432}\cos(12\theta_{\mu\nu}(\Omega t)),\\
Q\left(v_{\mu\nu}^{\pm},2\right)&=&\pm\frac{15 \sqrt{3}}{16}\sin\theta_{\mu\nu}(\Omega t)\mp\frac{15\sqrt{3}}{4}\sin(2\theta_{\mu\nu}(\Omega t))\pm\frac{3\sqrt{3}}{16}\sin(3\theta_{\mu\nu}(\Omega t)),\\
Q\left(v_{\mu\nu}^{\pm},3\right)&=&-\frac{1095\sqrt{3}}{256}\sin\theta_{\mu\nu}(\Omega t)+\frac{16239 \sqrt{3}}{1024}\sin(2\theta_{\mu\nu}(\Omega t))+\frac{855\sqrt{3}}{512}\sin(3\theta_{\mu\nu}(\Omega t))\\
&&\displaystyle+\frac{441\sqrt{3}}{256}\sin(4\theta_{\mu\nu}(\Omega t))-\frac{75\sqrt{3}}{512}\sin(5\theta_{\mu\nu}(\Omega t))+\frac{3\sqrt{3}}{1024}\sin(6\theta_{\mu\nu}(\Omega t)),\\
Q\left(v_{\mu\nu}^{\pm},4\right)&=&\pm\frac{1165791\sqrt{3}}{65536}\sin\theta_{\mu\nu}(\Omega t)\mp\frac{288213 \sqrt{3}}{4096}\sin(2\theta_{\mu\nu}(\Omega t))\mp\frac{122841\sqrt{3}}{32768}\sin(3\theta_{\mu\nu}(\Omega t))\\
&&\mp\frac{49515\sqrt{3}}{4096}\sin(4\theta_{\mu\nu}(\Omega t))\mp\frac{19995\sqrt{3}}{32768}\sin(5\theta_{\mu\nu}(\Omega t))\mp\frac{4905\sqrt{3}}{4096}\sin(6\theta_{\mu\nu}(\Omega t))\\
&&\pm\frac{18879\sqrt{3}}{131072}\sin(7\theta_{\mu\nu}(\Omega t))\mp\frac{45\sqrt{3}}{8192}\sin(8\theta_{\mu\nu}(\Omega t))\pm\frac{9\sqrt{3}}{131072}\sin(9\theta_{\mu\nu}(\Omega t)),\\
Q\left(v_{\mu\nu}^{\pm},5\right)&=&-\frac{180178337\sqrt{3}}{2097152}\sin\theta_{\mu\nu}(\Omega t)
+\frac{754445953\sqrt{3}}{2097152}\sin(2\theta_{\mu\nu}(\Omega t))
-\frac{11403261\sqrt{3}}{2097152}\sin(3\theta_{\mu\nu}(\Omega t))\\
&&+\frac{3237903959\sqrt{3}}{41943040}\sin(4\theta_{\mu\nu}(\Omega t))
-\frac{21404995\sqrt{3}}{4194304}\sin(5\theta_{\mu\nu}(\Omega t))+\frac{276997683\sqrt{3}}{20971520}\sin(6\theta_{\mu\nu}(\Omega t))\\
&&-\frac{479535\sqrt{3}}{4194304}\sin(7\theta_{\mu\nu}(\Omega t))+\frac{1944549\sqrt{3}}{2097152}\sin(8\theta_{\mu\nu}(\Omega t))-\frac{604125\sqrt{3}}{4194304}\sin(9\theta_{\mu\nu}(\Omega t))\\
&&+\frac{33765\sqrt{3}}{4194304}\sin(10\theta_{\mu\nu}(\Omega t))-\frac{825\sqrt{3}}{4194304}\sin(11\theta_{\mu\nu}(\Omega t))+\frac{15 \sqrt{3}}{8388608}\sin(12\theta_{\mu\nu}(\Omega t)).
\end{eqnarray*}
In the data analysis, by deducting the inherent variations of arm-lengths of Taiji triangle, one is able to acquire their accurate variations induced by GWs, where without doubt, based on these results, the variations of arm-lengths induced by the relativistic effect of the Sun's gravitational field and the perturbative effects of some celestial bodies need to be deducted as well.

The vertex angles of Taiji triangles between the relative radial vectors of SCs $\boldsymbol{r}_{\mu\lambda}^{\pm}$ and $\boldsymbol{r}_{\nu\lambda}^{\pm}\ (\mu\neq \nu)$, denoted by $\beta_{\mu\nu}^{\pm}$, are defined as
\begin{equation}\label{equ33}
\beta_{\mu\nu}^{\pm}:=\arccos\left(\hat{\boldsymbol{r}}_{\mu\lambda}^{\pm}\cdot\hat{\boldsymbol{r}}_{\nu\lambda}^{\pm}\right),
\end{equation}
where $\hat{\boldsymbol{r}}_{\mu\lambda}^{\pm}:=\boldsymbol{r}_{\mu\lambda}^{\pm}/l_{\mu\lambda}^{\pm}$ and  $\hat{\boldsymbol{r}}_{\nu\lambda}^{\pm}:=\boldsymbol{r}_{\nu\lambda}^{\pm}/l_{\nu\lambda}^{\pm}$
are the corresponding unit vectors, respectively. From Eq.~(\ref{equ33}), if the expansions of
$\hat{\boldsymbol{r}}_{\mu\lambda}^{\pm}$ and  $\hat{\boldsymbol{r}}_{\nu\lambda}^{\pm}$ are obtained, one is able to expand $\beta_{\mu\nu}^{\pm}$. Eq.~(\ref{equ25}) provides
\begin{eqnarray}
\label{equ35}\frac{1}{l_{\mu\nu}^{\pm}}=\frac{1}{eR\sqrt{a_{\mu\nu}^{\pm}(2)}}\left(\sum_{p=0}^{\infty}b_{\mu\nu}^{\pm}(p) e^p\right)^{-\frac{1}{2}},
\end{eqnarray}
and as before, according to Eqs.~(\ref{equC11})---(\ref{equC13}) and (\ref{equC16}), $(\sum_{p=0}^{\infty}b_{\mu\nu}^{\pm}(p) e^p)^{-1/2}$ can be expanded, namely,
\begin{eqnarray*}
\left(\sum_{p=0}^{\infty}b_{\mu\nu}^{\pm}(p) e^p\right)^{-\frac{1}{2}}=\sum_{p=0}^{\infty}\left(\delta_{0p}+\sum_{k=1}^{p}(-1)^{k}\frac{(2k-1)!!}{(2k)!!}b_{\mu\nu}^{(k)\pm}(p)\right)e^p,
\end{eqnarray*}
where the expression of $b_{\mu\nu}^{(k)\pm}(p)$ refer to Eq.~(\ref{equ27}).
Substituting above result to Eq.~(\ref{equ35}) gives
\begin{eqnarray}
\label{equ36}\frac{1}{l_{\mu\nu}^{\pm}}=\frac{1}{eR}\sum_{p=0}^{\infty}Q\left(\frac{1}{l_{\mu\nu}^{\pm}},p\right)e^p
\end{eqnarray}
with
\begin{eqnarray}
\label{equ37}Q\left(\frac{1}{l_{\mu\nu}^{\pm}},p\right):=\frac{1}{\sqrt{a_{\mu\nu}^{\pm}(2)}}\left(\delta_{0p}+\sum_{k=1}^{p}(-1)^{k}\frac{(2k-1)!!}{(2k)!!}
b_{\mu\nu}^{(k)\pm}(p)\right).
\end{eqnarray}

The combination of Eqs.~(\ref{equ17}) and (\ref{equ36}) brings about the expansions of $\hat{\boldsymbol{r}}_{\mu\nu}^{\pm}=(\hat{x}_{\mu\nu}^{\pm},\hat{y}_{\mu\nu}^{\pm},\hat{z}_{\mu\nu}^{\pm})$:
\begin{equation}\label{equ38}
\left\{\begin{array}{lll}
\displaystyle \hat{x}_{\mu\nu}^{\pm}&=&\displaystyle \sum_{n=0}^{\infty}Q\left(\hat{x}_{\mu\nu}^{\pm},n\right)e^n,\smallskip\\
\displaystyle \hat{y}_{\mu\nu}^{\pm}&=&\displaystyle \sum_{n=0}^{\infty}Q\left(\hat{y}_{\mu\nu}^{\pm},n\right)e^n,\smallskip\\
\displaystyle \hat{z}_{\mu\nu}^{\pm}&=&\displaystyle \sum_{n=0}^{\infty}Q\left(\hat{z}_{\mu\nu}^{\pm},n\right)e^n
\end{array}\right.
\end{equation}
with
\begin{equation}\label{equ39}
\left\{\begin{array}{lll}
\displaystyle Q\left(\hat{x}_{\mu\nu}^{\pm},n\right)&:=&\displaystyle \sum_{p=0}^{n}(\mp1)^{n-p+1}Q\left(x_{\mu\nu}^{\pm},n-p+1\right)Q\left(\frac{1}{l_{\mu\nu}^{\pm}},p\right),\smallskip\\
\displaystyle Q\left(\hat{y}_{\mu\nu}^{\pm},n\right)&:=&\displaystyle \sum_{p=0}^{n}(\mp1)^{n-p+1}Q\left(y_{\mu\nu}^{\pm},n-p+1\right)Q\left(\frac{1}{l_{\mu\nu}^{\pm}},p\right),\smallskip\\
\displaystyle Q\left(\hat{z}_{\mu\nu}^{\pm},n\right)&:=&\displaystyle \sum_{p=0}^{n}(\mp1)^{n-p}Q\left(z_{\mu\nu}^{\pm},n-p+1\right)Q\left(\frac{1}{l_{\mu\nu}^{\pm}},p\right),
\end{array}\right.
\end{equation}
and by using this result, one easily gets
\begin{eqnarray}
\label{equ40}B_{\mu\nu}^{\pm}:=\cos\beta_{\mu\nu}^{\pm}=\hat{\boldsymbol{r}}_{\mu\lambda}^{\pm}\cdot\hat{\boldsymbol{r}}_{\nu\lambda}^{\pm}=\sum_{p=0}^{\infty} Q\left(B_{\mu\nu}^{\pm},p\right)e^p
\end{eqnarray}
with
\begin{eqnarray}
\label{equ41}Q\left(B_{\mu\nu}^{\pm},p\right)
:=\sum_{k=0}^{p}\Big(Q\big(\hat{x}_{\mu\lambda}^{\pm},p-k\big)Q\big(\hat{x}_{\nu\lambda}^{\pm},k\big)
+Q\big(\hat{y}_{\mu\lambda}^{\pm},p-k\big)Q\big(\hat{y}_{\nu\lambda}^{\pm},k\big)
+Q\big(\hat{z}_{\mu\lambda}^{\pm},p-k\big)Q\big(\hat{z}_{\nu\lambda}^{\pm},k\big)\Big).
\end{eqnarray}
According to Eqs.~(\ref{equC11})---(\ref{equC13}) again,
\begin{eqnarray*}
\beta_{\mu\nu}^{\pm}=\arccos B_{\mu\nu}^{\pm}=\arccos\left(\sum_{p=0}^{\infty} Q\left(B_{\mu\nu}^{\pm},p\right)e^p\right)
\end{eqnarray*}
can be expanded, and then,
\begin{eqnarray}
\label{equ42}\beta_{\mu\nu}^{\pm}=\sum_{p=0}^{\infty}Q\left(\beta_{\mu\nu}^{\pm},p\right)e^p
\end{eqnarray}
with
\begin{eqnarray}
\label{equ43}Q\left(\beta_{\mu\nu}^{\pm},p\right)
:=\delta_{0p}\arccos\left(Q(B_{\mu\nu}^{\pm},0)\right)+\sum_{k=1}^{p}\frac{(\arccos)^{(k)}\left(Q(B_{\mu\nu}^{\pm},0)\right)}{k!}Q^{(k)}\left(B_{\mu\nu}^{\pm},p\right),
\end{eqnarray}
where
\begin{equation}\label{equ44}
Q^{(k)}\left(B_{\mu\nu}^{\pm},p\right)=\left\{\begin{array}{l}
\displaystyle Q\left(B_{\mu\nu}^{\pm},p\right),\quad k=1,\smallskip\\
\displaystyle \sum_{j_{k-1}=k-1}^{p-1}\ \sum_{j_{k-2}=k-2}^{j_{k-1}-1}\cdots\sum_{j_{2}=2}^{j_{3}-1}\sum_{j_{1}=1}^{j_{2}-1}
Q\left(B_{\mu\nu}^{\pm},p-j_{k-1}\right)Q\left(B_{\mu\nu}^{\pm},j_{k-1}-j_{k-2}\right)\cdots\\
\displaystyle\qquad\qquad\qquad\qquad\qquad\qquad\quad\times Q\left(B_{\mu\nu}^{\pm},j_{2}-j_{1}\right)Q\left(B_{\mu\nu}^{\pm},j_{1}\right),\quad k\geq2,
\end{array}\right.
\end{equation}
and the expression of $(\arccos)^{(k)}\ (k\geq1)$ refers to Eq.~(\ref{equC17}).

Next, the vertex angles of Taiji triangles will be discussed, and they also possess the symmetry:
\emph{Three components of the vertex angles of Taiji triangle in each configuration are also identical to each other up to a phase shift of $2\pi/3$ at every order, which is also independent on the tilt angle of Taiji plane relative to the ecliptic plane.} The proof
is lengthy, so the detailed process is put in Appendix B. From Eqs.~(\ref{equ31}) and (\ref{equ37}), $Q(1/l_{\mu\nu}^{\pm},0)=1/(a_{\mu\nu}^{\pm}(2))^{1/2}=1/Q(l_{\mu\nu}^{\pm},1)$, by which, one knows that
when $\phi^{\pm}=\pi/3$, $Q(1/l_{\mu\nu}^{\pm},0)=1/(2\sqrt{3})$, and then, Eqs.~(\ref{equ39}) and (\ref{equ41}) give
\begin{eqnarray*}
Q\left(B_{\mu\nu}^{\pm},0\right)
=\frac{1}{12}\left(Q\big(x_{\mu\lambda}^{\pm},1\big)Q\big(x_{\nu\lambda}^{\pm},1\big)
+Q\big(y_{\mu\lambda}^{\pm},1\big)Q\big(y_{\nu\lambda}^{\pm},1\big)
+Q\big(z_{\mu\lambda}^{\pm},1\big)Q\big(z_{\nu\lambda}^{\pm},1\big)\right).
\end{eqnarray*}
By further using Eqs.~(\ref{equ13}) and (\ref{equ18}), $Q(B_{\mu\nu}^{\pm},0)=1/2$ is obtained, and then, plugging this result into Eq.~(\ref{equ43}), one finally arrives at $Q(\beta_{\mu\nu}^{\pm},0)=\arccos(Q(B_{\mu\nu}^{\pm},0))=\pi/3$,
which shows that as expected, when $\phi^{\pm}=\pi/3$, Taiji triangles in both configurations are equilateral at the leading order terms of the vertex angles. Similarly to the cases of arm-lengths and their rates of change, the actual calculations of $Q(\beta_{\mu\nu}^{\pm},p)\ (p\geq1)$ imply that at the higher order terms, the vertex angles still undergo the inherent variations even when $\phi^{\pm}=\pi/3$. In the following, we will make use of the above algorithm to present the expansions of $\beta_{\mu\nu}^{\pm}$, and it will be proved that only if $\beta_{\mu\nu}^{\pm}$ are expanded to $e^4$ order, the obtained result is compatible with those of arm-lengths and their rates of change in Eq.~(\ref{equ32}). Suppose that $\boldsymbol{r}_{\mu\nu}^{\pm}$ have been expanded to $e^i$ order, the series expression of any related quantity $A$ should be truncated to $e^{\Lambda(A,i)}$ order, so that when $i\rightarrow\infty$, the truncated expression of $A$ can recover its original result, and by this rule, we have the following conclusions:
\begin{eqnarray}\label{equ45}
\Lambda\left(\big(\boldsymbol{r}_{\mu\nu}^{\pm}\big)^2,i\right)=i+1\Rightarrow\left\{\begin{array}{l}
\displaystyle \Lambda\big(l_{\mu\nu}^{\pm},i\big)=i\Rightarrow\Lambda\big(v_{\mu\nu}^{\pm},i\big)=i,\smallskip\\
\displaystyle \Lambda\big(1/l_{\mu\nu}^{\pm},i\big)=i-2\Rightarrow\Lambda\big(\hat{\boldsymbol{r}}_{\mu\nu}^{\pm},i\big)=i-1
\Rightarrow\Lambda\big(B_{\mu\nu}^{\pm},i\big)=i-1\Rightarrow\Lambda\big(\beta_{\mu\nu}^{\pm},i\big)=i-1.
\end{array}\right.
\end{eqnarray}
In the first step, following the process from Eq.~(\ref{equ24}) to Eq.~(\ref{equ25}), $(\sum_{p=0}^{\infty}b_{\mu\nu}^{\pm}(p) e^p)^{1/2}$ in Eq.~(\ref{equ25}) and $(\sum_{p=0}^{\infty}b_{\mu\nu}^{\pm}(p) e^p)^{-1/2}$ in Eq.~(\ref{equ35}) should be changed to be $(\sum_{p=0}^{i-1}b_{\mu\nu}^{\pm}(p) e^p)^{1/2}$ and $(\sum_{p=0}^{i-1}b_{\mu\nu}^{\pm}(p) e^p)^{-1/2}$, respectively, and in the spirit of Appendix C, one knows that only their further expansions to $e^{i-1}$ order are kept.
Therefore, from Eqs.~(\ref{equ25}) and (\ref{equ35}), $\Lambda(l_{\mu\nu}^{\pm},i)=i$ and $\Lambda(1/l_{\mu\nu}^{\pm},i)=i-2$. The above conclusion implies that if $i=5$, the following expansions of $\beta_{\mu\nu}^{\pm}$ to $e^4$ order are indeed compatible with those of arm-lengths and their rates of change in Eq.~(\ref{equ32}). As mentioned before, in order to consider the relativistic effect of the Sun's gravitational field and the perturbative effects of some celestial bodies, all the related results in the form of series
need to be truncated to necessary order, so the above conclusion~~(\ref{equ45}) plays an important role.
\begin{equation}\label{equ46}
\beta_{\mu\nu}^{\pm}=\frac{\pi}{3}+Q\left(\beta_{\mu\nu}^{\pm},1\right)e+Q\left(\beta_{\mu\nu}^{\pm},2\right)e^2+Q\left(\beta_{\mu\nu}^{\pm},3\right)e^3+Q\left(\beta_{\mu\nu}^{\pm},4\right)e^4,
\end{equation}
where from Eqs.~(\ref{equB1}), (\ref{equ13}), (\ref{equ18}), (\ref{equ24}), (\ref{equ25}), (\ref{equ27}), (\ref{equ37}), (\ref{equ39}), (\ref{equ41}), and (\ref{equ44}), Eq.~(\ref{equ43}) gives
\begin{eqnarray*}
Q\left(\beta_{\mu\nu}^{\pm},1\right)&=&\mp\frac{15 \sqrt{3}}{32}\cos\theta_{\mu\nu}(\Omega t)\pm\frac{15\sqrt{3}}{16}\cos(2\theta_{\mu\nu}(\Omega t)),\\
Q\left(\beta_{\mu\nu}^{\pm},2\right)&=&\frac{135\sqrt{3}}{128}\cos\theta_{\mu\nu}(\Omega t)-\frac{5997 \sqrt{3}}{2048}\cos(2\theta_{\mu\nu}(\Omega t))+\frac{447\sqrt{3}}{1024}\cos(4\theta_{\mu\nu}(\Omega t))+\frac{15\sqrt{3}}{512}\cos(5\theta_{\mu\nu}(\Omega t)),\\
Q\left(\beta_{\mu\nu}^{\pm},3\right)&=&\mp\frac{16899\sqrt{3}}{8192}\cos\theta_{\mu\nu}(\Omega t)\pm\frac{47307 \sqrt{3}}{4096}\cos(2\theta_{\mu\nu}(\Omega t))\mp\frac{6045\sqrt{3}}{2048}\cos(4\theta_{\mu\nu}(\Omega t))\\
&&\pm\frac{297\sqrt{3}}{16384}\cos(5\theta_{\mu\nu}(\Omega t))\pm\frac{897\sqrt{3}}{32768}\cos(7\theta_{\mu\nu}(\Omega t))\pm\frac{15\sqrt{3}}{16384}\cos(8\theta_{\mu\nu}(\Omega t)),\\
Q\left(\beta_{\mu\nu}^{\pm},4\right)&=&\frac{985237\sqrt{3}}{131072}\cos(\theta_{\mu\nu}(\Omega t))-\frac{30295387\sqrt{3}}{524288}\cos(2\theta_{\mu\nu}(\Omega t))+\frac{330064081 \sqrt{3}}{20971520}\cos(4\theta_{\mu\nu}(\Omega t))\\
&&+\frac{960503\sqrt{3}}{524288}\cos(5\theta_{\mu\nu}(\Omega t))-\frac{39585\sqrt{3}}{524288}\cos(7\theta_{\mu\nu}(\Omega t))-\frac{403803 \sqrt{3}}{2097152}\cos(8\theta_{\mu\nu}(\Omega t))\\
&&+\frac{1347\sqrt{3}}{1048576}\cos(10\theta_{\mu\nu}(\Omega t))+\frac{15\sqrt{3}}{524288}\cos(11\theta_{\mu\nu}(\Omega t)).
\end{eqnarray*}
One can check that when $p=1,2,3,4$,
\begin{eqnarray*}
Q\left(\beta_{12}^{\pm},p\right)+Q\left(\beta_{23}^{\pm},p\right)+Q\left(\beta_{31}^{\pm},p\right)=0,
\end{eqnarray*}
which is compatible with the result of $\beta_{12}^{\pm}+\beta_{23}^{\pm}+\beta_{31}^{\pm}=\pi$ in Euclidean geometry.
\subsection{Optimization of the orbits of Taiji SCs}

As indicated in Sec.~\ref{sec3.2}, for both configurations of Taiji, even when $\phi^{\pm}=\pi/3$, Taiji triangles are only equilateral at the leading terms of their kinematic indicators, and the higher order terms show that Taiji triangles undergo the inherent variations. One adverse effect brought about by such instability of Taiji triangle is
that the first generation TDI may work unsuccessfully, since it is only applicable for the stationary configuration, so that the laser frequency noise can not be suppressed effectively. In order to deal with this problem, one perhaps needs to turn to modified first generation TDI or further, the second generation TDI~\cite{Dhurandhar:2008yu,Tinto:2003vj,Vallisneri:2005ji,Tinto:2014lxa}. The application of the second generation TDI could be at the cost of possible difficulty in the data analysis~\cite{Dhurandhar:2008yu}, because the complex non-commuting time-delay operators are involved, and therefore, one should select a simpler TDI technique by optimizing the orbits of SCs, as the case of original LISA (presented in Refs.~\cite{Dhurandhar:2004rv,Nayak:2006zm,Dhurandhar:2008yu}). The smaller orbital eccentricity of Taiji SCs than that of the original LISA SCs means the more stable formation of Taiji than that of the original LISA, which will contribute to considering a simpler TDI strategy for Taiji by the optimization of orbits of SCs. Another adverse effect of the instability of Taiji triangle is the Doppler shift of the laser frequency, and optimizing the orbits of SCs also helps to reduce it.

By adjusting the tilt angle $\phi^{+}$ around $\pi/3$ at $e^1$ order, the orbits of SCs are optimized at the next leading orders of all the kinematic indicators in the original configuration of Taiji~\cite{Wu:2019thj}. According to the algorithm devised in the previous subsection, we will generalize this result in this subsection, and namely, by adjusting $\phi^{\pm}$ around $\pi/3$ to any order of $e$,  the orbits of SCs in both configurations of Taiji will be optimized, respectively, which means that Taiji triangles in both configurations can become as stable as possible with the different specific problem involved. To this end, suppose that $\phi^{\pm}$ have the forms of expansions around $\pi/3$ in $e$,
\begin{eqnarray}
\label{equ47}\phi^{\pm}:=\sum_{p=0}^{\infty}\gamma^{\pm}(p)e^p\qquad\text{with}\quad\gamma^{\pm}(0):=\frac{\pi}{3},
\end{eqnarray}
which means that we should modify the previous algorithm so that all the quantities involving $\phi^{\pm}$ can be reexpanded.

Let's start with the orbits of SCs, namely, $\boldsymbol{r}_{\kappa}^{\pm}=(x_{\kappa}^{\pm},y_{\kappa}^{\pm},z_{\kappa}^{\pm})\ (\kappa=1,2,3)$, and from  Eqs.~(\ref{equ2}), (\ref{equ5}), and (\ref{equ8})---(\ref{equ11}), one knows that their dependence on $\phi^{\pm}$ is originated from $\cos\varepsilon^{\pm}$ and $\sin\varepsilon^{\pm}$, and with the assumption~(\ref{equ47}), if $\cos\varepsilon^{\pm}$ and $\sin\varepsilon^{\pm}$ are reexpanded to infinite order of $e$, one can acquire the reexpansions of $\boldsymbol{r}_{\kappa}^{\pm}$. Technically, if $\cos\varepsilon^{\pm}$ and $\sin\varepsilon^{\pm}$ are reexpanded by redefining $Q(\cos\varepsilon^{\pm},n)$ and $Q(\sin\varepsilon^{\pm},n)$ in Eq.~(\ref{equA6}), while
Eq.~(\ref{equA5}) remains the same, one does not need to modify the remaining part of the previous algorithm to
obtain the reexpansions of $\boldsymbol{r}_{\kappa}^{\pm}$ and the further reexpansions of all the kinematic indicators of Taiji triangles. In Appendix D, according to the related conclusions in Appendix C, the modified expressions of $Q(\cos\varepsilon^{\pm},n)$ and $Q(\sin\varepsilon^{\pm},n)$ are derived, which are presented in Eq.~(\ref{equD12}).

Now, as mentioned above, according to the modified algorithm, all the kinematic indicators of Taiji triangles in both configurations can be reexpanded. Here, we take the reexpansions of $l_{\mu\nu}^{\pm}$ and $v_{\mu\nu}^{\pm}$ to $e^8$ order and the reexpansions of $\beta_{\mu\nu}^{\pm}$ to $e^7$ order as an example to explain how to optimize the orbits of SCs, and the corresponding results read
\begin{equation}\label{equ48}
\left\{\begin{array}{lllll}
\displaystyle l_{\mu\nu}^{\pm}&=&\displaystyle R\Big(2\sqrt{3}e+\sum_{p=2}^{8}Q\left(l_{\mu\nu}^{\pm},p\right)e^p\Big)&\quad\displaystyle\text{with}&\quad\displaystyle Q\left(l_{\mu\nu}^{\pm},p\right)=Q\left(l_{\mu\nu}^{\pm},p;\gamma^{\pm}(1),\cdots,\gamma^{\pm}(p-1)\right),\smallskip\\
\displaystyle v_{\mu\nu}^{\pm}&=&\displaystyle R\Omega\Big(0e+\sum_{p=2}^{8}Q\left(v_{\mu\nu}^{\pm},p\right)e^p\Big)&\quad\displaystyle\text{with}&\quad\displaystyle Q\left(v_{\mu\nu}^{\pm},p\right)=Q\left(v_{\mu\nu}^{\pm},p;\gamma^{\pm}(1),\cdots,\gamma^{\pm}(p-1)\right),\smallskip\\
\displaystyle \beta_{\mu\nu}^{\pm}&=&\displaystyle\frac{\pi}{3}+\sum_{p=1}^{7}Q\left(\beta_{\mu\nu}^{\pm},p\right)e^p&\quad\displaystyle\text{with}&\quad\displaystyle Q\left(\beta_{\mu\nu}^{\pm},p\right)=Q\left(\beta_{\mu\nu}^{\pm},p;\gamma^{\pm}(1),\cdots,\gamma^{\pm}(p)\right).
\end{array}\right.
\end{equation}
$Q(l_{\mu\nu}^{\pm},p)$, $Q(v_{\mu\nu}^{\pm},p)$, and $Q(\beta_{\mu\nu}^{\pm},p)$ are so lengthy as $p$ increases that their expressions do not need to be presented, but one should know that as before, $Q(l_{\mu\nu}^{\pm},p)$ and $Q(\beta_{\mu\nu}^{\pm},p)$ can be still written as the forms of the linear combination of $\cos(n_{1}\theta_{\mu\nu}(\Omega t))$, and $Q(v_{\mu\nu}^{\pm},p)$ can be still written as the form of the linear combination of $\sin(n_{2}\theta_{\mu\nu}(\Omega t))$, where both $n_{1}$ and $n_{2}$ are positive integers, and $\gamma^{\pm}(1)\cdots$ exist in the coefficients. The above choice about the truncated orders will ensure that the final determined orders of the optimized expressions of $l_{\mu\nu}^{\pm}$, $v_{\mu\nu}^{\pm}$, and $\beta_{\mu\nu}^{\pm}$ are the same as their previous those shown in Eqs.~(\ref{equ32}) and (\ref{equ46}), respectively.

Motivated by the idea in Ref.~\cite{Li:2008al}, if the following functions
\begin{eqnarray}
\label{equ49}Q\left(l_{\mu\nu}^{\pm},v_{\mu\nu}^{\pm},\beta_{\mu\nu}^{\pm}\right):=\omega_{l}^{\pm}D\left(l_{\mu\nu}^{\pm}\right)+\omega_{v}^{\pm}D\left(v_{\mu\nu}^{\pm}\right)+\omega_{\beta}^{\pm}D\left(\beta_{\mu\nu}^{\pm}\right)
\end{eqnarray}
take the minimums, the orbits of Taiji SCs in both configurations are optimized, respectively, where
\begin{equation}\label{equ50}
\left\{\begin{array}{lllll}
\displaystyle D\left(l_{\mu\nu}^{\pm}\right)&:=&\displaystyle\left<\left(\Delta
l_{\mu\nu}^{\pm}\right)^2\right>&=&\displaystyle\left<\left(l_{\mu\nu}^{\pm}-\left<l_{\mu\nu}^{\pm}\right>\right)^2\right>,\smallskip\\
\displaystyle D\left(v_{\mu\nu}^{\pm}\right)&:=&\displaystyle\left<\left(\Delta
v_{\mu\nu}^{\pm}\right)^2\right>&=&\displaystyle\left<\left(v_{\mu\nu}^{\pm}-\left<v_{\mu\nu}^{\pm}\right>\right)^2\right>,\smallskip\\
\displaystyle D\left(\beta_{\mu\nu}^{\pm}\right)&:=&\displaystyle\left<\left(\Delta
\beta_{\mu\nu}^{\pm}\right)^2\right>&=&\displaystyle\left<\left(\beta_{\mu\nu}^{\pm}-\left<\beta_{\mu\nu}^{\pm}\right>\right)^2\right>
\end{array}\right.
\end{equation}
are the variances of $l_{\mu\nu}^{\pm}$, $v_{\mu\nu}^{\pm}$, and $\beta_{\mu\nu}^{\pm}$ with
\begin{equation}\label{equ51}
\left\{\begin{array}{lllll}
\displaystyle \left<l_{\mu\nu}^{\pm}\right>&:=&\displaystyle\frac{\Omega}{2n\pi}\int_{t_{0}}^{t_{0}+2n\pi/\Omega}l_{\mu\nu}^{\pm}dt&=&\displaystyle R\left(2\sqrt{3}e+O\left(e^2\right)\right),\smallskip\\
\displaystyle \left<v_{\mu\nu}^{\pm}\right>&:=&\displaystyle\frac{\Omega}{2n\pi}\int_{t_{0}}^{t_{0}+2n\pi/\Omega}v_{\mu\nu}^{\pm}dt&=&\displaystyle
0,\smallskip\\
\displaystyle \left<\beta_{\mu\nu}^{\pm}\right>&:=&\displaystyle\frac{\Omega}{2n\pi}\int_{t_{0}}^{t_{0}+2n\pi/\Omega}\beta_{\mu\nu}^{\pm}dt&=&\displaystyle
\frac{\pi}{3}
\end{array}\right.
\end{equation}
as their averages within $n\ (n=1,2,3,\cdots)$ year, and $\omega_{l}^{\pm}$, $\omega_{v}^{\pm}$, and $\omega_{\beta}^{\pm}$ are their corresponding weights. To simplify calculation, define
\begin{eqnarray}
\label{equ52}\overline{\Delta l_{\mu\nu}^{\pm}}:=\displaystyle\frac{\Delta l_{\mu\nu}^{\pm}}{R},\qquad
\overline{\Delta v_{\mu\nu}^{\pm}}:=\displaystyle\frac{\Delta v_{\mu\nu}^{\pm}}{R\Omega},\qquad
\overline{\Delta\beta_{\mu\nu}^{\pm}}:=\displaystyle e\left(\Delta\beta_{\mu\nu}^{\pm}\right),
\end{eqnarray}
and then, from Eqs.~(\ref{equ50})---(\ref{equ52}),
\begin{eqnarray}
\label{equ53}Q\left(l_{\mu\nu}^{\pm},v_{\mu\nu}^{\pm},\beta_{\mu\nu}^{\pm}\right)
&=&\left<\omega_{l}^{\pm}\left(\Delta
l_{\mu\nu}^{\pm}\right)^2+\omega_{v}^{\pm}\left(\Delta
v_{\mu\nu}^{\pm}\right)^2+\omega_{\beta}^{\pm}\left(\Delta
\beta_{\mu\nu}^{\pm}\right)^2\right>\nonumber\\
&=&\left<\overline{\omega}_{l}^{\pm}\left(\overline{\Delta
l_{\mu\nu}^{\pm}}\right)^2+\overline{\omega}_{v}^{\pm}\left(\overline{\Delta
v_{\mu\nu}^{\pm}}\right)^2+\overline{\omega}_{\beta}^{\pm}\left(\overline{\Delta
\beta_{\mu\nu}^{\pm}}\right)^2\right>,
\end{eqnarray}
where
\begin{eqnarray}
\label{equ54}\overline{\omega}_{l}^{\pm}:=R^2\omega_{l}^{\pm},\qquad \overline{\omega}_{v}^{\pm}:=R^2\Omega^2\omega_{v}^{\pm},\qquad
\overline{\omega}_{\beta}^{\pm}:=\frac{\omega_{\beta}^{\pm}}{e^2}
\end{eqnarray}
are the reduced weights of $l_{\mu\nu}^{\pm}$, $v_{\mu\nu}^{\pm}$, and $\beta_{\mu\nu}^{\pm}$, respectively.
Let the superscript $[p]$ represents that the order of the corresponding term is $e^p$, and then,
Eqs.~(\ref{equ48}) and (\ref{equ52}) show
\begin{equation}\label{equ55}
\left\{\begin{array}{lll}
\displaystyle \overline{\Delta l_{\mu\nu}^{\pm}}&:=&\displaystyle\sum_{p=2}^{8}\left(\overline{\Delta l_{\mu\nu}^{\pm}}\right)^{[p]},\smallskip\\
\displaystyle \overline{\Delta v_{\mu\nu}^{\pm}}&:=&\displaystyle\sum_{p=2}^{8}\left(\overline{\Delta v_{\mu\nu}^{\pm}}\right)^{[p]},\smallskip\\
\displaystyle \overline{\Delta\beta_{\mu\nu}^{\pm}}&:=&\displaystyle\sum_{p=2}^{8}\left(\overline{\Delta \beta_{\mu\nu}^{\pm}}\right)^{[p]},
\end{array}\right.\Rightarrow
\left\{\begin{array}{lll}
\displaystyle \left(\overline{\Delta l_{\mu\nu}^{\pm}}\right)^2&=&\displaystyle\sum_{p=4}^{10}\left(\left(\overline{\Delta l_{\mu\nu}^{\pm}}\right)^2\right)^{[p]},\smallskip\\
\displaystyle \left(\overline{\Delta v_{\mu\nu}^{\pm}}\right)^2&=&\displaystyle\sum_{p=4}^{10}\left(\left(\overline{\Delta v_{\mu\nu}^{\pm}}\right)^2\right)^{[p]},\smallskip\\
\displaystyle \left(\overline{\Delta\beta_{\mu\nu}^{\pm}}\right)^2&=&\displaystyle\sum_{p=4}^{10}\left(\left(\overline{\Delta \beta_{\mu\nu}^{\pm}}\right)^2\right)^{[p]}.
\end{array}\right.
\end{equation}
Substituting this result to Eq.~(\ref{equ53}) gives
\begin{eqnarray}
\label{equ56}Q\left(l_{\mu\nu}^{\pm},v_{\mu\nu}^{\pm},\beta_{\mu\nu}^{\pm}\right)
&=&\sum_{p=4}^{10}Q^{[p]}\left(l_{\mu\nu}^{\pm},v_{\mu\nu}^{\pm},\beta_{\mu\nu}^{\pm}\right)
\end{eqnarray}
with
\begin{eqnarray}
\label{equ57}Q^{[p]}\left(l_{\mu\nu}^{\pm},v_{\mu\nu}^{\pm},\beta_{\mu\nu}^{\pm}\right)
&:=&\left<\overline{\omega}_{l}^{\pm}\left(\left(\overline{\Delta l_{\mu\nu}^{\pm}}\right)^2\right)^{[p]}+\overline{\omega}_{v}^{\pm}\left(\left(\overline{\Delta v_{\mu\nu}^{\pm}}\right)^2\right)^{[p]}+\overline{\omega}_{\beta}^{\pm}\left(\left(\overline{\Delta \beta_{\mu\nu}^{\pm}}\right)^2\right)^{[p]}\right>.
\end{eqnarray}

Above equations imply that if $Q^{[p]}(l_{\mu\nu}^{\pm},v_{\mu\nu}^{\pm},\beta_{\mu\nu}^{\pm})\ (p=4,5,\cdots)$
take the minimums, $Q(l_{\mu\nu}^{\pm},v_{\mu\nu}^{\pm},\beta_{\mu\nu}^{\pm})$ will take their minimums, and then after a tedious calculation, the following results are obtained:
\begin{equation}\label{equ58}
\left\{\begin{array}{lll}
\displaystyle \gamma^{\pm}(1)&=&\displaystyle\pm\frac{15}{8\sqrt{3}},\smallskip\\
\displaystyle \gamma^{\pm}(2)&=&\displaystyle-\frac{\mathcal{P}_{2}\big(\overline{\omega}_{l}^{\pm},\overline{\omega}_{v}^{\pm},\overline{\omega}_{\beta}^{\pm}\big)}{4\overline{\omega}_{l}^{\pm}+16\overline{\omega}_{v}^{\pm}+\overline{\omega}_{\beta}^{\pm}},\smallskip\\
\displaystyle \gamma^{\pm}(3)&=&\displaystyle\mp\frac{\mathcal{P}_{3}\big(\overline{\omega}_{l}^{\pm},\overline{\omega}_{v}^{\pm},\overline{\omega}_{\beta}^{\pm}\big)}{4\overline{\omega}_{l}^{\pm}+16\overline{\omega}_{v}^{\pm}+\overline{\omega}_{\beta}^{\pm}},\smallskip\\
\displaystyle
\gamma^{\pm}(4)&=&\displaystyle\frac{\mathcal{P}_{4}\big(\overline{\omega}_{l}^{\pm},\overline{\omega}_{v}^{\pm},\overline{\omega}_{\beta}^{\pm}\big)}
{\big(4\overline{\omega}_{l}^{\pm}+16\overline{\omega}_{v}^{\pm}+\overline{\omega}_{\beta}^{\pm}\big)^3}
\end{array}\right.
\end{equation}
with
\begin{eqnarray*}
\mathcal{P}_{2}\big(\overline{\omega}_{l}^{\pm},\overline{\omega}_{v}^{\pm},\overline{\omega}_{\beta}^{\pm}\big)&=&\displaystyle\frac{10922\overline{\omega}_{l}^{\pm}+17028\overline{\omega}_{v}^{\pm}+1317 \overline{\omega}_{\beta}^{\pm}}{1024\sqrt{3}},\\
\mathcal{P}_{3}\big(\overline{\omega}_{l}^{\pm},\overline{\omega}_{v}^{\pm},\overline{\omega}_{\beta}^{\pm}\big)&=&\frac{7(11114\overline{\omega}_{l}^{\pm}+17796\overline{\omega}_{v}^{\pm}+1365 \overline{\omega}_{\beta}^{\pm})}{4096\sqrt{3}},\\
\mathcal{P}_{4}\big(\overline{\omega}_{l}^{\pm},\overline{\omega}_{v}^{\pm},\overline{\omega}_{\beta}^{\pm}\big)&=&\frac{1}{1572864\sqrt{3}}\Big(1677590048 \big(\overline{\omega}_{l}^{\pm}\big)^3+12778457960\big(\overline{\omega}_{l}^{\pm}\big)^2\big(\overline{\omega}_{v}^{\pm}\big)\nonumber\\
&&+28501108176\big(\overline{\omega}_{l}^{\pm}\big)\big(\overline{\omega}_{v}^{\pm}\big)^2+16914868416\big(\overline{\omega}_{v}^{\pm}\big)^3+821189105 \big(\overline{\omega}_{l}^{\pm}\big)^2\big(\overline{\omega}_{\beta}^{\pm}\big)\nonumber\\
&&+2932122822\big(\overline{\omega}_{l}^{\pm}\big) \big(\overline{\omega}_{v}^{\pm}\big)\big(\overline{\omega}_{\beta}^{\pm}\big)+1888107408\big(\overline{\omega}_{v}^{\pm}\big)^2\big(\overline{\omega}_{\beta}^{\pm}\big)+79001022\big(\overline{\omega}_{l}^{\pm}\big) \big(\overline{\omega}_{\beta}^{\pm}\big)^2\nonumber\\
&&+35773092 \big(\overline{\omega}_{v}^{\pm}\big)\big(\overline{\omega}_{\beta}^{\pm}\big)^2-866259\big(\overline{\omega}_{\beta}^{\pm}\big)^3\Big).
\end{eqnarray*}
In fact, results in Eq.~(\ref{equ58}) are derived one after another when taking the minimums of $Q^{[p]}(l_{\mu\nu}^{\pm},v_{\mu\nu}^{\pm},\beta_{\mu\nu}^{\pm})$ for $p=4,6,8,10$, and when $p=5,7,9$, $Q^{[p]}(l_{\mu\nu}^{\pm},v_{\mu\nu}^{\pm},\beta_{\mu\nu}^{\pm})=0$, which are trivial. Then, from Eq.~(\ref{equ48}),
$l_{\mu\nu}^{\pm}$ and $v_{\mu\nu}^{\pm}$ are determined to $e^5$ order, and $\beta_{\mu\nu}^{\pm}$ are determined to $e^4$ order, and after omitting their undetermined parts, one finally arrives at
\begin{equation}\label{equ59}
\left\{\begin{array}{lll}
\displaystyle l_{\mu\nu}^{\pm}&=&\displaystyle
R\Big(2\sqrt{3}e+\sum_{p=2}^{5}Q\left(l_{\mu\nu}^{\pm},p\right)e^p\Big),\smallskip\\
\displaystyle v_{\mu\nu}^{\pm}&=&\displaystyle
R\Omega\Big(0e+\sum_{p=2}^{5}Q\left(v_{\mu\nu}^{\pm},p\right)e^p\Big),\smallskip\\
\displaystyle \beta_{\mu\nu}^{\pm}&=&\displaystyle
\frac{\pi}{3}+\sum_{p=1}^{4}Q\left(\beta_{\mu\nu}^{\pm},p\right)e^p,
\end{array}\right.
\end{equation}
where by plugging Eq.~(\ref{equ58}) into the expressions of $Q(l_{\mu\nu}^{\pm},p)$, $Q(v_{\mu\nu}^{\pm},p)$, and $Q(\beta_{\mu\nu}^{\pm},p)$ in Eq.~(\ref{equ48}), there are
\begin{eqnarray*}
Q\left(l_{\mu\nu}^{\pm},2\right)&=&\mp\frac{15\sqrt{3}}{16}\cos\theta_{\mu\nu}(\Omega t)\mp\frac{\sqrt{3}}{16}\cos(3\theta_{\mu\nu}(\Omega t)),\\
Q\left(l_{\mu\nu}^{\pm},3\right)&=&-\frac{\mathcal{Q}^{l}_{30}\big(\overline{\omega}_{l}^{\pm},\overline{\omega}_{v}^{\pm},\overline{\omega}_{\beta}^{\pm}\big)}
{4\overline{\omega}_{l}^{\pm}+16\overline{\omega}_{v}^{\pm}+\overline{\omega}_{\beta}^{\pm}}+\frac{\mathcal{Q}^{l}_{32}\big(\overline{\omega}_{l}^{\pm},\overline{\omega}_{v}^{\pm},\overline{\omega}_{\beta}^{\pm}\big)}
{4\overline{\omega}_{l}^{\pm}+16\overline{\omega}_{v}^{\pm}+\overline{\omega}_{\beta}^{\pm}}\cos(2\theta_{\mu\nu}(\Omega t))+\frac{9\sqrt{3}}{1024}\cos(4\theta_{\mu\nu}(\Omega t))\\
&&-\frac{\sqrt{3}}{2048}\cos(6\theta_{\mu\nu}(\Omega t)),\\
Q\left(l_{\mu\nu}^{\pm},4\right)&=&\pm\frac{\mathcal{Q}^{l}_{41}\big(\overline{\omega}_{l}^{\pm},\overline{\omega}_{v}^{\pm},\overline{\omega}_{\beta}^{\pm}\big)}{4\overline{\omega}_{l}^{\pm}+16\overline{\omega}_{v}^{\pm}+\overline{\omega}_{\beta}^{\pm}}\cos(\theta_{\mu\nu}(\Omega t))\pm\frac{\mathcal{Q}^{l}_{43}\big(\overline{\omega}_{l}^{\pm},\overline{\omega}_{v}^{\pm},\overline{\omega}_{\beta}^{\pm}\big)}
{4\overline{\omega}_{l}^{\pm}+16\overline{\omega}_{v}^{\pm}+\overline{\omega}_{\beta}^{\pm}}\cos(3\theta_{\mu\nu}(\Omega t))\\
&&\pm\frac{\mathcal{Q}^{l}_{45}\big(\overline{\omega}_{l}^{\pm},\overline{\omega}_{v}^{\pm},\overline{\omega}_{\beta}^{\pm}\big)}{4\overline{\omega}_{l}^{\pm}+16\overline{\omega}_{v}^{\pm}+\overline{\omega}_{\beta}^{\pm}}\cos(5\theta_{\mu\nu}(\Omega t))\pm\frac{3 \sqrt{3}}{131072}\cos(7\theta_{\mu\nu}(\Omega t))\mp\frac{\sqrt{3}}{131072}\cos(9\theta_{\mu\nu}(\Omega t)),
\end{eqnarray*}
\begin{eqnarray*}
Q\left(l_{\mu\nu}^{\pm},5\right)&=&\frac{\mathcal{Q}^{l}_{50}\big(\overline{\omega}_{l}^{\pm},\overline{\omega}_{v}^{\pm},\overline{\omega}_{\beta}^{\pm}\big)}{\big(4\overline{\omega}_{l}^{\pm}+16\overline{\omega}_{v}^{\pm}+\overline{\omega}_{\beta}^{\pm}\big)^3}-\frac{\mathcal{Q}^{l}_{52}\big(\overline{\omega}_{l}^{\pm},\overline{\omega}_{v}^{\pm},\overline{\omega}_{\beta}^{\pm}\big)}
{\big(4\overline{\omega}_{l}^{\pm}+16\overline{\omega}_{v}^{\pm}+\overline{\omega}_{\beta}^{\pm}\big)^3}\cos(2\theta_{\mu\nu}(\Omega t))\\
&&-\frac{\mathcal{Q}^{l}_{54}\big(\overline{\omega}_{l}^{\pm},\overline{\omega}_{v}^{\pm},\overline{\omega}_{\beta}^{\pm}\big)}{\big(4\overline{\omega}_{l}^{\pm}+16\overline{\omega}_{v}^{\pm}+\overline{\omega}_{\beta}^{\pm}\big)^2}\cos(4\theta_{\mu\nu}(\Omega t))-\frac{\mathcal{Q}^{l}_{56}\big(\overline{\omega}_{l}^{\pm},\overline{\omega}_{v}^{\pm},\overline{\omega}_{\beta}^{\pm}\big)}
{4\overline{\omega}_{l}^{\pm}+16\overline{\omega}_{v}^{\pm}+\overline{\omega}_{\beta}^{\pm}}\cos(6\theta_{\mu\nu}(\Omega t))\\
&&+\frac{\mathcal{Q}^{l}_{58}\big(\overline{\omega}_{l}^{\pm},\overline{\omega}_{v}^{\pm},\overline{\omega}_{\beta}^{\pm}\big)}{4\overline{\omega}_{l}^{\pm}+16\overline{\omega}_{v}^{\pm}+\overline{\omega}_{\beta}^{\pm}}\cos(8\theta_{\mu\nu}(\Omega t))-\frac{3\sqrt{3}}{8388608}\cos(10\theta_{\mu\nu}(\Omega t))\\
&&-\frac{5\sqrt{3}}{33554432}\cos(12\theta_{\mu\nu}(\Omega t)),\\
Q\left(v_{\mu\nu}^{\pm},2\right)&=&\pm\frac{15\sqrt{3}}{16}\sin\theta_{\mu\nu}(\Omega t)\pm\frac{3\sqrt{3}}{16}\sin(3\theta_{\mu\nu}(\Omega t)),\\
Q\left(v_{\mu\nu}^{\pm},3\right)&=&-\frac{2\mathcal{Q}^{l}_{32}\big(\overline{\omega}_{l}^{\pm},\overline{\omega}_{v}^{\pm},\overline{\omega}_{\beta}^{\pm}\big)}
{4\overline{\omega}_{l}^{\pm}+16\overline{\omega}_{v}^{\pm}+\overline{\omega}_{\beta}^{\pm}}\sin(2\theta_{\mu\nu}(\Omega t))-\frac{9\sqrt{3}}{256}\sin(4\theta_{\mu\nu}(\Omega t))+\frac{3\sqrt{3}}{1042}\sin(6\theta_{\mu\nu}(\Omega t)),\\
Q\left(v_{\mu\nu}^{\pm},4\right)&=&\mp\frac{\mathcal{Q}^{l}_{41}\big(\overline{\omega}_{l}^{\pm},\overline{\omega}_{v}^{\pm},\overline{\omega}_{\beta}^{\pm}\big)}{4\overline{\omega}_{l}^{\pm}+16\overline{\omega}_{v}^{\pm}+\overline{\omega}_{\beta}^{\pm}}\sin(\theta_{\mu\nu}(\Omega t))\mp\frac{3\mathcal{Q}^{l}_{43}\big(\overline{\omega}_{l}^{\pm},\overline{\omega}_{v}^{\pm},\overline{\omega}_{\beta}^{\pm}\big)}
{4\overline{\omega}_{l}^{\pm}+16\overline{\omega}_{v}^{\pm}+\overline{\omega}_{\beta}^{\pm}}\sin(3\theta_{\mu\nu}(\Omega t))\\
&&\mp\frac{5\mathcal{Q}^{l}_{45}\big(\overline{\omega}_{l}^{\pm},\overline{\omega}_{v}^{\pm},\overline{\omega}_{\beta}^{\pm}\big)}{4\overline{\omega}_{l}^{\pm}+16\overline{\omega}_{v}^{\pm}+\overline{\omega}_{\beta}^{\pm}}\sin(5\theta_{\mu\nu}(\Omega t))\mp\frac{21\sqrt{3}}{131072}\sin(7\theta_{\mu\nu}(\Omega t))\pm\frac{9\sqrt{3}}{131072}\sin(9\theta_{\mu\nu}(\Omega t)),\\
Q\left(v_{\mu\nu}^{\pm},5\right)&=&\frac{2\mathcal{Q}^{l}_{52}\big(\overline{\omega}_{l}^{\pm},\overline{\omega}_{v}^{\pm},\overline{\omega}_{\beta}^{\pm}\big)}
{\big(4\overline{\omega}_{l}^{\pm}+16\overline{\omega}_{v}^{\pm}+\overline{\omega}_{\beta}^{\pm}\big)^3}\sin(2\theta_{\mu\nu}(\Omega t))+\frac{4\mathcal{Q}^{l}_{54}\big(\overline{\omega}_{l}^{\pm},\overline{\omega}_{v}^{\pm},\overline{\omega}_{\beta}^{\pm}\big)}{\big(4\overline{\omega}_{l}^{\pm}+16\overline{\omega}_{v}^{\pm}+\overline{\omega}_{\beta}^{\pm}\big)^2}\sin(4\theta_{\mu\nu}(\Omega t))\\
&&+\frac{6\mathcal{Q}^{l}_{56}\big(\overline{\omega}_{l}^{\pm},\overline{\omega}_{v}^{\pm},\overline{\omega}_{\beta}^{\pm}\big)}
{4\overline{\omega}_{l}^{\pm}+16\overline{\omega}_{v}^{\pm}+\overline{\omega}_{\beta}^{\pm}}\sin(6\theta_{\mu\nu}(\Omega t))-\frac{8\mathcal{Q}^{l}_{58}\big(\overline{\omega}_{l}^{\pm},\overline{\omega}_{v}^{\pm},\overline{\omega}_{\beta}^{\pm}\big)}{4\overline{\omega}_{l}^{\pm}+16\overline{\omega}_{v}^{\pm}+\overline{\omega}_{\beta}^{\pm}}\sin(8\theta_{\mu\nu}(\Omega t))\\
&&+\frac{15\sqrt{3}}{4194304}\sin(10\theta_{\mu\nu}(\Omega t))+\frac{15\sqrt{3}}{8388608}\sin(12\theta_{\mu\nu}(\Omega t)),
\end{eqnarray*}
and
\begin{eqnarray*}
Q\left(\beta_{\mu\nu}^{\pm},1\right)&=&\mp\frac{15\sqrt{3}}{32}\cos\theta_{\mu\nu}(\Omega t),\\
Q\left(\beta_{\mu\nu}^{\pm},2\right)&=&
\frac{\mathcal{Q}^{\beta}_{22}\big(\overline{\omega}_{l}^{\pm},\overline{\omega}_{v}^{\pm},\overline{\omega}_{\beta}^{\pm}\big)}
{4\overline{\omega}_{l}^{\pm}+16\overline{\omega}_{v}^{\pm}+\overline{\omega}_{\beta}^{\pm}}\cos(2\theta_{\mu\nu}(\Omega t))-\frac{3\sqrt{3}}{1024}\cos(4\theta_{\mu\nu}(\Omega t)),\\
Q\left(\beta_{\mu\nu}^{\pm},3\right)&=&\pm\frac{\mathcal{Q}^{\beta}_{31}\big(\overline{\omega}_{l}^{\pm},\overline{\omega}_{v}^{\pm},\overline{\omega}_{\beta}^{\pm}\big)}{4\overline{\omega}_{l}^{\pm}+16\overline{\omega}_{v}^{\pm}+\overline{\omega}_{\beta}^{\pm}}\cos(\theta_{\mu\nu}(\Omega t))\pm\frac{\mathcal{Q}^{\beta}_{35}\big(\overline{\omega}_{l}^{\pm},\overline{\omega}_{v}^{\pm},\overline{\omega}_{\beta}^{\pm}\big)}{4\overline{\omega}_{l}^{\pm}+16\overline{\omega}_{v}^{\pm}+\overline{\omega}_{\beta}^{\pm}}\cos(5\theta_{\mu\nu}(\Omega t))\\
&&\mp\frac{3\sqrt{3}}{32768}\cos(7\theta_{\mu\nu}(\Omega t)),\\
Q\left(\beta_{\mu\nu}^{\pm},4\right)&=&-\frac{\mathcal{Q}^{\beta}_{42}\big(\overline{\omega}_{l}^{\pm},\overline{\omega}_{v}^{\pm},\overline{\omega}_{\beta}^{\pm}\big)}
{\big(4\overline{\omega}_{l}^{\pm}+16\overline{\omega}_{v}^{\pm}+\overline{\omega}_{\beta}^{\pm}\big)^3}\cos(2\theta_{\mu\nu}(\Omega t))+\frac{\mathcal{Q}^{\beta}_{44}\big(\overline{\omega}_{l}^{\pm},\overline{\omega}_{v}^{\pm},\overline{\omega}_{\beta}^{\pm}\big)}{\big(4\overline{\omega}_{l}^{\pm}+16\overline{\omega}_{v}^{\pm}+\overline{\omega}_{\beta}^{\pm}\big)^2}\cos(4\theta_{\mu\nu}(\Omega t))\\
&&+\frac{\mathcal{Q}^{\beta}_{48}\big(\overline{\omega}_{l}^{\pm},\overline{\omega}_{v}^{\pm},\overline{\omega}_{\beta}^{\pm}\big)}{4\overline{\omega}_{l}^{\pm}+16\overline{\omega}_{v}^{\pm}+\overline{\omega}_{\beta}^{\pm}}\cos(8\theta_{\mu\nu}(\Omega t))-\frac{3\sqrt{3}}{1048576}\cos(10\theta_{\mu\nu}(\Omega t))
\end{eqnarray*}
with
\begin{eqnarray*}
\mathcal{Q}^{l}_{30}\big(\overline{\omega}_{l}^{\pm},\overline{\omega}_{v}^{\pm},\overline{\omega}_{\beta}^{\pm}\big)&=&\displaystyle\frac{\sqrt{3}\big(10503\overline{\omega}_{l}^{\pm}+28682\overline{\omega}_{v}^{\pm}+1919 \overline{\omega}_{\beta}^{\pm}\big)}{512},\\
\mathcal{Q}^{l}_{32}\big(\overline{\omega}_{l}^{\pm},\overline{\omega}_{v}^{\pm},\overline{\omega}_{\beta}^{\pm}\big)&=&\displaystyle\frac{\sqrt{3} \big(17368\overline{\omega}_{l}^{\pm}+16152\overline{\omega}_{v}^{\pm}+1515\overline{\omega}_{\beta}^{\pm}\big)}{2048},\\
\mathcal{Q}^{l}_{41}\big(\overline{\omega}_{l}^{\pm},\overline{\omega}_{v}^{\pm},\overline{\omega}_{\beta}^{\pm}\big)&=&\displaystyle\frac{\sqrt{3}\big(1852288\overline{\omega}_{l}^{\pm}+3516792 \overline{\omega}_{v}^{\pm}+256701\overline{\omega}_{\beta}^{\pm}\big)}{65536},
\end{eqnarray*}
\begin{eqnarray*}
\mathcal{Q}^{l}_{43}\big(\overline{\omega}_{l}^{\pm},\overline{\omega}_{v}^{\pm},\overline{\omega}_{\beta}^{\pm}\big)&=&\displaystyle\frac{\sqrt{3} \big(4498\overline{\omega}_{l}^{\pm}+524532\overline{\omega}_{v}^{\pm}+27981\overline{\omega}_{\beta}^{\pm}\big)}{65536},\\
\mathcal{Q}^{l}_{45}\big(\overline{\omega}_{l}^{\pm},\overline{\omega}_{v}^{\pm},\overline{\omega}_{\beta}^{\pm}\big)&=&\displaystyle\frac{\sqrt{3}\big(24554\overline{\omega}_{l}^{\pm}+71556\overline{\omega}_{v}^{\pm}+4725 \overline{\omega}_{\beta}^{\pm}\big)}{65536},\\
\mathcal{Q}^{l}_{50}\big(\overline{\omega}_{l}^{\pm},\overline{\omega}_{v}^{\pm},\overline{\omega}_{\beta}^{\pm}\big)&=&\displaystyle\frac{\sqrt{3}}{16777216}\Big(47642365472 \big(\overline{\omega}_{l}^{\pm}\big)^3+399069232000\big(\overline{\omega}_{l}^{\pm}\big)^2\big(\overline{\omega}_{v}^{\pm}\big)\\
&&+26075374104 \big(\overline{\omega}_{l}^{\pm}\big)^2\big(\overline{\omega}_{\beta}^{\pm}\big)+1036985622144\big(\overline{\omega}_{l}^{\pm}\big) \big(\overline{\omega}_{v}^{\pm}\big)^2\\
&&+128928982880\big(\overline{\omega}_{l}^{\pm}\big)\big(\overline{\omega}_{v}^{\pm}\big) \big(\overline{\omega}_{\beta}^{\pm}\big)+4088092496\big(\overline{\omega}_{l}^{\pm}\big)\big(\overline{\omega}_{\beta}^{\pm}\big)^2\\
&&+811946166784 \big(\overline{\omega}_{v}^{\pm}\big)^3+148384407456\big(\overline{\omega}_{v}^{\pm}\big)^2\big(\overline{\omega}_{\beta}^{\pm}\big)\\
&&+9032642280 \big(\overline{\omega}_{v}^{\pm}\big)\big(\overline{\omega}_{\beta}^{\pm}\big)^2+184675777\big(\overline{\omega}_{\beta}^{\pm}\big)^3\Big),\\
\mathcal{Q}^{l}_{52}\big(\overline{\omega}_{l}^{\pm},\overline{\omega}_{v}^{\pm},\overline{\omega}_{\beta}^{\pm}\big)&=&\displaystyle\frac{\sqrt{3}}{4194304}\Big(6331721632 \big(\overline{\omega}_{l}^{\pm}\big)^3+46245994048\big(\overline{\omega}_{l}^{\pm}\big)^2\big(\overline{\omega}_{v}^{\pm}\big)\\
&&+3046484592 \big(\overline{\omega}_{l}^{\pm}\big)^2\big(\overline{\omega}_{\beta}^{\pm}\big)+97440130048\big(\overline{\omega}_{l}^{\pm}\big)\big(\overline{\omega}_{v}^{\pm}\big)^2\\
&&+11015200048\big(\overline{\omega}_{l}^{\pm}\big)\big(\overline{\omega}_{v}^{\pm}\big) \big(\overline{\omega}_{\beta}^{\pm}\big)+327885768\big(\overline{\omega}_{l}^{\pm}\big)\big(\overline{\omega}_{\beta}^{\pm}\big)^2\\
&&+55054799872 \big(\overline{\omega}_{v}^{\pm}\big)^3+7430284320\big(\overline{\omega}_{v}^{\pm}\big)^2\big(\overline{\omega}_{\beta}^{\pm}\big)\\
&&+282997416 \big(\overline{\omega}_{v}^{\pm}\big)\big(\overline{\omega}_{\beta}^{\pm}\big)^2+2487199\big(\overline{\omega}_{\beta}^{\pm}\big)^3\Big),\\
\mathcal{Q}^{l}_{54}\big(\overline{\omega}_{l}^{\pm},\overline{\omega}_{v}^{\pm},\overline{\omega}_{\beta}^{\pm}\big)&=&\displaystyle\frac{\sqrt{3}}{167772160}\Big(3563227584 \big(\overline{\omega}_{l}^{\pm}\big)^2+16611195072\big(\overline{\omega}_{l}^{\pm}\big)\big(\overline{\omega}_{v}^{\pm}\big)\\
&&+1150966632 \big(\overline{\omega}_{l}^{\pm}\big)\big(\overline{\omega}_{\beta}^{\pm}\big)+23648250944\big(\overline{\omega}_{v}^{\pm}\big)^2\\
&&+3137566528 \big(\overline{\omega}_{v}^{\pm}\big)\big(\overline{\omega}_{\beta}^{\pm}\big)+104999579\big(\overline{\omega}_{\beta}^{\pm}\big)^2\Big),\\
\mathcal{Q}^{l}_{56}\big(\overline{\omega}_{l}^{\pm},\overline{\omega}_{v}^{\pm},\overline{\omega}_{\beta}^{\pm}\big)&=&\frac{7 \sqrt{3}\big(1823132\overline{\omega}_{l}^{\pm}+6759328\overline{\omega}_{v}^{\pm}+427513\overline{\omega}_{\beta}^{\pm}\big)}{41943040},\\
\mathcal{Q}^{l}_{58}\big(\overline{\omega}_{l}^{\pm},\overline{\omega}_{v}^{\pm},\overline{\omega}_{\beta}^{\pm}\big)&=&\frac{3 \sqrt{3}\big(38392\overline{\omega}_{l}^{\pm}+100248\overline{\omega}_{v}^{\pm}+6771\overline{\omega}_{\beta}^{\pm}\big)}{16777216},\\
\mathcal{Q}^{\beta}_{22}\big(\overline{\omega}_{l}^{\pm},\overline{\omega}_{v}^{\pm},\overline{\omega}_{\beta}^{\pm}\big)&=&\displaystyle\frac{\sqrt{3}\big(4987\overline{\omega}_{l}^{\pm}+6618\overline{\omega}_{v}^{\pm}+540 \overline{\omega}_{\beta}^{\pm}\big)}{1024},\\
\mathcal{Q}^{\beta}_{31}\big(\overline{\omega}_{l}^{\pm},\overline{\omega}_{v}^{\pm},\overline{\omega}_{\beta}^{\pm}\big)&=&\displaystyle\frac{3\sqrt{3} \big(33422\overline{\omega}_{l}^{\pm}+53708\overline{\omega}_{v}^{\pm}+4115\overline{\omega}_{\beta}^{\pm}\big)}{16384},\\
\mathcal{Q}^{\beta}_{35}\big(\overline{\omega}_{l}^{\pm},\overline{\omega}_{v}^{\pm},\overline{\omega}_{\beta}^{\pm}\big)&=&\displaystyle\frac{\sqrt{3} \big(17114\overline{\omega}_{l}^{\pm}+41796\overline{\omega}_{v}^{\pm}+2865\overline{\omega}_{\beta}^{\pm}\big)}{65536},\\
\mathcal{Q}^{\beta}_{42}\big(\overline{\omega}_{l}^{\pm},\overline{\omega}_{v}^{\pm},\overline{\omega}_{\beta}^{\pm}\big)&=&\displaystyle\frac{\sqrt{3}}{1048576}\Big(692633616 \big(\overline{\omega}_{l}^{\pm}\big)^3+5166147320\big(\overline{\omega}_{l}^{\pm}\big)^2\big(\overline{\omega}_{v}^{\pm}\big)\\
&&+336981015 \big(\overline{\omega}_{l}^{\pm}\big)^2\big(\overline{\omega}_{\beta}^{\pm}\big)+11296361392\big(\overline{\omega}_{l}^{\pm}\big)\big(\overline{\omega}_{v}^{\pm}\big)^2\\
&&+1236584274\big(\overline{\omega}_{l}^{\pm}\big) \big(\overline{\omega}_{v}^{\pm}\big)\big(\overline{\omega}_{\beta}^{\pm}\big)+35603882\big(\overline{\omega}_{l}^{\pm}\big) \big(\overline{\omega}_{\beta}^{\pm}\big)^2\\
&&+6855639872\big(\overline{\omega}_{v}^{\pm}\big)^3+891106656\big(\overline{\omega}_{v}^{\pm}\big)^2\big(\overline{\omega}_{\beta}^{\pm}\big)\\
&&+30716436 \big(\overline{\omega}_{v}^{\pm}\big)\big(\overline{\omega}_{\beta}^{\pm}\big)^2+160544\big(\overline{\omega}_{\beta}^{\pm}\big)^3\Big),\\
\mathcal{Q}^{\beta}_{44}\big(\overline{\omega}_{l}^{\pm},\overline{\omega}_{v}^{\pm},\overline{\omega}_{\beta}^{\pm}\big)&=&\displaystyle\frac{\sqrt{3}}{41943040}\Big(220359572 \big(\overline{\omega}_{l}^{\pm}\big)^2-415245424 \big(\overline{\omega}_{l}^{\pm}\big)\big(\overline{\omega}_{v}^{\pm}\big)\\
&&-5303164\big(\overline{\omega}_{l}^{\pm}\big)\big(\overline{\omega}_{\beta}^{\pm}\big)-1632956848 \big(\overline{\omega}_{v}^{\pm}\big)^2\\
&&-188904056\big(\overline{\omega}_{v}^{\pm}\big)\big(\overline{\omega}_{\beta}^{\pm}\big)-5108353 \big(\overline{\omega}_{\beta}^{\pm}\big)^2\Big),\\
\mathcal{Q}^{\beta}_{48}\big(\overline{\omega}_{l}^{\pm},\overline{\omega}_{v}^{\pm},\overline{\omega}_{\beta}^{\pm}\big)&=&\displaystyle\frac{\sqrt{3}\big(8575\overline{\omega}_{l}^{\pm}+20970\overline{\omega}_{v}^{\pm}+1437 \overline{\omega}_{\beta}^{\pm}\big)}{1048576}.
\end{eqnarray*}

As depicted above, the truncated orders of $l_{\mu\nu}^{\pm}$, $v_{\mu\nu}^{\pm}$, and $\beta_{\mu\nu}^{\pm}$ in Eq.~(\ref{equ48}) result in $Q(l_{\mu\nu}^{\pm},v_{\mu\nu}^{\pm},\beta_{\mu\nu}^{\pm})$ should be truncated to $e^{10}$ order, from which, only $\gamma^{\pm}(p)\ (p=1,2,3,4)$ are provided, and consequently, the final determined orders of the optimized expressions of $l_{\mu\nu}^{\pm}$, $v_{\mu\nu}^{\pm}$, and $\beta_{\mu\nu}^{\pm}$ are the same as
their previous those shown in Eqs.~(\ref{equ32}) and (\ref{equ46}), respectively. Although the above optimized expressions of all the kinematic indicators seem lengthy, when their weights, namely, $\omega_{l}^{\pm}(\overline{\omega}_{l}^{\pm})$, $\omega_{v}^{\pm}(\overline{\omega}_{v}^{\pm})$, and $\omega_{\beta}^{\pm}(\overline{\omega}_{\beta}^{\pm})$ are given, all the above complex coefficients are degenerated into the corresponding numbers, so compared with their original expressions, the optimized those are indeed simpler and more compact. Physically speaking, with these optimized expressions of all the kinematic indicators in Eq.~(\ref{equ59}), $Q(l_{\mu\nu}^{\pm},v_{\mu\nu}^{\pm},\beta_{\mu\nu}^{\pm})$ reach their minimums, which means that
a set of reasonably determined weights can result in that Taiji triangles in both configurations become the most stable. Further, without doubt, following the above algorithm about optimization, no matter what the truncated orders of all the kinematic indicators are, Taiji triangles in both configurations can become as stable as possible, and therefore, above algorithm applies to the optimization of the inherent orbital variations of SCs involving any specific problem. Thus, as mentioned before, after such optimization of the orbits of SCs, the more stable formation of Taiji may contribute to selecting a simpler TDI technique to suppress the laser frequency noise and reducing the adverse effect brought by the Doppler shift of the laser frequency. Moreover, when considering the post-Newtonian effects of the Sun's gravitational field and the perturbative effects of some celestial bodies, the above algorithm can be readily generalized so that the more stable formation of Taiji can be obtained.
\section{Summary and discussions}

The space-based GW detectors like LISA~\cite{Dhurandhar:2004rv,Nayak:2006zm} or later Taiji~\cite{xuefei2011,Gong:2014mca,Hu:2017mde,Wu:2018clg} are becoming increasingly important, because the ground-based detectors are unable to detect GWs below $0.1$ Hz~\cite{Danzmann:1997hm,Harms:2013raa}. Like LISA,
Taiji is composed of three identical SCs orbiting the Sun and forming an  equilateral triangle whose arm-length is about $3\times10^6$ km. Taiji will observe GWs covering the range from $0.1$ mHz to $1.0$ Hz by using coherent laser beams exchanged between three SCs. In this paper, a new configuration for the orbits of Taiji SCs is proposed by finding the new relationship between the inclination $\varepsilon$ of the orbits of SCs with respect to the ecliptic plane and the orbital eccentricity $e$. The original configuration, designed for LISA~\cite{Dhurandhar:2004rv,Nayak:2006zm,Dhurandhar:2008yu,Pucacco:2010mn}, is studied as one part of the prestudy of Taiji~\cite{Wu:2019thj}. The orbits of SC$\kappa\ (\kappa=1,2,3)$ at every order in these two configurations
are symmetric about either $z$ axis or $x$-$y$ plane in the heliocentric coordinate system, which embodies the duality between them. In view that the trailing angle of Taiji constellation following the Earth from the viewpoint of the Sun can take values of $\pm\pi/9$, where the negative value means that the constellation
is preceding the Earth, and that in each case, Taiji has two symmetric orbits of SCs about the ecliptic plane, these two configurations, in practice, provide eight kinds of potential orbit schemes for Taiji.

For the unperturbed Keplerian orbits of SCs in both configurations of Taiji, an algorithm is devised to expand them to infinite order of $e$ in the heliocentric coordinate system. When the post-Newtonian effects of the Sun's gravitational field and the perturbative effects of some celestial bodies from Jupiter and the Moon etc. are considered, the unperturbed Keplerian orbits of SCs should be truncated to necessary order and then viewed as the zeroth-order approximation of the corresponding perturbative solution. Therefore, the algorithm lays the foundation for discussion of relativistic and perturbative effects on Taiji. Further, based on the algorithm, all the kinematic indicators of Taiji triangles in both configurations are also expanded to infinite order of $e$, where as a preliminary example, the expressions of arm-lengths and their rates of change to $e^5$ order, and the expressions of vertex angles to $e^4$ order are presented when $\phi^{\pm}=\pi/3$. These results imply that even $\phi^{\pm}=\pi/3$, Taiji triangles in both configurations are equilateral only up to their leading order terms, where to the higher order terms, Taiji triangles undergo the inherent variations.  Such inherent variation of Taiji could lower its sensitivity~\cite{Dhurandhar:2008yu}, so the inherent variation of Taiji triangle is significant in the data analysis,
e.g., the inherent variations of arm-lengths need to be deducted so as to acquire their accurate variations induced by GWs. By using the above algorithm, an accurate knowledge of the inherent variations of Taiji triangles in two configurations can be obtained. Moreover, with the above algorithm, it is proved that for both configurations of Taiji, three components of every kinematic indicator are identical to each other up to a phase shift of $2\pi/3$ at every order, which is independent on the value of the tilt angle of Taiji plane relative to the ecliptic plane.

The first generation TDI may not suppress the laser frequency noise effectively, because of the instability of Taiji triangle resulted from its inherent variation. The application of the second generation TDI~\cite{Dhurandhar:2008yu,Tinto:2003vj,Vallisneri:2005ji,Tinto:2014lxa} could possibly cause difficulty in the data analysis~\cite{Dhurandhar:2008yu} due to the complex non-commuting time-delay operators. Therefore, it is necessary to consider a simple TDI strategy for Taiji. In this paper, by adjusting $\phi^{\pm}$ around $\pi/3$ to any order of $e$, the orbits of SCs in both configurations of Taiji are optimized, respectively, which, as the case of original LISA (presented in Refs.~\cite{Dhurandhar:2004rv,Nayak:2006zm,Dhurandhar:2008yu}), may contribute to Taiji's selecting a simpler TDI technique. Technically, under the assumption~(\ref{equ47}), by slightly modifying the above algorithm, all the kinematic indicators of Taiji triangles in both configurations are first reexpanded, and their expressions certainly include the parameters $\gamma^{\pm}(1),\gamma^{\pm}(2)\cdots$ in assumption~(\ref{equ47}). Then, if a set of reasonably determined weights is given, by taking the minimums of $Q\left(l_{\mu\nu}^{\pm},v_{\mu\nu}^{\pm},\beta_{\mu\nu}^{\pm}\right)$, the parameters $\gamma^{\pm}(1),\gamma^{\pm}(2)\cdots$ can be derived one after another, and with them, the optimized expressions of all the kinematic indicators can be further obtained. Compared with their previous expressions, the optimized those are indeed simpler and more compact. Thus, following the above algorithm about optimization, Taiji triangles in both configurations can become as stable as possible with the different specific problem involved. As a preliminary example, the results of optimizing all the kinematic indicators in both configurations by adjusting $\phi^{\pm}$ around $\pi/3$ to $e^4$ order are provided in the present paper.
When the post-Newtonian effects of the Sun's gravitational field and the perturbative effects of some celestial bodies are considered, the above algorithm can be readily generalized so that the more stable
formation of Taiji can be obtained.

As mentioned in our previous paper~\cite{Wu:2019thj}, LISA and Taiji might be in operation at the same time for a period in the future, and based on the new configuration in this paper, there are more combinations available to be chosen. Moreover, these various combinations could be used to design the next generation space-based GW detector, which
may need more SCs to form a better configuration in order to improve the sensitivity and angle resolution of detecting GWs. The algorithm devised in the present paper actually applies to any space-based GW detector like LISA in triangular configuration, and the expansions of the unperturbed Keplerian orbits of SCs to infinite order of $e$ are essentially their complete series solutions. With these solutions, all the kinematic indicators can also be expressed in the form of series, which is the main idea of the algorithm. Moreover, by following the slightly modified algorithm about optimization, Taiji triangles in both configurations can become as stable as possible with the different specific problem involved. As far as we know, these results have not been given before for Taiji or LISA, so the results in the present paper may be useful for their development. Further, based on this algorithm, as mentioned earlier, the relativistic effect of the Sun's gravitational field and the perturbative effects of some celestial bodies can be taken into account, and thus, the analytic framework used to calculate the practical solutions of the orbits of SCs can be constructed in the following task, where in this framework, the series solutions in this paper need to be viewed as the zeroth-order approximation of the corresponding perturbative solution.
\begin{acknowledgements}
This work was supported, in part, by the Strategic Priority Research Program of the Chinese Academy of Sciences, Grants No. XDB23030100 and No. XDB23040000, by the National Natural Science Foundation of China (NSFC) under Grants No.~11690022 and No. 11635009, and by the Ministry of Science and Technology of the People's Republic of China
(2015CB856703).
\end{acknowledgements}

\appendix\label{appendix}
\section{Derivation of Eq.~(\ref{equ12}) and the barycenters of three SCs in both configurations of Taiji}
By reasonably inducing the result about the expansion of Kepler's equation in Ref.~\cite{moulton1960}, from Eq.~(\ref{equ11}), the expansions of $\cos\psi_{\kappa}^{\pm}$ and $\sin\psi_{\kappa}^{\pm}$ to infinite order of $e$ are, respectively,
\begin{equation}\label{equA1}
\left\{\begin{array}{lll}
\displaystyle \cos\psi_{\kappa}^{\pm}&=&\displaystyle \pm\frac{e}{2}-\sum_{n=0}^{\infty}\frac{(\mp1)^n}{n!}Q\left(\cos\psi_{\kappa},n\right)e^n,\smallskip\\
\displaystyle \sin\psi_{\kappa}^{\pm}&=&\displaystyle \sum_{n=0}^{\infty}\frac{(\mp1)^n}{(n+1)!}Q\left(\sin\psi_{\kappa},n\right)e^n
\end{array}\right.
\end{equation}
with $\kappa=1,2,3$ and
\begin{equation}\label{equA2}
\left\{\begin{array}{lll}
\displaystyle Q\left(\cos\psi_{\kappa},n\right)&:=&\displaystyle \sum_{k=0}^{\left[\frac{n}{2}\right]}\frac{(-1)^{k+1}}{2^n}C_{n+1}^k(n+1-2k)^{n-1}
\cos\left((n+1-2k)\sigma_{\kappa}\right),\smallskip\\
\displaystyle Q\left(\sin\psi_{\kappa},n\right)&:=&\displaystyle \sum_{k=0}^{\left[\frac{n}{2}\right]}\frac{(-1)^{k}}{2^n}C_{n+1}^k(n+1-2k)^{n}
\sin\left((n+1-2k)\sigma_{\kappa}\right).
\end{array}\right.
\end{equation}
To expand $\cos\varepsilon^{\pm}$ and $\sin\varepsilon^{\pm}$, the Taylor expansions of $\alpha^{\pm}$ need to be dealt with firstly. From Eqs.~(\ref{equ4}), (\ref{equ7}), and $\alpha^{\pm}=\sqrt{3}d^{\pm}/(2R)$, one can derive
\begin{eqnarray}
\label{equA3}\alpha^{\pm}=\frac{\sqrt{3}}{2}\sum_{n=1}^{\infty}(\pm1)^{n+1}Q\left(\alpha^{\pm},n\right)e^n
\end{eqnarray}
with
\begin{eqnarray}
\label{equA4}Q\left(\alpha^{\pm},n\right):=\sum_{k=0}^{\left[\frac{n}{2}\right]}\frac{(-1)^{n-k-1}(2n-2k-3)!!}{2^kk!(n-2k)!\left(\cos\phi^{\pm}\right)^{2n-2k-1}}.
\end{eqnarray}
Then, substituting above result to Eqs.~(\ref{equ5}) and (\ref{equ8}) gives the Taylor expansions of $\cos\varepsilon^{\pm}$ and $\sin\varepsilon^{\pm}$, respectively,
\begin{equation}\label{equA5}
\left\{\begin{array}{lll}
\displaystyle \cos\varepsilon^{\pm}&=&\displaystyle 1+\sum_{n=1}^{\infty}(\pm1)^n Q\left(\cos\varepsilon^{\pm},n\right)e^n,\smallskip\\
\displaystyle \sin\varepsilon^{\pm}&=&\displaystyle \sum_{n=1}^{\infty}(\pm1)^{n+1}Q\left(\sin\varepsilon^{\pm},n\right)e^n
\end{array}\right.
\end{equation}
with
\begin{equation}\label{equA6}
\left\{\begin{array}{lll}
\displaystyle Q\left(\cos\varepsilon^{\pm},n\right)&:=&\displaystyle (-1)^n+\sum_{k=0}^{n-1}(-1)^k Q\left(\alpha^{\pm},n-k\right)\cos\phi^{\pm},\smallskip\\
\displaystyle Q\left(\sin\varepsilon^{\pm},n\right)&:=&\displaystyle \sum_{k=0}^{n-1}(-1)^k Q\left(\alpha^{\pm},n-k\right)\sin\phi^{\pm}.
\end{array}\right.
\end{equation}

The combination of Eqs.~(\ref{equA1})---(\ref{equA6}) and the Taylor expansion of $\sqrt{1-e^2}$, namely,
\begin{eqnarray*}
\sqrt{1-e^2}=1-\sum_{k=2}^{\infty}\frac{(k-3)!!}{k!!}\cos^2\frac{k\pi}{2}\ e^k
\end{eqnarray*}
with $(-1)!!:=1$ can bring about Eq.~(\ref{equ12}). Then, the barycenters of three SCs in both configurations of Taiji are trivially derived by
\begin{equation}\label{equA7}
\boldsymbol{r}_{c}^{\pm}=\frac{1}{3}\left(\boldsymbol{r}_{1}^{\pm}+\boldsymbol{r}_{2}^{\pm}+\boldsymbol{r}_{3}^{\pm}\right)=
\left(x_{c}^{\pm}, y_{c}^{\pm}, z_{c}^{\pm}\right)
\end{equation}
with
\begin{equation}\label{equA8}
\left\{\begin{array}{l}
\displaystyle x_{c}^{\pm}=R\cos(\Omega t)+R\sum_{n=1}^{\infty}(\mp1)^n Q\left(x_{c}^{\pm},n\right)e^n,\smallskip\\
\displaystyle y_{c}^{\pm}=R\sin(\Omega t)+R\sum_{n=1}^{\infty}(\mp1)^n Q\left(y_{c}^{\pm},n\right)e^n,\smallskip\\
\displaystyle z_{c}^{\pm}=R\sum_{n=1}^{\infty}(\mp1)^{n-1} Q\left(z_{c}^{\pm},n\right)e^n,
\end{array}\right.
\end{equation}
where
\begin{equation}\label{equA9}
\left\{\begin{array}{lll}
\displaystyle Q\left(x_{c}^{\pm},n\right)&:=&\displaystyle \frac{1}{3}\sum_{\substack{k=0\\ k\neq1}}^{n}\sum_{j=0}^{\left[\frac{n-k}{2}\right]}\bigg(f_s^{\pm}(n,k,j)\sum_{\kappa=1}^3\cos\left((n-k+1-2j)\sigma_{\kappa}-\rho_{\kappa}\right)+g_s^{\pm}(n,k,j)\smallskip\\ &&\displaystyle\qquad\qquad\qquad\qquad\qquad\qquad\qquad\times\sum_{\kappa=1}^3\cos\left((n-k+1-2j)\sigma_{\kappa}+\rho_{\kappa}\right)\bigg)\smallskip,\\
\displaystyle Q\left(y_{c}^{\pm},n\right)&:=&\displaystyle -\frac{1}{3}\sum_{\substack{k=0\\ k\neq1}}^{n}\sum_{j=0}^{\left[\frac{n-k}{2}\right]}\bigg(f_s^{\pm}(n,k,j)\sum_{\kappa=1}^3\sin\left((n-k+1-2j)\sigma_{\kappa}-\rho_{\kappa}\right)-g_s^{\pm}(n,k,j)\smallskip\\
\displaystyle &&\displaystyle\qquad\qquad\qquad\qquad\qquad\qquad\qquad\times\sum_{\kappa=1}^3\sin\left((n-k+1-2j)\sigma_{\kappa}+\rho_{\kappa}\right)\bigg)\smallskip,\\
\displaystyle Q\left(z_{c}^{\pm},n\right)&:=&\displaystyle -C_v^{\pm}(n)-\frac{1}{3}\sum_{k=1}^{n}\sum_{j=0}^{\left[\frac{n-k}{2}\right]}h_s^{\pm}(n,k,j)\sum_{\kappa=1}^3\cos\left((n-k+1-2j)\sigma_{\kappa}\right).
\end{array}\right.
\end{equation}

\section{Expressions of $Q((\boldsymbol{r}_{\mu\nu}^{\pm})^2,n)$ as the functions of $t$ and the proof of the symmetry of $\beta^{\pm}_{\mu\nu}$}

By substituting Eqs.~(\ref{equ13}) and (\ref{equ18}) to Eq.~(\ref{equ22}), $Q((\boldsymbol{r}_{\mu\nu}^{\pm})^2,n)$ can be rewritten as the functions of $t$, namely,
\begin{eqnarray}
&&\label{equB1} Q\left(\left(\boldsymbol{r}_{\mu\nu}^{\pm}\right)^2,n\right)=
\sum_{p=1}^{n-1}\bigg[2C_{h}^{\pm}(n-p)C_{h}^{\pm}(p)\left(1-\cos\Big((\mu-\nu)\frac{2\pi}{3}\Big)\right)\nonumber\\
&+&2\sum_{\substack{k=0\\ k\neq1}}^{n-p}\sum_{j=0}^{\big[\frac{n-p-k}{2}\big]}\sum_{\substack{k'=0\\ k'\neq1}}^{p}\sum_{j'=0}^{\big[\frac{p-k'}{2}\big]}\bigg(f_s^{\pm}(n-p,k,j)f_s^{\pm}(p,k',j')\Big(\mathcal{C}^{F}_{\mu\nu}(\Omega t;\xi_{npkj}-\chi_{pk'j'},\xi_{npkj}-\chi_{pk'j'})\nonumber\\
&&\qquad\qquad\qquad\qquad\qquad\qquad-\mathcal{C}^{F}_{\mu\nu}(\Omega t;\xi_{npkj}+\chi_{pk'j'}+2,\xi_{npkj}-\chi_{pk'j'})\Big)\nonumber\\
&&\qquad\qquad\qquad\qquad\qquad\quad+g_s^{\pm}(n-p,k,j)g_s^{\pm}(p,k',j')\Big(\mathcal{C}^{F}_{\mu\nu}(\Omega t;\xi_{npkj}-\chi_{pk'j'},\xi_{npkj}-\chi_{pk'j'})\nonumber\\
&&\qquad\qquad\qquad\qquad\qquad\qquad-\mathcal{C}^{F}_{\mu\nu}(\Omega t;\xi_{npkj}+\chi_{pk'j'}-2,\xi_{npkj}-\chi_{pk'j'})\Big)\nonumber\\
&&\qquad\qquad\qquad\qquad\qquad\quad-2f_s^{\pm}(n-p,k,j)g_s^{\pm}(p,k',j')\Big(\mathcal{C}^{F}_{\mu\nu}(\Omega t;\xi_{npkj}-\chi_{pk'j'}+2,\xi_{npkj}+\chi_{pk'j'})\nonumber\\
&&\qquad\qquad\qquad\qquad\qquad\qquad-\mathcal{C}^{F}_{\mu\nu}(\Omega t;\xi_{npkj}+\chi_{pk'j'},\xi_{npkj}+\chi_{pk'j'})\Big)\bigg)\nonumber\\
&+&4\sum_{k=1}^{n-p}\sum_{j=0}^{\big[\frac{n-p-k}{2}\big]}\sum_{k'=1}^{p}\sum_{j'=0}^{\big[\frac{p-k'}{2}\big]}h_s^{\pm}(n-p,k,j)h_s^{\pm}(p,k',j')
\mathcal{S}^{F}_{\mu\nu}(\Omega t;\xi_{npkj},\xi_{npkj})\mathcal{S}^{F}_{\mu\nu}(\Omega t;\chi_{pk'j'},\chi_{pk'j'})\nonumber\\
&+&4\sum_{\substack{k=0\\ k\neq1}}^{n-p}\sum_{j=0}^{\big[\frac{n-p-k}{2}\big]}C_{h}^{\pm}(p)\bigg(f_s^{\pm}(n-p,k,j)
\Big(\mathcal{C}^{F}_{\mu\nu}(\Omega t;\xi_{npkj},\xi_{npkj})-\mathcal{C}^{F}_{\mu\nu}(\Omega t;\xi_{npkj}+2,\xi_{npkj})\Big)\nonumber\\
&&\qquad\qquad\qquad\qquad\ \ -g_s^{\pm}(n-p,k,j)\Big(\mathcal{C}^{F}_{\mu\nu}(\Omega t;\xi_{npkj}-2,\xi_{npkj})-\mathcal{C}^{F}_{\mu\nu}(\Omega t;\xi_{npkj},\xi_{npkj})\Big)\bigg)\bigg]
\end{eqnarray}
with
\begin{equation}\label{equB2}
\left\{\begin{array}{lll}
\displaystyle \mathcal{C}^{F}_{\mu\nu}(\Omega t;\epsilon,\eta)&:=&\displaystyle \cos\left(\epsilon(\mu-\nu)\frac{\pi}{3}\right)\cos\bigg(\eta\Big(\Omega t-(\mu+\nu-2)\frac{\pi}{3}\Big)\bigg),\smallskip\\
\displaystyle \mathcal{S}^{F}_{\mu\nu}(\Omega t;\epsilon,\eta)&:=&\displaystyle \sin\left(\epsilon(\mu-\nu)\frac{\pi}{3}\right)\sin\bigg(\eta\Big(\Omega t-(\mu+\nu-2)\frac{\pi}{3}\Big)\bigg),
\end{array}\right.
\end{equation}
and
\begin{equation*}
\left\{\begin{array}{lll}
\displaystyle \xi_{npkj}&:=&\displaystyle n-p-k-2j+1,\smallskip\\
\displaystyle \chi_{pk'j'}&:=&\displaystyle p-k'-2j'+1.
\end{array}\right.
\end{equation*}
This result can be used to
prove the symmetry of arm-lengths and their rates of change as done in Sec.~\ref{sec3.2}. Here,
we will prove that $\beta_{\mu\nu}^{\pm}$ also possess the same symmetry, and namely, their three components
are identical to each other up to a phase shift of $2\pi/3$ at every order, which does not depend on $\phi^{\pm}$. Technically, one only needs to prove that $Q(\beta_{\mu\nu}^{\pm},p)$ in Eq.~(\ref{equ42})
can be expressed as $G(\theta_{\mu\nu}(\Omega t))$, where $G$ is the corresponding function of a single variable, and $\theta_{\mu\nu}(\Omega t)$ is defined in Eq.~(\ref{equ23}). Firstly, from Eq.~(\ref{equ33}), $\beta_{\mu\nu}^{\pm}$ can be rewritten as the following form:
\begin{equation}\label{equB3}
\beta_{\mu\nu}^{\pm}=\arccos\frac{\boldsymbol{r}_{\mu\lambda}^{\pm}\cdot\boldsymbol{r}_{\nu\lambda}^{\pm}}{l_{\mu\lambda}^{\pm} l_{\nu\lambda}^{\pm}}\qquad\text{with}\quad \left(l_{\mu\lambda}^{\pm}l_{\nu\lambda}^{\pm}\right)^2=\left(\boldsymbol{r}_{\mu\lambda}^{\pm}\cdot\boldsymbol{r}_{\nu\lambda}^{\pm}\right)^2+\left(\boldsymbol{r}_{\mu\lambda}^{\pm}\times\boldsymbol{r}_{\nu\lambda}^{\pm}\right)^2,
\end{equation}
where
\begin{equation}\label{equB4}
\left\{\begin{array}{cll}
\displaystyle \boldsymbol{r}_{\mu\lambda}^{\pm}\cdot\boldsymbol{r}_{\nu\lambda}^{\pm}&=&\displaystyle x_{\mu\lambda}^{\pm}x_{\nu\lambda}^{\pm}+y_{\mu\lambda}^{\pm}y_{\nu\lambda}^{\pm}+z_{\mu\lambda}^{\pm}z_{\nu\lambda}^{\pm},\smallskip\\
\displaystyle \boldsymbol{r}_{\mu\lambda}^{\pm}\times\boldsymbol{r}_{\nu\lambda}^{\pm}&=&\displaystyle
\left(\big(\boldsymbol{r}_{\mu\lambda}^{\pm}\times\boldsymbol{r}_{\nu\lambda}^{\pm}\big)_{x},\big(\boldsymbol{r}_{\mu\lambda}^{\pm}\times\boldsymbol{r}_{\nu\lambda}^{\pm}\big)_{y},
\big(\boldsymbol{r}_{\mu\lambda}^{\pm}\times\boldsymbol{r}_{\nu\lambda}^{\pm}\big)_{z}\right)\smallskip\\
\displaystyle &=&\displaystyle \left(y_{\mu\lambda}^{\pm}z_{\nu\lambda}^{\pm}-y_{\nu\lambda}^{\pm}z_{\mu\lambda}^{\pm},z_{\mu\lambda}^{\pm}x_{\nu\lambda}^{\pm}-z_{\nu\lambda}^{\pm}x_{\mu\lambda}^{\pm},x_{\mu\lambda}^{\pm}y_{\nu\lambda}^{\pm}-x_{\nu\lambda}^{\pm}y_{\mu\lambda}^{\pm}\right).
\end{array}\right.
\end{equation}
Eq.~(\ref{equB3}) shows that one should begin to
deal with $\boldsymbol{r}_{\mu\lambda}^{\pm}\cdot\boldsymbol{r}_{\nu\lambda}^{\pm}$ and $\boldsymbol{r}_{\mu\lambda}^{\pm}\times\boldsymbol{r}_{\nu\lambda}^{\pm}$, and then,
from Eqs.~(\ref{equ17}) and (\ref{equ18}),
\begin{equation}\label{equB5}
\left\{\begin{array}{cll}
\displaystyle \boldsymbol{r}_{\mu\lambda}^{\pm}\cdot\boldsymbol{r}_{\nu\lambda}^{\pm}&=&\displaystyle R^2\sum_{n=2}^{\infty}(\mp1)^n Q\left(\boldsymbol{r}_{\mu\lambda}^{\pm}\cdot\boldsymbol{r}_{\nu\lambda}^{\pm},n\right)e^n,\smallskip\\
\displaystyle \big(\boldsymbol{r}_{\mu\lambda}^{\pm}\times\boldsymbol{r}_{\nu\lambda}^{\pm}\big)_{x}&=&\displaystyle R^2\sum_{n=2}^{\infty}(\mp1)^{n-1} Q\left(\big(\boldsymbol{r}_{\mu\lambda}^{\pm}\times\boldsymbol{r}_{\nu\lambda}^{\pm}\big)_{x},n\right)e^n,\smallskip\\
\displaystyle \big(\boldsymbol{r}_{\mu\lambda}^{\pm}\times\boldsymbol{r}_{\nu\lambda}^{\pm}\big)_{y}&=&\displaystyle R^2\sum_{n=2}^{\infty}(\mp1)^{n-1} Q\left(\big(\boldsymbol{r}_{\mu\lambda}^{\pm}\times\boldsymbol{r}_{\nu\lambda}^{\pm}\big)_{y},n\right)e^n,\smallskip\\
\displaystyle \big(\boldsymbol{r}_{\mu\lambda}^{\pm}\times\boldsymbol{r}_{\nu\lambda}^{\pm}\big)_{z}&=&\displaystyle R^2\sum_{n=2}^{\infty}(\mp1)^{n} Q\left(\big(\boldsymbol{r}_{\mu\lambda}^{\pm}\times\boldsymbol{r}_{\nu\lambda}^{\pm}\big)_{z},n\right)e^n
\end{array}\right.
\end{equation}
with
\begin{equation}\label{equB6}
\left\{\begin{array}{clll}
\displaystyle Q\left(\boldsymbol{r}_{\mu\lambda}^{\pm}\cdot\boldsymbol{r}_{\nu\lambda}^{\pm},n\right)&:=&\displaystyle\sum_{k=1}^{n-1}\Big(Q\big(x_{\mu\lambda}^{\pm},n-k\big)Q\big(x_{\nu\lambda}^{\pm},k\big)
+Q\big(y_{\mu\lambda}^{\pm},n-k\big)Q\big(y_{\nu\lambda}^{\pm},k\big)+Q\big(z_{\mu\lambda}^{\pm},n-k\big)Q\big(z_{\nu\lambda}^{\pm},k\big)\Big),\smallskip\\
\displaystyle Q\left(\big(\boldsymbol{r}_{\mu\lambda}^{\pm}\times\boldsymbol{r}_{\nu\lambda}^{\pm}\big)_{x},n\right)&:=&\displaystyle \sum_{k=1}^{n-1}\Big(Q\big(y_{\mu\lambda}^{\pm},n-k\big)Q\big(z_{\nu\lambda}^{\pm},k\big)-Q\big(y_{\nu\lambda}^{\pm},n-k\big)Q\big(z_{\mu\lambda}^{\pm},k\big)\Big),\smallskip\\
\displaystyle Q\left(\big(\boldsymbol{r}_{\mu\lambda}^{\pm}\times\boldsymbol{r}_{\nu\lambda}^{\pm}\big)_{y},n\right)&:=&\displaystyle \sum_{k=1}^{n-1}\Big(Q\big(z_{\mu\lambda}^{\pm},n-k\big)Q\big(x_{\nu\lambda}^{\pm},k\big)-Q\big(z_{\nu\lambda}^{\pm},n-k\big)Q\big(x_{\mu\lambda}^{\pm},k\big)\Big),\smallskip\\
\displaystyle Q\left(\big(\boldsymbol{r}_{\mu\lambda}^{\pm}\times\boldsymbol{r}_{\nu\lambda}^{\pm}\big)_{z},n\right)&:=&\displaystyle \sum_{k=1}^{n-1}\Big(Q\big(x_{\mu\lambda}^{\pm},n-k\big)Q\big(y_{\nu\lambda}^{\pm},k\big)-Q\big(x_{\nu\lambda}^{\pm},n-k\big)Q\big(y_{\mu\lambda}^{\pm},k\big)\Big),
\end{array}\right.
\end{equation}
and further,
\begin{equation}\label{equB7}
\left\{\begin{array}{cll}
\displaystyle \big(\boldsymbol{r}_{\mu\lambda}^{\pm}\cdot\boldsymbol{r}_{\nu\lambda}^{\pm}\big)^2&\phantom{}=&\displaystyle R^4\sum_{n=4}^{\infty}(\mp1)^n Q\Big(\big(\boldsymbol{r}_{\mu\lambda}^{\pm}\cdot\boldsymbol{r}_{\nu\lambda}^{\pm}\big)^2,n\Big)e^n,\smallskip\\
\displaystyle \big(\boldsymbol{r}_{\mu\lambda}^{\pm}\times\boldsymbol{r}_{\nu\lambda}^{\pm}\big)_{hs}
&:=&\displaystyle \left(\big(\boldsymbol{r}_{\mu\lambda}^{\pm}\times\boldsymbol{r}_{\nu\lambda}^{\pm}\big)_{x}\right)^2+\left(\big(\boldsymbol{r}_{\mu\lambda}^{\pm}\times\boldsymbol{r}_{\nu\lambda}^{\pm}\big)_{y}\right)^2= R^4\sum_{n=4}^{\infty}(\mp1)^n Q\Big(\big(\boldsymbol{r}_{\mu\lambda}^{\pm}\times\boldsymbol{r}_{\nu\lambda}^{\pm}\big)_{hs},n\Big)e^n,\smallskip\\
\displaystyle \big(\boldsymbol{r}_{\mu\lambda}^{\pm}\times\boldsymbol{r}_{\nu\lambda}^{\pm}\big)_{vs}&:=&\displaystyle
\left(\big(\boldsymbol{r}_{\mu\lambda}^{\pm}\times\boldsymbol{r}_{\nu\lambda}^{\pm}\big)_{z}\right)^2=
R^4\sum_{n=4}^{\infty}(\mp1)^n Q\Big(\big(\boldsymbol{r}_{\mu\lambda}^{\pm}\times\boldsymbol{r}_{\nu\lambda}^{\pm}\big)_{vs},n\Big)e^n
\end{array}\right.
\end{equation}
with
\begin{equation}\label{equB8}
\left\{\begin{array}{cll}
\displaystyle Q\Big(\big(\boldsymbol{r}_{\mu\lambda}^{\pm}\cdot\boldsymbol{r}_{\nu\lambda}^{\pm}\big)^2,n\Big)
&:=&\displaystyle\sum_{k=2}^{n-2}Q\Big(\boldsymbol{r}_{\mu\lambda}^{\pm}\cdot\boldsymbol{r}_{\nu\lambda}^{\pm},n-k\Big)
Q\Big(\boldsymbol{r}_{\mu\lambda}^{\pm}\cdot\boldsymbol{r}_{\nu\lambda}^{\pm},k\Big),\smallskip\\
\displaystyle Q\Big(\big(\boldsymbol{r}_{\mu\lambda}^{\pm}\times\boldsymbol{r}_{\nu\lambda}^{\pm}\big)_{hs},n\Big)
&:=&\displaystyle\sum_{k=2}^{n-2}\bigg(Q\Big(\big(\boldsymbol{r}_{\mu\lambda}^{\pm}\times\boldsymbol{r}_{\nu\lambda}^{\pm}\big)_{x},n-k\Big)Q\Big(\big(\boldsymbol{r}_{\mu\lambda}^{\pm}\times\boldsymbol{r}_{\nu\lambda}^{\pm}\big)_{x},k\Big)\smallskip\\
&&\qquad+Q\Big(\big(\boldsymbol{r}_{\mu\lambda}^{\pm}\times\boldsymbol{r}_{\nu\lambda}^{\pm}\big)_{y},n-k\Big)Q\Big(\big(\boldsymbol{r}_{\mu\lambda}^{\pm}\times\boldsymbol{r}_{\nu\lambda}^{\pm}\big)_{y},k\Big)\bigg),\smallskip\\
\displaystyle Q\Big(\big(\boldsymbol{r}_{\mu\lambda}^{\pm}\times\boldsymbol{r}_{\nu\lambda}^{\pm}\big)_{vs},n\Big)
&:=&\displaystyle\sum_{k=2}^{n-2}Q\left(\big(\boldsymbol{r}_{\mu\lambda}^{\pm}\times\boldsymbol{r}_{\nu\lambda}^{\pm}\big)_{z},n-k\right)
Q\left(\big(\boldsymbol{r}_{\mu\lambda}^{\pm}\times\boldsymbol{r}_{\nu\lambda}^{\pm}\big)_{z},k\right).
\end{array}\right.
\end{equation}
Then, by Eq.~(\ref{equB3}), one directly gets
\begin{eqnarray}
\label{equB9}\left(l_{\mu\lambda}^{\pm}l_{\nu\lambda}^{\pm}\right)^2=\displaystyle R^4\sum_{n=4}^{\infty} Q\Big(\big(l_{\mu\lambda}^{\pm}l_{\nu\lambda}^{\pm}\big)^2,n\Big)e^n
\end{eqnarray}
with
\begin{eqnarray}
\label{equB10}Q\Big(\big(l_{\mu\lambda}^{\pm}l_{\nu\lambda}^{\pm}\big)^2,n\Big):=\displaystyle (\mp1)^n Q\Big(\big(\boldsymbol{r}_{\mu\lambda}^{\pm}\cdot\boldsymbol{r}_{\nu\lambda}^{\pm}\big)^2,n\Big)+
(\mp1)^n Q\Big(\big(\boldsymbol{r}_{\mu\lambda}^{\pm}\times\boldsymbol{r}_{\nu\lambda}^{\pm}\big)_{hs},n\Big)+
(\mp1)^n Q\Big(\big(\boldsymbol{r}_{\mu\lambda}^{\pm}\times\boldsymbol{r}_{\nu\lambda}^{\pm}\big)_{vs},n\Big),
\end{eqnarray}
where from Eqs.~(\ref{equ26}), (\ref{equ28}), and (\ref{equ31}), there are
\begin{eqnarray}
\label{equB11}Q\Big(\big(l_{\mu\lambda}^{\pm}l_{\nu\lambda}^{\pm}\big)^2,4\Big)=Q\big(l_{\mu\lambda}^{\pm},1\big)^2Q\big(l_{\nu\lambda}^{\pm},1\big)^2=
a_{\mu\lambda}^{\pm}(2)a_{\nu\lambda}^{\pm}(2)>0.
\end{eqnarray}
Eq.~(\ref{equB9}) provides
\begin{eqnarray}
\label{equB12}\frac{1}{l_{\mu\lambda}^{\pm}l_{\nu\lambda}^{\pm}}=\frac{1}{e^2R^2}\bigg(\sum_{p=0}^{\infty} c^{\pm}_{\mu\nu\lambda}(p)e^{p}\bigg)^{-\frac{1}{2}}\qquad\text{with}\quad
c^{\pm}_{\mu\nu\lambda}(p):=Q\Big(\big(l_{\mu\lambda}^{\pm}l_{\nu\lambda}^{\pm}\big)^2,p+4\Big)
\end{eqnarray}
by $p:=n-4$. From Eqs.~(\ref{equB11}) and (\ref{equB12}), $(c^{\pm}_{\mu\nu\lambda}(0))^{-1/2}=(a_{\mu\lambda}^{\pm}(2)a_{\nu\lambda}^{\pm}(2))^{-1/2}\neq0$, and then, according to Eqs.~(\ref{equC11})---(\ref{equC13}) and (\ref{equC16}), $(\sum_{p=0}^{\infty}c^{\pm}_{\mu\nu\lambda}(p)e^{p})^{-1/2}$ in Eq.~(\ref{equB12}) can be expanded, and then,
substituting the obtained result to Eq.~(\ref{equB12}) gives
\begin{eqnarray}
\label{equB13}\frac{1}{l_{\mu\lambda}^{\pm}l_{\nu\lambda}^{\pm}}=\frac{1}{e^2R^2}\sum_{p=0}^{\infty} Q\bigg(\frac{1}{l_{\mu\lambda}^{\pm}l_{\nu\lambda}^{\pm}},p\bigg)e^{p}
\end{eqnarray}
with
\begin{eqnarray}
\label{equB14}Q\bigg(\frac{1}{l_{\mu\lambda}^{\pm}l_{\nu\lambda}^{\pm}},p\bigg):=\delta_{0p}\frac{1}{\sqrt{a_{\mu\lambda}^{\pm}(2)a_{\nu\lambda}^{\pm}(2)}}+\sum_{k=1}^{p}(-1)^{k}\frac{(2k-1)!!}{(2k)!!}
\frac{c^{(k)\pm}_{\mu\nu\lambda}(p)}{\left(\sqrt{a_{\mu\lambda}^{\pm}(2)a_{\nu\lambda}^{\pm}(2)}\right)^{2k+1}},
\end{eqnarray}
where
\begin{equation}\label{equB15}
c^{(k)\pm}_{\mu\nu\lambda}(p)=\left\{\begin{array}{l}
\displaystyle c^{\pm}_{\mu\nu\lambda}(p),\quad k=1,\smallskip\\
\displaystyle \sum_{j_{k-1}=k-1}^{p-1}\ \sum_{j_{k-2}=k-2}^{j_{k-1}-1}\cdots\sum_{j_{2}=2}^{j_{3}-1}\sum_{j_{1}=1}^{j_{2}-1}
c^{\pm}_{\mu\nu\lambda}(p-j_{k-1})c^{\pm}_{\mu\nu\lambda}(j_{k-1}-j_{k-2})\cdots c^{\pm}_{\mu\nu\lambda}(j_{2}-j_{1})c^{\pm}_{\mu\nu\lambda}(j_{1}),\quad k\geq2.
\end{array}\right.
\end{equation}

Thus, from Eqs.~(\ref{equB5}) and (\ref{equB13}), one can derive
\begin{eqnarray}
\label{equB16}B_{\mu\nu}^{\pm}=\cos\beta_{\mu\nu}^{\pm}=\frac{\boldsymbol{r}_{\mu\lambda}^{\pm}\cdot\boldsymbol{r}_{\nu\lambda}^{\pm}}{l_{\mu\lambda}^{\pm} l_{\nu\lambda}^{\pm}}=\sum_{p=0}^{\infty}Q\left(B_{\mu\nu}^{\pm},p\right)e^p
\end{eqnarray}
with
\begin{eqnarray}
\label{equB17}Q\left(B_{\mu\nu}^{\pm},p\right)=\sum_{k=0}^{p}(\mp1)^{p-k}Q\left(\boldsymbol{r}_{\mu\lambda}^{\pm}\cdot\boldsymbol{r}_{\nu\lambda}^{\pm},p-k+2\right)Q\bigg(\frac{1}{l_{\mu\lambda}^{\pm}l_{\nu\lambda}^{\pm}},k\bigg).
\end{eqnarray}

By substituting Eqs.~(\ref{equ13}) and (\ref{equ18}) to Eqs.~(\ref{equB6}) and (\ref{equB8}), $Q(\boldsymbol{r}_{\mu\lambda}^{\pm}\cdot\boldsymbol{r}_{\nu\lambda}^{\pm},n)$, $Q((\boldsymbol{r}_{\mu\lambda}^{\pm}\times\boldsymbol{r}_{\nu\lambda}^{\pm})_{z},n)$, and $Q((\boldsymbol{r}_{\mu\lambda}^{\pm}\times\boldsymbol{r}_{\nu\lambda}^{\pm})_{hs},n)$ can be rewritten as the functions of $t$, respectively,
\begin{eqnarray}
&&Q\left(\boldsymbol{r}_{\mu\lambda}^{\pm}\cdot\boldsymbol{r}_{\nu\lambda}^{\pm},n\right)=\sum_{p=1}^{n-1}
\bigg[4C_{h}^{\pm}(n-p)C_{h}^{\pm}(p)\mathcal{C}^{G}_{\mu\nu}(\Omega t;1,1)\nonumber\\
&+&4\sum_{\substack{k=0\\ k\neq1}}^{n-p}\sum_{j=0}^{\big[\frac{n-p-k}{2}\big]}\sum_{\substack{k'=0\\ k'\neq1}}^{p}\sum_{j'=0}^{\big[\frac{p-k'}{2}\big]}\Big(f_s^{\pm}(n-p,k,j)f_s^{\pm}(p,k',j')
\mathcal{C}^{G}_{\mu\nu}(\Omega t;\xi_{npkj}+1,\chi_{pk'j'}+1)\nonumber\\
&&\qquad\qquad\qquad\qquad\qquad\ +g_s^{\pm}(n-p,k,j)g_s^{\pm}(p,k',j')\mathcal{C}^{G}_{\mu\nu}(\Omega t;\xi_{npkj}-1,\chi_{pk'j'}-1)\nonumber\\
&&\qquad\qquad\qquad\qquad\qquad\ +f_s^{\pm}(n-p,k,j)g_s^{\pm}(p,k',j')\mathcal{C}^{G}_{\mu\nu}(\Omega t;\xi_{npkj}+1,-\chi_{pk'j'}+1)\nonumber\\
&&\qquad\qquad\qquad\qquad\qquad\ +g_s^{\pm}(n-p,k,j)f_s^{\pm}(p,k',j')
\mathcal{C}^{G}_{\mu\nu}(\Omega t;\xi_{npkj}-1,-\chi_{pk'j'}-1)\Big)\nonumber\\
&+&2\sum_{k=1}^{n-p}\sum_{j=0}^{\big[\frac{n-p-k}{2}\big]}\sum_{k'=1}^{p}\sum_{j'=0}^{\big[\frac{p-k'}{2}\big]}h_s^{\pm}(n-p,k,j)h_s^{\pm}(p,k',j')
\Big(\mathcal{C}^{G}_{\mu\nu}(\Omega t;\xi_{npkj},\chi_{pk'j'})\nonumber\\
&&\qquad\qquad\qquad\qquad\qquad\qquad\qquad\qquad\qquad\qquad\quad\ \ +\mathcal{C}^{G}_{\mu\nu}(\Omega t;\xi_{npkj},-\chi_{pk'j'})\Big)\nonumber\\
&+&4\sum_{\substack{k=0\\ k\neq1}}^{n-p}\sum_{j=0}^{\big[\frac{n-p-k}{2}\big]}C_{h}^{\pm}(p)\Big(f_s^{\pm}(n-p,k,j)\mathcal{C}^{G}_{\mu\nu}(\Omega t;\xi_{npkj}+1,1)\nonumber\\
&&\qquad\qquad\qquad\qquad\ +g_s^{\pm}(n-p,k,j)\mathcal{C}^{G}_{\mu\nu}(\Omega t;-\xi_{npkj}+1,1)\Big)\nonumber\\
&+&4\sum_{\substack{k'=0\\ k'\neq1}}^{p}\sum_{j'=0}^{\big[\frac{p'-k}{2}\big]}C_{h}^{\pm}(n-p)\Big(f_s^{\pm}(p,k',j')\mathcal{C}^{G}_{\mu\nu}(\Omega t;1,\chi_{pk'j'}+1)\nonumber\\
&&\label{equB18}\qquad\qquad\qquad\qquad\qquad+g_s^{\pm}(p,k',j')\mathcal{C}^{G}_{\mu\nu}(\Omega t;1,-\chi_{pk'j'}+1)\Big)\bigg],\\
&&Q\Big(\big(\boldsymbol{r}_{\mu\lambda}^{\pm}\times\boldsymbol{r}_{\nu\lambda}^{\pm}\big)_{z},n\Big)=\sum_{p=1}^{n-1}
\bigg[4C_{h}^{\pm}(n-p)C_{h}^{\pm}(p)\mathcal{S}^{G}_{\mu\nu}(\Omega t;1,1)\nonumber\\
&+&4\sum_{\substack{k=0\\ k\neq1}}^{n-p}\sum_{j=0}^{\big[\frac{n-p-k}{2}\big]}\sum_{\substack{k'=0\\ k'\neq1}}^{p}\sum_{j'=0}^{\big[\frac{p-k'}{2}\big]}\Big(f_s^{\pm}(n-p,k,j)f_s^{\pm}(p,k',j')
\mathcal{S}^{G}_{\mu\nu}(\Omega t;\xi_{npkj}+1,\chi_{pk'j'}+1)\nonumber\\
&&\qquad\qquad\qquad\qquad\qquad\ -g_s^{\pm}(n-p,k,j)g_s^{\pm}(p,k',j')
\mathcal{S}^{G}_{\mu\nu}(\Omega t;\xi_{npkj}-1,\chi_{pk'j'}-1)\nonumber\\
&&\qquad\qquad\qquad\qquad\qquad\ +f_s^{\pm}(n-p,k,j)g_s^{\pm}(p,k',j')
\mathcal{S}^{G}_{\mu\nu}(\Omega t;\xi_{npkj}+1,-\chi_{pk'j'}+1)\nonumber\\
&&\qquad\qquad\qquad\qquad\qquad\ -g_s^{\pm}(n-p,k,j)f_s^{\pm}(p,k',j')
\mathcal{S}^{G}_{\mu\nu}(\Omega t;\xi_{npkj}-1,-\chi_{pk'j'}-1)\Big)\nonumber\\
&+&4\sum_{\substack{k=0\\ k\neq1}}^{n-p}\sum_{j=0}^{\big[\frac{n-p-k}{2}\big]}C_{h}^{\pm}(p)\Big(f_s^{\pm}(n-p,k,j)\mathcal{S}^{G}_{\mu\nu}(\Omega t;\xi_{npkj}+1,1)\nonumber\\
&&\qquad\qquad\qquad\qquad\ +g_s^{\pm}(n-p,k,j)\mathcal{S}^{G}_{\mu\nu}(\Omega t;-\xi_{npkj}+1,1)\Big)\nonumber\\
&+&4\sum_{\substack{k'=0\\ k'\neq1}}^{p}\sum_{j'=0}^{\big[\frac{p-k'}{2}\big]}C_{h}^{\pm}(n-p)\Big(f_s^{\pm}(p,k',j')\mathcal{S}^{G}_{\mu\nu}(\Omega t;1,\chi_{pk'j'}+1)\nonumber\\
&&\label{equB19}\qquad\qquad\qquad\qquad\quad\ \ +g_s^{\pm}(p,k',j')\mathcal{S}^{G}_{\mu\nu}(\Omega t;1,-\chi_{pk'j'}+1)\Big)\bigg],
\end{eqnarray}
and
\begin{eqnarray}
&&\label{equB20} Q\Big(\big(\boldsymbol{r}_{\mu\lambda}^{\pm}\times\boldsymbol{r}_{\nu\lambda}^{\pm}\big)_{hs},n\Big)=\nonumber\\
&&\sum_{s=2}^{n-2}\bigg[16\sum_{p=1}^{n-s-1}\sum_{k'=1}^{p}\sum_{j'=0}^{\big[\frac{p-k'}{2}\big]}\sum_{p_{s}=1}^{s-1}\sum_{k_{s}'=1}^{p_{s}}\sum_{j_{s}'=0}^{\big[\frac{p_{s}-k_{s}'}{2}\big]}
C_{h}^{\pm}(n-s-p)h_s^{\pm}(p,k',j')C_{h}^{\pm}(s-p_{s})h_s^{\pm}(p_{s},k_{s}',j_{s}')\nonumber\\
&&\qquad\qquad\times\mathcal{N}^{H}_{\mu\nu}(\Omega t;\varsigma(-1,0,0),\varsigma(-1,0,0),\varsigma(p,k',j'),\varsigma(p_{s},k_{s}',j_{s}'))\nonumber\\
&+&16\sum_{p=1}^{n-s-1}\sum_{\substack{k=0\\ k\neq1}}^{n-s-p}\sum_{j=0}^{\big[\frac{n-s-p-k}{2}\big]}\sum_{k'=1}^{p}\sum_{j'=0}^{\big[\frac{p-k'}{2}\big]}
\sum_{p_{s}=1}^{s-1}\sum_{\substack{k_{s}=0\\ k_{s}\neq1}}^{s-p_{s}}\sum_{j_{s}=0}^{\big[\frac{s-p_{s}-k_{s}}{2}\big]}\sum_{k_{s}'=1}^{p_{s}}\sum_{j_{s}'=0}^{\big[\frac{p_{s}-k_{s}'}{2}\big]}
h_s^{\pm}(p,k',j')h_s^{\pm}(p_{s},k_{s}',j_{s}')\nonumber\\
&&\qquad\qquad\times\Big(f_s^{\pm}(n-s-p,k,j)f_s^{\pm}(s-p_{s},k_{s},j_{s})\nonumber\\
&&\qquad\qquad\quad\ \times\mathcal{N}^{H}_{\mu\nu}(\Omega t;\varsigma(n-s-p,k,j),\varsigma(s-p_{s},k_{s},j_{s}),\varsigma(p,k',j'),\varsigma(p_{s},k_{s}',j_{s}'))\nonumber\\
&&\qquad\qquad\quad\ +g_s^{\pm}(n-s-p,k,j)g_s^{\pm}(s-p_{s},k_{s},j_{s})\nonumber\\
&&\qquad\qquad\quad\ \times \mathcal{N}^{H}_{\mu\nu}(\Omega t;-\varsigma(n-s-p,k,j),-\varsigma(s-p_{s},k_{s},j_{s}),\varsigma(p,k',j'),\varsigma(p_{s},k_{s}',j_{s}'))\nonumber\\
&&\qquad\qquad\quad\ +2f_s^{\pm}(n-s-p,k,j)g_s^{\pm}(s-p_{s},k_{s},j_{s})\nonumber\\
&&\qquad\qquad\quad\ \times\mathcal{N}^{H}_{\mu\nu}(\Omega t;\varsigma(n-s-p,k,j),-\varsigma(s-p_{s},k_{s},j_{s}),\varsigma(p,k',j'),\varsigma(p_{s},k_{s}',j_{s}'))\Big)\nonumber\\
&+&32\sum_{p=1}^{n-s-1}\sum_{k'=1}^{p}\sum_{j'=0}^{\big[\frac{p-k'}{2}\big]}
\sum_{p_{s}=1}^{s-1}\sum_{\substack{k_{s}=0\\ k_{s}\neq1}}^{s-p_{s}}\sum_{j_{s}=0}^{\big[\frac{s-p_{s}-k_{s}}{2}\big]}\sum_{k_{s}'=1}^{p_{s}}\sum_{j_{s}'=0}^{\big[\frac{p_{s}-k_{s}'}{2}\big]}
C_{h}^{\pm}(n-s-p)h_s^{\pm}(p,k',j')h_s^{\pm}(p_{s},k_{s}',j_{s}')\nonumber\\
&&\qquad\qquad\times\Big(f_s^{\pm}(s-p_{s},k_{s},j_{s})
\mathcal{N}^{H}_{\mu\nu}(\Omega t;\varsigma(-1,0,0),\varsigma(s-p_{s},k_{s},j_{s}),\varsigma(p,k',j'),\varsigma(p_{s},k_{s}',j_{s}'))\nonumber\\
&&\qquad\qquad\quad+g_s^{\pm}(s-p_{s},k_{s},j_{s})
\mathcal{N}^{H}_{\mu\nu}(\Omega t;\varsigma(-1,0,0),-\varsigma(s-p_{s},k_{s},j_{s}),\varsigma(p,k',j'),\varsigma(p_{s},k_{s}',j_{s}'))\Big)\bigg]
\end{eqnarray}
with
\begin{equation*}
\varsigma(p,k,j):=p-k-2j+1,
\end{equation*}
where
\begin{equation}\label{equB21}
\left\{\begin{array}{lll}
\displaystyle \mathcal{C}^{G}_{\mu\nu}(\Omega t;\rho,\sigma)&:=&\displaystyle \sin\left(\rho(\mu-\lambda)\frac{\pi}{3}\right)\sin\left(\sigma(\nu-\lambda)\frac{\pi}{3}\right)
\cos\bigg(\rho\left(\Omega t-(\mu+\lambda-2)\frac{\pi}{3}\right)-\sigma\left(\Omega t-(\nu+\lambda-2)\frac{\pi}{3}\right)\bigg),\smallskip\\
\displaystyle \mathcal{S}^{G}_{\mu\nu}(\Omega t;\rho,\sigma)&:=&\displaystyle \sin\left(\rho(\mu-\lambda)\frac{\pi}{3}\right)\sin\left(\sigma(\nu-\lambda)\frac{\pi}{3}\right)
\sin\bigg(\rho\left(\Omega t-(\mu+\lambda-2)\frac{\pi}{3}\right)-\sigma\left(\Omega t-(\nu+\lambda-2)\frac{\pi}{3}\right)\bigg),
\end{array}\right.
\end{equation}
\begin{eqnarray}
\mathcal{N}^{H}_{\mu\nu}(\Omega t;\rho,\rho_{s},\sigma,\sigma_{s})&:=&\mathcal{K}^{H}_{\mu\nu}(\Omega t;\rho+1,\rho_{s}+1,\sigma,\sigma_{s},\rho-\rho_{s},0,0)\sin\bigg(\sigma\left(\Omega t-(\nu+\lambda-2)\frac{\pi}{3}\right)\bigg)\nonumber\\
&&\times\sin\bigg(\sigma_{s}\left(\Omega t-(\nu+\lambda-2)\frac{\pi}{3}\right)\bigg)+\mathcal{K}^{H}_{\mu\nu}(\Omega t;\sigma,\sigma_{s},\rho+1,\rho_{s}+1,0,\rho-\rho_{s},0)\nonumber\\
&&\times\sin\bigg(\sigma\left(\Omega t-(\mu+\lambda-2)\frac{\pi}{3}\right)\bigg)\sin\bigg(\sigma_{s}\left(\Omega t-(\mu+\lambda-2)\frac{\pi}{3}\right)\bigg)\nonumber\\
&\phantom{:}-&\mathcal{K}^{H}_{\mu\nu}(\Omega t;\rho+1,\sigma_{s},\rho_{s}+1,\sigma,\rho,-\rho_{s},-1)\sin\bigg(\sigma_{s}\left(\Omega t-(\mu+\lambda-2)\frac{\pi}{3}\right)\bigg)\nonumber\\
&&\times\sin\bigg(\sigma\left(\Omega t-(\nu+\lambda-2)\frac{\pi}{3}\right)\bigg)-\mathcal{K}^{H}_{\mu\nu}(\Omega t;\rho_{s}+1,\sigma,\rho+1,\sigma_{s},-\rho_{s},\rho,1)\nonumber\\
&&\label{equB22}\times\sin\bigg(\sigma\left(\Omega t-(\mu+\lambda-2)\frac{\pi}{3}\right)\bigg)\sin\bigg(\sigma_{s}\left(\Omega t-(\nu+\lambda-2)\frac{\pi}{3}\right)\bigg)
\end{eqnarray}
with
\begin{eqnarray*}
&&\mathcal{K}^{H}_{\mu\nu}(\Omega t;\iota_{1},\iota_{2},\iota_{3},\iota_{4},\vartheta_{1},\vartheta_{2},\vartheta_{3}):=\sin\left(\iota_{1}(\mu-\lambda)\frac{\pi}{3}\right)\sin\left(\iota_{2}(\mu-\lambda)\frac{\pi}{3}\right)
\sin\left(\iota_{3}(\nu-\lambda)\frac{\pi}{3}\right)\nonumber\\
&&\times\sin\left(\iota_{4}(\nu-\lambda)\frac{\pi}{3}\right)\cos\bigg(\vartheta_{1}\left(\Omega t-(\mu+\lambda-2)\frac{\pi}{3}\right)+\vartheta_{2}\left(\Omega t-(\nu+\lambda-2)\frac{\pi}{3}\right)+\vartheta_{3}(\mu-\nu)\frac{\pi}{3}\bigg).
\end{eqnarray*}
$\boldsymbol{r}_{\mu\lambda}^{\pm}\cdot\boldsymbol{r}_{\nu\lambda}^{\pm}$, $(\boldsymbol{r}_{\mu\lambda}^{\pm}\times\boldsymbol{r}_{\nu\lambda}^{\pm})_{hs}$, and $(\boldsymbol{r}_{\mu\lambda}^{\pm}\times\boldsymbol{r}_{\nu\lambda}^{\pm})_{vs}$
show that all of them are symmetric about $\mu,\nu$, so from
$(\boldsymbol{r}_{\mu\lambda}^{\pm}\times\boldsymbol{r}_{\nu\lambda}^{\pm})^2=(\boldsymbol{r}_{\mu\lambda}^{\pm}\times\boldsymbol{r}_{\nu\lambda}^{\pm})_{hs}+(\boldsymbol{r}_{\mu\lambda}^{\pm}\times\boldsymbol{r}_{\nu\lambda}^{\pm})_{vs}$
and Eq.~(\ref{equB3}), one only needs to consider $(\mu,\nu)\in\{(1,2),(2,3),(3,1)\}$ when proving the symmetry of $\beta^{\pm}_{\mu\nu}$. All the terms of
$Q(\boldsymbol{r}_{\mu\lambda}^{\pm}\cdot\boldsymbol{r}_{\nu\lambda}^{\pm},n)$, $Q((\boldsymbol{r}_{\mu\lambda}^{\pm}\times\boldsymbol{r}_{\nu\lambda}^{\pm})_{z},n)$, and $Q((\boldsymbol{r}_{\mu\lambda}^{\pm}\times\boldsymbol{r}_{\nu\lambda}^{\pm})_{hs},n)$
contain, respectively, $\mathcal{C}^{G}_{\mu\nu}(\Omega t;\rho,\sigma)$, $\mathcal{S}^{G}_{\mu\nu}(\Omega t;\rho,\sigma)$, and
$\mathcal{N}^{H}_{\mu\nu}(\Omega t;\rho,\rho_{s},\sigma,\sigma_{s})$
whose expressions refer to Eqs.~(\ref{equB21}) and (\ref{equB22}), where $\rho,\sigma,\rho_{s},\sigma_{s}$ are integers. By a direct calculation, if $(\mu,\nu)\in\{(1,2),(2,3),(3,1)\}$,
there are
\begin{eqnarray*}
\mathcal{C}^{G}_{\mu\nu}(\Omega t;\rho,\sigma)&=&\cos\Big((\rho-\sigma)\theta_{\mu\nu}(\Omega t)+(2\rho-\sigma)\frac{\pi}{3}\Big)\sin \Big(\frac{\rho\pi}{3}\Big)\sin\Big(\frac{2\sigma\pi}{3}\Big),\\
\mathcal{S}^{G}_{\mu\nu}(\Omega t;\rho,\sigma)&=&\sin\Big((\rho-\sigma)\theta_{\mu\nu}(\Omega t)+(2\rho-\sigma)\frac{\pi}{3}\Big)\sin \Big(\frac{\rho\pi}{3}\Big)\sin\Big(\frac{2\sigma\pi}{3}\Big),
\end{eqnarray*}
and
\begin{eqnarray*}
\mathcal{N}^{H}_{\mu\nu}(\Omega t;\rho,\rho_{s},\sigma,\sigma_{s})&:=&\mathcal{M}^{H}_{\mu\nu}(\Omega t;1,1,2,2,2,0,1,1)+\mathcal{M}^{H}_{\mu\nu}(\Omega t;2,2,1,1,1,0,2,2)\nonumber\\
&\phantom{:}+&\mathcal{M}^{H}_{\mu\nu}(\Omega t;2,1,1,2,2,-1,2,1)+\mathcal{M}^{H}_{\mu\nu}(\Omega t;1,2,2,1,1,1,1,2)
\end{eqnarray*}
with
\begin{eqnarray*}
&&\mathcal{M}^{H}_{\mu\nu}(\Omega t;\zeta_{1},\zeta_{2},\zeta_{3},\zeta_{4},\zeta_{5},\zeta_{6},\zeta_{7},\zeta_{8}):=\sin\Big(\zeta_{1}(\rho+1)\frac{\pi}{3}\Big)\sin\Big(\zeta_{2}(\rho_{s}+1)\frac{\pi}{3}\Big)
\sin\Big(\zeta_{3}\frac{\sigma\pi}{3}\Big)\sin\Big(\zeta_{4}\frac{\sigma_{s}\pi}{3}\Big)\\
&&\cos\bigg((\rho-\rho_{s})\Big(\theta_{\mu\nu}(\Omega t)+\zeta_{5}\frac{\pi}{3}\Big)+\zeta_{6}(\rho-2)\frac{\pi}{3}\bigg)\sin\bigg(\sigma\Big(\theta_{\mu\nu}(\Omega t)+\zeta_{7}\frac{\pi}{3}\Big)\bigg)\sin\bigg(\sigma_{s}\Big(\theta_{\mu\nu}(\Omega t)+\zeta_{8}\frac{\pi}{3}\Big)\bigg),
\end{eqnarray*}
and then, Eq.~(\ref{equB8}) implies that all of $Q(\boldsymbol{r}_{\mu\lambda}^{\pm}\cdot\boldsymbol{r}_{\nu\lambda}^{\pm},n)$, $Q((\boldsymbol{r}_{\mu\lambda}^{\pm}\cdot\boldsymbol{r}_{\nu\lambda}^{\pm})^2,n)$, $Q((\boldsymbol{r}_{\mu\lambda}^{\pm}\times\boldsymbol{r}_{\nu\lambda}^{\pm})_{hs},n)$, and $Q((\boldsymbol{r}_{\mu\lambda}^{\pm}\times\boldsymbol{r}_{\nu\lambda}^{\pm})_{vs},n)$ can be expressed as
the functions of $\theta_{\mu\nu}(\Omega t)$, respectively. Therefore, from Eqs.~(\ref{equB10})---(\ref{equB12}), (\ref{equB14}), (\ref{equB15}), and (\ref{equB17}), $Q\left(B_{\mu\nu}^{\pm},p\right)$ can also be expressed as the functions of $\theta_{\mu\nu}(\Omega t)$, and further with Eqs.~(\ref{equ43}) and (\ref{equ44}), one finally concludes that $Q(\beta_{\mu\nu}^{\pm},p)$ in Eq.~(\ref{equ42})
can be expressed as $G(\theta_{\mu\nu}(\Omega t))$.

\section{Expansion of $f(\sum_{p=0}^{\infty}d(p)e^p)$ with $f(d(0))\neq0$ to infinite order of $e$}

To expand $f(\sum_{p=0}^{\infty}d(p)e^p)$ with $f(d(0))\neq0$ to infinite order of $e$, the expansion of $f(\sum_{p=0}^{i}d(p)e^p)$ with $i$ as any positive integer to $e^i$ order should be first taken into account, and then, there should be
\begin{eqnarray}
\label{equC1}f\left(\sum_{p=0}^{\infty}d(p)e^p\right)=\lim_{i\rightarrow\infty}f\left(\sum_{p=0}^{i}d(p)e^p\right).
\end{eqnarray}
$f(\sum_{p=0}^{i}d(p)e^p)$ can be rewritten as the following form,
\begin{eqnarray}
\label{equC2}f\left(\sum_{p=0}^{i}d(p)e^p\right)=f\left(d(0)+\Delta\right)\qquad \text{with}\quad \Delta:=\sum_{p=1}^{i}d(p)e^p,
\end{eqnarray}
and because $\Delta$ is small, the further Taylor expansion gives
\begin{eqnarray}
\label{equC3}f\left(\sum_{p=0}^{i}d(p)e^p\right)=\sum_{k=0}^{i}\frac{f^{(k)}\left(d(0)\right)}{k!}\Delta^{k},
\end{eqnarray}
where $f^{(k)}$ is the $k$th derivative of $f$ with $f^{(0)}=f$. Eq.~(\ref{equC3}) implies that the expansion of $\Delta^{k}\ (1\leq k\leq i)$ to $e^i$ order needs to be dealt with.

Next, by induction, we will derive $\Delta_{i}^{k}=(\sum_{p=1}^{i}d(p)e^p)^{k}_{i}\ (1\leq k\leq i)$, where the subscript $i$ means that only the expansion of $\Delta^{k}$ to $e^i$ order is kept. For $k=1$ and $k=2$, the expansions are trivial,
\begin{eqnarray}
\label{equC4}\left(\sum_{p=1}^{i}d(p)e^p\right)_{i}^{1}&=&\sum_{p=1}^{i}d^{(1)}(p)e^p\qquad \text{with}\quad d^{(1)}(p):=d(p),\\
\label{equC5}\left(\sum_{p=1}^{i}d(p)e^p\right)_{i}^{2}&=&\sum_{p=2}^{i}\left(d(p-1)d(1)+d(p-2)d(2)+\cdots+d(1)d(p-1)\right)e^p\nonumber\\
&=&\sum_{p=2}^{i}d^{(2)}(p)e^p\qquad \text{with}\quad d^{(2)}(p):=\sum_{j_{1}=1}^{p-1}d(p-j_{1})d(j_{1}).
\end{eqnarray}
Then, Eqs.~(\ref{equC4}) and (\ref{equC5}) can be used to derive the expansion for $k=3$, namely,
\begin{eqnarray}
\label{equC6}\left(\sum_{p=1}^{i}d(p)e^p\right)_{i}^{3}&=&\left[\left(\sum_{p=1}^{i}d(p)e^p\right)^{1}\left(\sum_{p=1}^{i}d(p)e^p\right)^{2}\right]_{i}\nonumber\\
&=&\left[\left(\sum_{p=1}^{i}d^{(1)}(p)e^p\right)\left(\sum_{p=2}^{i}d^{(2)}(p)e^p\right)\right]_{i}\nonumber\\
&=&\sum_{p=3}^{i}\left(d^{(1)}(p-2)d^{(2)}(2)+d^{(1)}(p-3)d^{(2)}(3)+\cdots+d^{(1)}(1)d^{(2)}(p-1)\right)e^p\nonumber\\
&=&\sum_{p=3}^{i}d^{(3)}(p)e^p
\end{eqnarray}
with
\begin{eqnarray}
\label{equC7} d^{(3)}(p):=\sum_{j_{2}=2}^{p-1}d^{(1)}(p-j_{2})d^{(2)}(j_{2})=\sum_{j_{2}=2}^{p-1}\sum_{j_{1}=1}^{j_{2}-1}d(p-j_{2})d(j_{2}-j_{1})d(j_{1}).
\end{eqnarray}
Similarly, the expansion for $k=4$ can be further derived, and there are
\begin{eqnarray}
\label{equC8}\left(\sum_{p=1}^{i}d(p)e^p\right)_{i}^{4}&=&\sum_{p=4}^{i}d^{(4)}(p)e^p
\end{eqnarray}
with
\begin{eqnarray}
\label{equC9} d^{(4)}(p):=\sum_{j_{3}=3}^{p-1}\sum_{j_{2}=2}^{j_{3}-1}\sum_{j_{1}=1}^{j_{2}-1}d(p-j_{3})
d(j_{3}-j_{2})d(j_{2}-j_{1})d(j_{1}).
\end{eqnarray}
Repeating the same procedure, the expansions for $5\leq k\leq i$ can also be obtained, and then, one arrives at
\begin{eqnarray}
\label{equC10}\Delta_{i}^{k}=\left(\sum_{p=1}^{i}d(p)e^p\right)_{i}^{k}=\sum_{p=k}^{i}d^{(k)}(p)e^p
\end{eqnarray}
with
\begin{equation}\label{equC11}
d^{(k)}(p):=\left\{\begin{array}{l}
\displaystyle d(p),\quad k=1,\smallskip\\
\displaystyle \sum_{j_{k-1}=k-1}^{p-1}\ \sum_{j_{k-2}=k-2}^{j_{k-1}-1}\cdots\sum_{j_{2}=2}^{j_{3}-1}\sum_{j_{1}=1}^{j_{2}-1}
d(p-j_{k-1})d(j_{k-1}-j_{k-2})\cdots d(j_{2}-j_{1})d(j_{1}),\quad 2\leq k\leq i.
\end{array}\right.
\end{equation}

Plugging Eq.~(\ref{equC10}) into Eq.~(\ref{equC3}) and setting $i\rightarrow\infty$ give
\begin{eqnarray}
\label{equC12}f\left(\sum_{p=0}^{\infty}d(p)e^p\right)=\sum_{p=0}^{\infty}Q\left(f,p\right)e^p
\end{eqnarray}
with
\begin{eqnarray}
\label{equC13}Q\left(f,p\right):=\delta_{0p}f\left(d(0)\right)+\sum_{k=1}^{p}\frac{f^{(k)}\left(d(0)\right)}{k!}d^{(k)}(p).
\end{eqnarray}

For a definite function $f$, in order to apply the above result, one needs to know the expression of $f^{(k)}$, and here, some typical examples are given.
\begin{itemize}
\item $f(\zeta)=\zeta^{m}$ with $m$ as any real number.
\end{itemize}
One can easily prove the following formula.
\begin{eqnarray}
\label{equC14}f^{(k)}(\zeta)&=&k!C_{m}^{k}\zeta^{m-k}
\end{eqnarray}
with $C_{m}^{k}:=m(m-1)\cdots(m-k+1)/k!$ as the generalized binomial coefficients, where if $m$ is any positive integer
and $k>m$, $C_{m}^{k}=0$. The following two formulas can be derived by this result.
\begin{eqnarray}
\label{equC15}f^{(k)}(\zeta)&=&(-1)^{k-1}\frac{(2k-3)!!}{2^k}\zeta^{-\frac{2k-1}{2}},\qquad\text{for}\quad f(\zeta)=\sqrt{\zeta},\\
\label{equC16}f^{(k)}(\zeta)&=&(-1)^{k}\frac{(2k-1)!!}{2^k}\zeta^{-\frac{2k+1}{2}},\qquad\ \ \ \text{for}\quad f(\zeta)=\frac{1}{\sqrt{\zeta}},
\end{eqnarray}
where $(-3)!!:=-1$.
\begin{itemize}
\item $f(\zeta)=\arccos(\zeta).$
\end{itemize}
By induction, one can obtain the following formula.
\begin{eqnarray}
\label{equC17}f^{(k)}(\zeta)&=&-\sum_{p=0}^{\left[\frac{k-1}{2}\right]}\frac{(2k-2p-3)!!(k-1)!}{(2p)!!(k-2p-1)!}\frac{\zeta^{k-2p-1}}{(1-\zeta^2)^{\frac{2k-1}{2}-p}},\qquad k\geq1.
\end{eqnarray}
Now, we will prove it. If $k=1$, Eq.~(\ref{equC17}) gives $f^{(1)}(\zeta)=-1/(1-\zeta^2)^{1/2}$, which holds.
Suppose that when $k=l\geq1$, Eq.~(\ref{equC17}) holds, namely,
\begin{eqnarray*}
f^{(l)}(\zeta)=-\sum_{p=0}^{\left[\frac{l-1}{2}\right]}\frac{a^{l}_{l-1-2p}\zeta^{l-2p-1}}{(1-\zeta^2)^{\frac{2l-1}{2}-p}},\quad a^{l}_{l-1-2p}:=\frac{(2l-2p-3)!!(l-1)!}{(2p)!!(l-2p-1)!}.
\end{eqnarray*}
Then,
\begin{eqnarray*}
f^{(l+1)}(\zeta)&=&-\sum_{p=0}^{\left[\frac{l-1}{2}\right]}\frac{(2l-2p-1)a^{l}_{l-1-2p}\zeta^{l-2p}}{(1-\zeta^2)^{\frac{2l-1}{2}-p+1}}-\sum_{p=0}^{\left[\frac{l-1}{2}\right]}\frac{(l-2p-1)a^{l}_{l-1-2p}\zeta^{l-2p-2}}{(1-\zeta^2)^{\frac{2l-1}{2}-p}}\\
&=&-\frac{(2l-1)a^{l}_{l-1}\zeta^{l}}{(1-\zeta^2)^{\frac{2l+1}{2}}}
-\sum_{p=1}^{\left[\frac{l-1}{2}\right]}\frac{(2l-2p-1)a^{l}_{l-1-2p}\zeta^{l-2p}}{(1-\zeta^2)^{\frac{2l+1}{2}-p}}-\sum_{p'=1}^{\left[\frac{l-1}{2}\right]+1}\frac{(l-2p'+1)a^{l}_{l-1-2p'+2}\zeta^{l-2p'}}{(1-\zeta^2)^{\frac{2l+1}{2}-p'}}\\
&=&-\frac{(2l-1)a^{l}_{l-1}\zeta^{l}}{(1-\zeta^2)^{\frac{2l+1}{2}}}
-\sum_{p=1}^{\left[\frac{l-1}{2}\right]}\frac{\left((2l-2p-1)a^{l}_{l-1-2p}+(l-2p+1)a^{l}_{l-1-2(p-1)}\right)\zeta^{l-2p}}{(1-\zeta^2)^{\frac{2l+1}{2}-p}}\\
&&-\frac{(l-2p+1)a^{l}_{l-1-2(p-1)}\zeta^{l-2p}}{(1-\zeta^2)^{\frac{2l+1}{2}-p}}\bigg|_{p=\left[\frac{l-1}{2}\right]+1}\\
&=&-\sum_{p=0}^{\left[\frac{l-1}{2}\right]}\frac{(2l-2p-1)!!\ l!}{(2p)!!(l-2p)!}\frac{\zeta^{l-2p}}{(1-\zeta^2)^{\frac{2l+1}{2}-p}}-\frac{(l-2p+1)a^{l}_{l-1-2(p-1)}\zeta^{l-2p}}{(1-\zeta^2)^{\frac{2l+1}{2}-p}}\bigg|_{p=\left[\frac{l-1}{2}\right]+1}\\
&=&-\sum_{p=0}^{\left[\frac{l}{2}\right]}\frac{(2l-2p-1)!!\ l!}{(2p)!!(l-2p)!}\frac{\zeta^{l-2p}}{(1-\zeta^2)^{\frac{2l+1}{2}-p}},
\end{eqnarray*}
where in the second step, $p':=p+1$, in the last second step,
\begin{eqnarray*}
(2l-1)!!&=&(2l-1)a^{l}_{l-1},\\
\frac{(2l-2p-1)!!\ l!}{(2p)!!(l-2p)!}&=&(2l-2p-1)a^{l}_{l-1-2p}+(l-2p+1)a^{l}_{l-1-2(p-1)}
\end{eqnarray*}
have been used, and in the last step,
when $l$ is even, by using $[(l-1)/2]+1=[l/2]=l/2$, one gets
\begin{eqnarray*}
-\frac{(l-2p+1)a^{l}_{l-1-2(p-1)}\zeta^{l-2p}}{(1-\zeta^2)^{\frac{2l+1}{2}-p}}\bigg|_{p=\left[\frac{l-1}{2}\right]+1}=
-\frac{(2l-2p-1)!!\ l!}{(2p)!!(l-2p)!}\frac{\zeta^{l-2p}}{(1-\zeta^2)^{\frac{2l+1}{2}-p}}\bigg|_{p=\left[\frac{l}{2}\right]},
\end{eqnarray*}
and when $l$ is odd, $[(l-1)/2]=[l/2]\Leftrightarrow[(l-1)/2]+1=[l/2]+1$, which means that the above term vanishes.
\begin{itemize}
\item $f(\zeta)=\sec(\zeta).$
\end{itemize}
In Ref.~\cite{Anotonio2014}, the following formula is presented.
\begin{eqnarray}
\label{equC18}f^{(k)}(\zeta)=\sec\zeta\left(\sqrt{-1}\right)^k\sum _{j=0}^k(-1)^{j}j!\sum_{p=j}^{k} C_{k}^{p}S(p,j)2^{p-j}\big(\sqrt{-1}\tan\zeta+1\big)^j,
\end{eqnarray}
where $S(p,j)$ denotes the Stirling number of the second kind.
\begin{itemize}
\item $f(\zeta)=\cos(\zeta).$
\end{itemize}
The following formula is readily derived.
\begin{eqnarray}
\label{equC19}f^{(k)}(\zeta)=(-1)^{\left[\frac{k+1}{2}\right]}\left(\frac{1+(-1)^k}{2}\cos\zeta+\frac{1-(-1)^k}{2}\sin\zeta\right).
\end{eqnarray}
\begin{itemize}
\item $f(\zeta)=\sin(\zeta).$
\end{itemize}
From Eq.~(\ref{equC19}), one can get the following formula.
\begin{eqnarray}
\label{equC20}f^{(k)}(\zeta)=(-1)^{\left[\frac{k}{2}\right]}\left(\frac{1-(-1)^{k}}{2}\cos\zeta+\frac{1+(-1)^{k}}{2}\sin\zeta\right).
\end{eqnarray}

\section{Reexpansions of $\cos\varepsilon^{\pm}$ and $\sin\varepsilon^{\pm}$ based on Eq.~(\ref{equ47})}

Eqs.~(\ref{equ5}) and (\ref{equ8}) show that $\alpha^{\pm}$ need to be first reexpanded, and then, one can acquire the reexpansions of $\cos\varepsilon^{\pm}$ and $\sin\varepsilon^{\pm}$. From Eq.~(\ref{equA4}), the expansions of $\sec\phi^{\pm}=\sec(\sum_{p=0}^{\infty}\gamma^{\pm}(p)e^p)$ need to be taken into account, and with Eqs.~(\ref{equC11})---(\ref{equC13}) and (\ref{equC18}), there is
\begin{eqnarray}
\label{equD1}\sec\phi^{\pm}=\sum_{p=0}^{\infty}Q\left(\sec\phi^{\pm},p\right)e^p
\end{eqnarray}
with
\begin{eqnarray}
\label{equD2}Q\left(\sec\phi^{\pm},p\right)
:=2\delta_{0p}+\sum_{k=1}^{p}\frac{(\sec)^{(k)}\left(\pi/3\right)}{k!}\gamma^{(k)\pm}(p),
\end{eqnarray}
where
\begin{equation}\label{equD3}
\gamma^{(k)\pm}(p)=\left\{\begin{array}{l}
\displaystyle \gamma^{\pm}(p),\quad k=1,\smallskip\\
\displaystyle \sum_{j_{k-1}=k-1}^{p-1}\ \sum_{j_{k-2}=k-2}^{j_{k-1}-1}\cdots\sum_{j_{2}=2}^{j_{3}-1}\sum_{j_{1}=1}^{j_{2}-1}
\gamma^{\pm}(p-j_{k-1})\gamma^{\pm}(j_{k-1}-j_{k-2})\cdots\gamma^{\pm}(j_{2}-j_{1})\gamma^{\pm}(j_{1}),\quad k\geq2.
\end{array}\right.
\end{equation}
Then, Eq.~(\ref{equA4}) shows that $\sec^{m}\phi^{\pm}=(\sum_{p=0}^{\infty}Q(\sec\phi^{\pm},p)e^p)^{m}$ with $m=2n-2k-1$ should be expanded, and using Eqs.~(\ref{equC11})---(\ref{equC14}) gives
\begin{eqnarray}
\label{equD4}\sec^{m}\phi^{\pm}=\sum_{p=0}^{\infty}Q\left(\sec^{m}\phi^{\pm},p\right)e^p
\end{eqnarray}
with
\begin{eqnarray}
\label{equD5}Q\left(\sec^{m}\phi^{\pm},p\right)
:=2^{m}\delta_{0p}+\sum_{l=1}^{p}C_{m}^{l}2^{m-l}Q^{(l)}(\sec\phi^{\pm},p),
\end{eqnarray}
where $Q(\sec\phi^{\pm},0)=2$ and
\begin{equation}\label{equD6}
Q^{(l)}(\sec\phi^{\pm},p)=\left\{\begin{array}{l}
\displaystyle Q(\sec\phi^{\pm},p),\quad l=1,\smallskip\\
\displaystyle \sum_{j_{l-1}=l-1}^{p-1}\ \sum_{j_{l-2}=l-2}^{j_{l-1}-1}\cdots\sum_{j_{2}=2}^{j_{3}-1}\sum_{j_{1}=1}^{j_{2}-1}
Q\left(\sec\phi^{\pm},p-j_{l-1}\right)Q\left(\sec\phi^{\pm},j_{l-1}-j_{l-2}\right)\cdots\\
\displaystyle\qquad\qquad\qquad\qquad\qquad\qquad\times Q\left(\sec\phi^{\pm},j_{2}-j_{1}\right)Q\left(\sec\phi^{\pm},j_{1}\right),\quad l\geq2.
\end{array}\right.
\end{equation}
By substituting Eq.~(\ref{equD4}) to Eqs.~(\ref{equA3}) and (\ref{equA4}), $\alpha^{\pm}$ are reexpanded by only modifying the expressions of $Q(\alpha^{\pm},n)$, namely,
\begin{eqnarray}
\label{equD7}Q\left(\alpha^{\pm},n\right):=\sum_{s=0}^{n-1}(\pm1)^s\sum_{k=0}^{\left[\frac{n-s}{2}\right]}(-1)^{n-s-k-1}
\frac{(2n-2s-2k-3)!!}{2^kk!(n-s-2k)!}Q\left(\left(\sec\phi^{\pm}\right)^{2n-2s-2k-1},s\right).
\end{eqnarray}

One also needs to expand $\cos\phi^{\pm}=\cos(\sum_{p=0}^{\infty}\gamma^{\pm}(p)e^p)$ and $\sin\phi^{\pm}=\sin(\sum_{p=0}^{\infty}\gamma^{\pm}(p)e^p)$ in Eq.~(\ref{equA6}), which is easy to deal with
by use of Eqs.~(\ref{equC11})---(\ref{equC13}) again, and here, we directly present their results together,
\begin{equation}\label{equD8}
\left\{\begin{array}{lll}
\displaystyle \cos\phi^{\pm}&=&\displaystyle \sum_{p=0}^{\infty}Q\left(\cos\phi^{\pm},p\right)e^p,\smallskip\\
\displaystyle \sin\phi^{\pm}&=&\displaystyle \sum_{p=0}^{\infty}Q\left(\sin\phi^{\pm},p\right)e^p
\end{array}\right.
\end{equation}
with
\begin{equation}\label{equD9}
\left\{\begin{array}{lll}
\displaystyle Q\left(\cos\phi^{\pm},p\right)
&:=&\displaystyle \frac{1}{2}\delta_{0p}+\sum_{k=1}^{p}\frac{(\cos)^{(k)}\left(\pi/3\right)}{k!}\gamma^{(k)\pm}(p),\smallskip\\
\displaystyle Q\left(\sin\phi^{\pm},p\right)
&:=&\displaystyle \frac{\sqrt{3}}{2}\delta_{0p}+\sum_{k=1}^{p}\frac{(\sin)^{(k)}\left(\pi/3\right)}{k!}\gamma^{(k)\pm}(p),
\end{array}\right.
\end{equation}
where the expressions of $\gamma^{(k)\pm}(p)$ refer to Eq.~(\ref{equD3}), and the expressions of $(\cos)^{(k)}$ and $(\sin)^{(k)}$ refer to Eqs.~(\ref{equC19}) and (\ref{equC20}), respectively. The combination of Eqs.~(\ref{equD4}) and (\ref{equD8}) provides
\begin{equation}\label{equD10}
\left\{\begin{array}{lllll}
\displaystyle \Phi_{c}^{\pm}(m)&:=&\displaystyle \sec^{m}\phi^{\pm}\cos\phi^{\pm}&=&\displaystyle \sum_{p=0}^{\infty}Q\left(\Phi_{c}^{\pm}(m),p\right)e^p,\smallskip\\
\displaystyle \Phi_{s}^{\pm}(m)&:=&\displaystyle \sec^{m}\phi^{\pm}\sin\phi^{\pm}&=&\displaystyle \sum_{p=0}^{\infty}Q\left(\Phi_{s}^{\pm}(m),p\right)e^p
\end{array}\right.
\end{equation}
with
\begin{equation}\label{equD11}
\left\{\begin{array}{lll}
\displaystyle Q\left(\Phi_{c}^{\pm}(m),p\right)
&:=&\displaystyle \sum_{k=0}^{p}Q\left(\sec^{m}\phi^{\pm},p-k\right)Q\left(\cos\phi^{\pm},k\right),\smallskip\\
\displaystyle Q\left(\Phi_{s}^{\pm}(m),p\right)
&:=&\displaystyle \sum_{k=0}^{p}Q\left(\sec^{m}\phi^{\pm},p-k\right)Q\left(\sin\phi^{\pm},k\right),
\end{array}\right.
\end{equation}
and then, plugging them into Eqs.~(\ref{equA5}) and (\ref{equA6}) gives the reexpansions of $\cos\varepsilon^{\pm}$ and $\sin\varepsilon^{\pm}$ by only modifying the expressions of $Q(\cos\varepsilon^{\pm},n)$ and $Q(\sin\varepsilon^{\pm},n)$, namely,
\begin{equation}\label{equD12}
\left\{\begin{array}{lll}
\displaystyle Q\left(\cos\varepsilon^{\pm},n\right)&:=&\displaystyle (-1)^n+\sum_{s=0}^{n-1}(\pm1)^s\sum_{k=0}^{n-s-1}\sum_{p=0}^{\left[\frac{n-s-k}{2}\right]}
(-1)^{n-s-p-1}
\frac{(2n-2s-2k-2p-3)!!}{2^pp!(n-s-k-2p)!}\smallskip\\
\displaystyle&&\qquad\qquad\qquad\qquad\qquad\qquad\qquad\displaystyle\times Q\left(\Phi_{c}^{\pm}(2n-2s-2k-2p-1),s\right),\smallskip\\
\displaystyle Q\left(\sin\varepsilon^{\pm},n\right)&:=&\displaystyle \sum_{s=0}^{n-1}(\pm1)^s\sum_{k=0}^{n-s-1}\sum_{p=0}^{\left[\frac{n-s-k}{2}\right]}
(-1)^{n-s-p-1}
\frac{(2n-2s-2k-2p-3)!!}{2^pp!(n-s-k-2p)!}\smallskip\\
\displaystyle &&\qquad\qquad\qquad\qquad\qquad\qquad\qquad\displaystyle\times Q\left(\Phi_{s}^{\pm}(2n-2s-2k-2p-1),s\right).
\end{array}\right.
\end{equation}

\end{document}